\documentclass[11pt,a4paper]{article}
\pdfoutput=1
\usepackage{jheppub}
\usepackage{slashed}
\usepackage{blkarray}
\usepackage{leftidx}
\usepackage[makeroom]{cancel}
\usepackage[normalem]{ulem}

\makeatletter
\def\@fpheader{\relax}
\makeatother

\usepackage{tensor}

\usepackage{graphicx}
\usepackage{amsmath,amsfonts,amssymb}
\usepackage{url}
\usepackage{mathabx}
\usepackage{tabularx}
\DeclareMathOperator{\MyProd}{\scalebox{1.4}{$\mathrm{I\kern-0.2ex I}$}}
\DeclareMathOperator{\Tr}{Tr}

\usepackage[dvipsnames]{xcolor}

\usepackage{amsthm}
\usepackage{slashed}
\usepackage{mathtools}
\usepackage{float}
\usepackage{empheq}
\usepackage{bm}
\usepackage{framed}
\usepackage{comment}
\usepackage{xcolor}

\definecolor{Mblue}{rgb}{0.37,0.51,0.71}
\definecolor{Morange}{rgb}{0.88,0.61,0.14}
\definecolor{Mgreen}{rgb}{0.56,0.69,0.19}

\def\ss{\subsection}
\def\sss{\subsubsection}
\def\pg{\paragraph}

\def\nt{\notag}

\def\wt{\widetilde}

\def\R{\mathbb{R}}

\def\cD{\mathcal{D}}

\def\cL{\mathcal{L}}
\def\cM {\mathcal{M}}
\def\cN{\mathcal{N}}
\def\cO{\mathcal{O}}

\def\cQ {\mathcal{Q}}
\def\cR{\mathcal{R}}
\def\cS{\mathcal{S}}

\def\dg {\dagger}
\def\p{\partial}

\def\sD{\slashed{D}}

\def\/{\over}

\def\t{\theta}
\def\s{\sigma}
\def\e{\epsilon}
\def\ve{\varepsilon}

\def\a{\alpha}
\def\b{\beta}
\def\d{\delta}
\def\k{\kappa}
\def\g {\gamma}
\def\la {\lambda}
\def\w {\omega}

\def\l{\ell}
\def\mn{{\mu\nu}}
\def\rs{{\rho\sigma}}

\def\n {\nabla}
\def\L{\Lambda}
\def\D{\Delta}

\def\Om {\Omega}

\def\ra{\rightarrow}

\def\Tr{\mathrm{Tr}}

\def\r{\mathrm}

\def\_{\hspace{2cm}}
\def\-{\\\notag}
\def\={&=&}

\newcommand\be{\begin{equation}}
\newcommand\ee{\end{equation}}

\newcommand{\bea}{\begin{eqnarray}}
\newcommand{\eea}{\end{eqnarray}}

\newcommand{\bpm}{\begin{pmatrix}}
\newcommand{\epm}{\end{pmatrix}}

\newcommand{\bit}{\begin{itemize}}
\newcommand{\eit}{\end{itemize}}

\newcommand{\ben}{\begin{enumerate}}
\newcommand{\een}{\end{enumerate}}

\newcommand\bsp{\begin{split}}
\newcommand\esp{\end{split}}

\def\le{\left}
\def\ri{\right}

\def\bpsi{\bar{\psi}}

\def\bchi{\bar{\chi}}
\def\bla{\bar{\la}}

\def\l{\ell}

\def\qq{\qquad}

\def\cos{\r{cos}}
\def\sin{\r{sin}}

\def\FF{F_\mn F^\mn}

\def\LFull{\ell}
\def\LS{r_0}

\def\hg{\hat{\g}}

\preprint{LCTP-21-31}

\title{
Non-topological logarithmic corrections in minimal gauged supergravity
}

\author[a]{Marina David}

\emailAdd{mmdavid@umich.edu}

\author[b]{Victor Godet}

\emailAdd{victor.godet@icts.res.in}

\author[a,d]{Zhihan Liu}

\emailAdd{billyliu@umich.edu}

\author[a, c]{Leopoldo A. Pando Zayas}

\emailAdd{lpandoz@umich.edu}

\affiliation[a]{Leinweber Center for Theoretical Physics, University of Michigan, Ann Arbor, MI 48109, U.S.A.}

\affiliation[b]{International Centre for Theoretical Sciences (ICTS-TIFR),\\ Tata Institute of Fundamental Research,
Shivakote, Hesaraghatta, Bangalore 560089, India}

\affiliation[c]{The Abdus Salam International Centre for Theoretical Physics, 34014 Trieste, Italy}

\affiliation[d]{Department of Applied Mathematics and Theoretical Physics, University of Cambridge, Wilberforce Road, Cambridge, CB3 0WA, UK}

\abstract{We compute the logarithmic correction to the entropy of asymptotically AdS$_4$ black holes in minimal ${\cal N}=2$ gauged supergravity.  We show that for extremal black holes the logarithmic correction computed in the near horizon geometry agrees with the result in the full geometry up to zero mode contributions, thus clarifying where the quantum degrees of freedom lie in AdS spacetimes. In contrast to flat space, we observe that the logarithmic correction for supersymmetric  black holes can be non-topological in AdS as it is controlled by additional four-derivative terms other than the Euler density. The available microscopic data and results in 11d supergravity  indicate that the full logarithmic correction is topological,  which suggests that the topological nature of logarithmic corrections could serve as a diagnosis of whether a low-energy gravity theory admits an ultraviolet completion.
}


\arxivnumber{}

\begin{document}

\maketitle

\section{Introduction and summary}

One of the central results  of Einstein gravity states that  the  entropy of black holes is universally given by a quarter of the area of the event horizon. When quantum effects are taken into consideration, corrections to the entropy area law arise. Of particular interest is a special type of logarithmic correction which can be computed in the low-energy effective theory (the infrared) and should be reproduced by any candidate ultraviolet complete description of the gravitational theory. Such an infrared window into the black hole microstates was emphasized by Ashoke Sen and collaborators  in the context of asymptotically flat black holes and matched with the microscopic results when available  \cite{Banerjee:2010qc,Banerjee:2011jp, Sen:2012cj, Sen:2012dw, Sen:2012kpz}.
In this paper we explain how to compute such logarithmic corrections to the entropy of asymptotically  AdS$_4$ black holes. Although our methods apply broadly, our main focus here is in the simplest theory that admits supersymmetric black hole solutions: minimal $\cN=2$  gauged supergravity.

There are a number of recent results that precipitate the timing of our investigations. In the context of the AdS/CFT correspondence,  a microscopic foundation for the  entropy of   AdS$_4$ black holes has now been provided. The first explicit computation was performed for static magnetically charged black holes by Benini, Hristov and Zaffaroni in \cite{Benini:2015eyy} (see \cite{Zaffaroni:2019dhb} for a review). More recently,  a microscopic description for the entropy of rotating, electrically charged AdS$_4$ black holes has been provided via the superconformal index \cite{Choi:2019zpz} and by a related localization computation \cite{Nian:2019pxj}. Similar results were obtained in \cite{Bobev:2019zmz,Benini:2019dyp}. These developments built upon the original counting for rotating, electrically charged, asymptotically AdS$_5$ black holes via the superconformal index of ${\cal N}=4$  maximally supersymmetric Yang-Mills \cite{Cabo-Bizet:2018ehj,Choi:2018hmj,Benini:2018ywd} which has also been extended to black holes in AdS$_6$ \cite{Choi:2019miv} and AdS$_7$ \cite{Kantor:2019lfo,Nahmgoong:2019hko}. These impressive results set the stage for precision entropy counting and we hope that this manuscript will serve as a handbook for computations on the gravitational side.

We use the heat kernel method to compute the logarithmic corrections to the entropy of AdS$_4$ black holes. This technique has been applied to obtain the logarithmic corrections for various flat space black holes  and can be applied to extremal as well as non-extremal black holes. Our work builds on previous one-loop results in gauged supergravity, with the work of Christensen and Duff \cite{Christensen:1979iy}, focusing on pure gravity with a non-vanishing cosmological constant, as well as that of Fradkin and Tseytlin \cite{Fradkin:1982bd}, focusing on $O(N)$ supergravity theories. 

We are particularly interested in the result for BPS black holes since microscopic countings are available  only in the BPS case. In flat space, the BPS logarithmic corrections are \emph{topological}, {\it i.e.}, they are controlled by the topological Euler density. As a result, the coefficient of the logarithm is \textit{universal} ~-~a pure number that does not depend on the black hole parameters. This universality property was explored for flat space black holes in \cite{Charles:2015eha,Charles:2017dbr,Castro:2018hsc,Larsen:2018cts}. In this paper, we show that AdS$_4$ black holes have a richer set of logarithmic corrections, which can be non-topological in the BPS case.

\subsection{Summary of results}

In this manuscript,  we take a practical, bottom-up approach to the question of logarithmic contributions in four dimensions.  Our main result is the computation of logarithmic correction in $\mathcal{N}=2$ minimal  gauged supergravity. We also present results for minimally coupled fields as well as for the  Einstein-Maxwell theory with a negative cosmological constant. 

The black hole we are interested is the AdS-Kerr-Newman geometry \cite{carter1968, Plebanski:1976gy, Caldarelli:1999xj}. In the extremal case, we will also consider the near horizon geometry which includes a warped circle fibration over AdS$_2$. Our results also give the logarithmic correction to the free energy of thermal AdS$_4$. They can also be applied to the hyperbolic black hole \cite{Casini:2011kv} from which we obtain the logarithmic corrections to the corresponding entanglement entropy.

The microcanonical entropy of the black hole is given by
\be
S = \frac{A}{4 G} +  C \log \la + \dots~, \qq (\la\to+\infty)
\ee
where $A$ is the area of the horizon and the subleading logarithmic term is the explicit quantum correction we seek. We are interested in the coefficient of the $\log \la$ in the ``isometric'' scaling regime where all length scales (in Planck units) are multiplied by $\la$ and we take $\la\to+\infty$.

The logarithmic correction receives two types of contributions
\be 
C=C_\r{local}+C_\r{global}~.
\ee
The global contribution $C_\r{global}$ is an integer that captures the contribution from the zero modes and from the change of ensemble from canonical to microcanonical. The more interesting local contribution, $C_\r{local}$, receives contributions from the non-zero modes and can be computed using the heat kernel expansion. It is given by an integral over the Euclidean spacetime
\begin{align}
	C_{\text{local}} \equiv \int d^d x \sqrt{ g} \, a_4(x)~,
\end{align}
where the so-called fourth Seeley-DeWitt coefficient is a sum of four-derivative terms
\be\label{a4general}
a_4(x) =  - a_\r{E} E_4  + c\, W^2 + b_1 R^2 + b_2 R \FF,
\ee
evaluated on the background. The backgrounds we consider are solutions of Einstein-Maxwell theory with a negative cosmological constant. Using the equations of motion, a general four-derivative expression such as $a_4(x)$ can always be decomposed in the above basis. The expression of Euler, $E_4$, and the Weyl tensor squared, $ W^2$, are given in \eqref{E4 and W2 definition}. The heat kernel expansion provides a way to compute these coefficients from any two-derivative action using the formula \eqref{intro:a4}. The results are summarized for the theories studied in this paper in Table \ref{tab:results}.

\begin{table}
\centering\arraycolsep=4pt\def\arraystretch{1.4}
\begin{tabular}{|c|c|c|c|c|} 
\hline
Multiplet  & $a_\r{E}$  & $c$  & $b_1$ & $b_2$  \cr
\hline
\hline
{\rm Free scalar {\color{black}$(m\ell=-2)$} } & ${1\/360}$ & ${1\/120}$ & ${1\/288}(\D(\D-3) {\color{black} + 2} )^2$ & $0$ \cr
\hline
{\rm Free {\color{black}Dirac spinor $(m=0)$} } & $-{11\/360}$ & $ {\color{black}-}{1\/20}$ &  ${1\/72} \le( \D-\tfrac32\ri)^2\le(\le( \D-\tfrac32\ri)^2-2 \ri)$  & $0$  \cr
\hline 
{\rm Free vector {\color{black}$(m=0)$}} & ${31\/180} $ & $ {1\/10}$ & $0$  & $0$  \cr
\hline
{\rm Free gravitino {\color{black}$(m=0)$}} & $- {229\/720} $ & $  - {77\/120} $ & $- {1\/9} $  & $0$  \cr
\hline
{\rm Einstein-Maxwell}  & ${53\/45}$ & $ {137\/60} $ & $-{13\/36} $ & $0$  \cr
\hline 
{\rm $\cN=2$ gravitini}  & $-{589\/360}$ & $ -{137\/60}$ & $0$ & $ {13\/18}$  \cr
\hline
{\rm $\cN=2$ gravity multiplet} &  $-{11\/24} $ & $0$  & $-{13\/36} $ & $ {13\/18} $ \cr
\hline
\end{tabular} 
\caption{Results for the Seeley-DeWitt coefficient $a_4$ responsible for the logarithmic corrections. The results for $a_0$ and $a_2$ are given in Table \ref{tab:a0a2} in Appendix \ref{app:renorm}. {\color{black} The spin statistics are included in this table with corresponding $(-1)^F$. Note that the $\mathcal{N}=2$ gravity multiplet includes a gravitino with mass $m\ell=1$ rather than the free gravitino with $m=0$.} }\label{tab:results}
\end{table}

\newpage

Our final result for the Seeley-DeWitt coefficient of minimal $\cN=2$  gauged  supergravity takes the form
\be 
\label{eqn: HKC}
(4\pi)^2a_4(x) = {11\/24} E_4 -{13\/36} R^2 + {13\/18} R \FF~.
\ee
Evaluating this expression on the BPS Kerr-Newman black hole, we obtain
\be\label{ClocalSummary}
    C_\r{local} = {11\/6} -{26\/3}{a(\l^2-4\l-a^2)\/(\l-a)(a^2+ 6 a\l + \l^2)},
\ee
where $a = J/M$ is the rotation parameter and $\ell$ is the AdS$_4$ radius. 
The integer corrections, $C_\r{global}$, are summarized in Table \ref{tab:global}. We observe that the logarithmic correction for a BPS black hole in gauged supergravity has a richer structure than in flat space: {\it the logarithmic correction is non-topological, i.e., its coefficient is not a pure number but depends on black hole parameters.} 


By computing the Seeley-DeWitt coefficient $a_4(x)$ for minimal $\mathcal{N}=2$ gauged supergravity, we find that the non-topological contribution comes from the additional four-derivative terms $R^2$ and $R F_{\mu\nu}F^{\mu\nu}$. In the flat space limit, these terms both vanish and the logarithmic correction becomes topological and gives $C_\r{local} = {11\/6}$. This was shown in \cite{Charles:2015eha,Charles:2017dbr} and is a non-trivial consequence of supersymmetry.

We suspect that the non-topological piece can be interpreted as a contribution from the AdS boundary. It is possible to interpret the logarithmic correction as the Atiyah-Singer index of an appropriate supercharge \cite{Hristov:2021zai}. We surmise that the non-topological term should correspond to the $\eta$-invariant which is a correction due to the presence of a boundary.

We note that according to microscopic computations \cite{Liu:2017vll,Benini:2019dyp,Gang:2019uay,PandoZayas:2020iqr,Liu:2018bac}, we expect the full logarithmic entropy correction to be topological. Such expectation has been confirmed in various 11d supergravity computations \cite{Liu:2017vbl,Benini:2019dyp,Gang:2019uay,PandoZayas:2020iqr}. There is, however,  no contradiction because the 4d minimal gauged supergravity is by itself not the low-energy effective theory of a UV complete theory as matter multiplets, arising from Kaluza-Klein reduction, need to be included. Nonetheless, our result shows that supersymmetry is not enough to guarantee a topological logarithmic correction. This observation suggests that the topological nature of the logarithmic correction could be used to indicate which low-energy theories admit a UV completion\footnote{The possibility of using the topological nature of logarithmic correction for such questions was emphasized to us by Alejandra Castro and was discussed in \cite{AlejandraKITP}.}.

The rest of the paper is organized as follows. We review the heat kernel formalism for the computation of logarithmic corrections to black hole entropy in section \ref{Sec:LogReview} with special emphasis on the scaling limit that we should use in AdS. After presenting the black hole backgrounds and their near horizon geometries in section \ref{Sec:Backgrounds}, we present the main results for the logarithmic corrections in section \ref{Sec:EMLambda} and \ref{Sec:N2Sugra}. We discuss the Einstein-Maxwell theory with a negative cosmological constant in section \ref{Sec:EMLambda}. In section \ref{Sec:N2Sugra} we discuss the corrections for  minimal ${\cal N}=2$  gauged supergravity.  We conclude with a discussion in section \ref{Sec:Discussion}. The topic is intrinsically quite technical and we relegate a number of partial results to a series of appendices to aid the reader interested in technical details.

\section{Logarithmic corrections in AdS$_4$}\label{Sec:LogReview}

In this section, we review logarithmic corrections to black hole entropy and the heat kernel method for their computation \cite{Sen:2012kpz, Sen:2012cj, Sen:2012dw}. This method has been chiefly applied to asymptotically flat  black holes. We also explain how to apply it to asymptotically AdS black holes.

\subsection{Euclidean quantum gravity}

 We consider theories of Einstein gravity in $D$ dimensions coupled to matter fields. We restrict to theories with a scaling property so that purely bosonic terms have two derivatives,  terms with two fermions have one derivative and terms with four fermions have no derivative. This covers a wide range of theories, such as Einstein gravity with minimally coupled scalars, fermions and gauge fields, but also a variety of supergravity theories at a generic point in the moduli space. We also allow for the presence of a  cosmological constant.

We now consider a black hole solution in this theory. To define the quantum entropy of the black hole, we use the fact that this black hole appears as a saddle-point of the Euclidean path integral
\be
Z(\b, \mu_\a) = \int D\Psi \,e^{-S_E(\Psi)}~,
\ee
where $\cS_E $ is the Euclidean action and the integration is done while fixing the temperature $\b$,  thermodynamically conjugate to the mass, $M$, and appropriate chemical potentials $\mu^\a$ associated to the $\r{U}(1)$ charges $q_\a$. 

Upon studying the black hole solutions in this paper, we probe the Euclidean spacetimes via a continuation to imaginary time and analytically continue the action. For the case of the Kerr solutions, these quasi-Euclidean metrics are complex and do indeed give appropriate thermodynamics, see for example \cite{Gibbons:1976ue}. Our computations focus on the small fluctuations around the complex saddle points and we do not expect the subtleties of analytic continuation to affect these quantum corrections. Therefore, for the sake of this paper, we consider these quasi-Euclidean solutions (which we call Euclidean) as is, and leave the subtleties of spacetimes with complex metrics for future study. We simply comment that complex solutions in Euclidean gravity is an evolving subject and we refer the reader to a few examples in the literature \cite{Louko:1995jw, Sorkin:2009ka} as well as more recent discussions on this matter \cite{Kontsevich:2021dmb, Witten:2021nzp}.

The black hole entropy is given by the Legendre transform
\be
S = \log Z+ \beta M + \sum_\a \mu^\a q_\a~.
\ee
At leading order, the classical approximation $\log Z = -S_E^\r{classical}$ is the Euclidean on-shell action. It is a classic  result of Gibbons and Hawking  \cite{Gibbons:1976ue} that the transform  leads to the Bekenstein-Hawking entropy  formula
\be
S = {\r{Area}(q_\a)\/4 G} + \dots~.
\ee
At one-loop order around the saddle-point, we obtain
\be
Z(\b, \mu_\a) \sim {1\/\sqrt{\det \cQ}} e^{-S_E^\r{classical}},
\ee
where $\cQ = {\d^2 S_E \/ \d\Psi^2}$ is the quadratic operator for the fluctuating fields on the background. This expression is divergent and needs to be regulated. The one-loop correction to the black hole entropy is 
\be
\d S = -\frac{1}{2} \log \det \cQ.
\ee

\subsection{Scaling regime}\label{sec:scaling}

The result for the logarithmic correction is highly sensitive to the precise scaling regime we consider. To isolate the logarithmic correction, we  consider a reference configuration with fixed length scales $\l_i^{(0)}$. In the example of AdS-Schwarzschild, these length scales can be taken to be the AdS$_4$ radius $\l$ and the horizon size $r_+$. We then consider a rescaled configuration where all length scales are multiplied by the \emph{same} factor $\la\gg1$: $\l_i = \la \l_i^{(0)}$. We are then interested in the coefficient of $\log\la$ in the one-loop correction to the entropy of the rescaled configuration.

This scaling regime is ``isometric'' because it only magnifies the geometry without deforming it. As a result, the eigenvalues of $\cQ$ are given by
\be \label{logs:eigenchange}
\k_n = \la^2\k_n^{(0)},
\ee
where $\k_n^{(0)}$ are the eigenvalues of the reference configuration. As explained in the next section, this relation is important to ensure that the logarithmic correction depends only on the small $s$ expansion of the heat kernel.

For more general scaling regimes, there will not  be any simple relation between the eigenvalues of the scaled versus reference configuration, because the geometry gets deformed. In this case, the logarithmic correction cannot be computed by the heat kernel expansion and would require knowledge of the heat kernel at general values of $s$. For a background with $k$ independent length scales $\l_1,\dots,\l_k$ (in Planck units), the general logarithmic correction would take the form
\be
S = {A\/4G}+ \sum_{i=1}^k C_i\log\l_i+\dots~,
\ee
with an independent coefficient $C_i$ for each independent length scale $\l_i$. In these terms, the heat kernel expansion can only give us  the sum
\be
C=\sum_{i=1}^k C_i~,
\ee
without being able to access the individual $C_i$. Indeed, $C$ is the coefficient of $\log \la$ if we write $\l_i = \la\l_i^{(0)}$ with $\l_i^{(0)}$ fixed.

Let us now contrast this regime with the flat space regime of \cite{Sen:2012kpz,Sen:2012cj,Sen:2012dw}. In flat space, we do not rescale the mass $m$ of massive fields. As a consequence, it can be shown that massive fields do not contribute to the logarithmic correction  of flat space black holes. In AdS, the prescription is to fix the conformal dimension, or equivalently the combination $m \l$, so we get non-trivial logarithmic corrections for massive fields as a function of the conformal dimension. Clearly we can see that in the flat space limit $\l\to+\infty$, only fields with $m=0$ can contribute and in that limit, we actually reproduce the scaling regime of \cite{Sen:2012kpz,Sen:2012cj,Sen:2012dw}. We indeed see that we reproduce known results for flat space black holes by taking the flat space limit of our results.

It can also be shown that higher loops do not contribute to the logarithmic correction as they are suppressed by positive powers of $\la$  \cite{Sen:2012dw}. Summarizing, the logarithmic correction to the entropy arises only at one-loop from the two-derivative Lagrangian and can be unambiguously computed in the low-energy effective theory.

For extremal black holes, we need to be more careful. In particular,  the thermal circle is infinite which naively makes the Euclidean on-shell action divergent. To obtain a well-defined $\b\to+\infty$ limit, we remove a divergence that can be viewed as an infinite shift in the ground state energy. This can be made precise using the quantum entropy function \cite{Sen:2008vm} in which the quantum entropy is defined using the AdS$_2$/CFT$_1$ correspondence in the near horizon geometry. This procedure was used, for example, in \cite{Banerjee:2010qc, Banerjee:2011jp,Sen:2012cj}.

\subsection{Heat kernel expansion}\label{sec:HKexpansion}

We will now describe the main technical tool which makes possible the exact computation of the logarithmic correction for a variety of black holes: the heat kernel expansion \cite{Vassilevich:2003xt,Fursaev:2011zz,Percacci:2017fkn}.

The one-loop correction to the partition function  decomposes as a contribution $Z_{\text{nz}}$ from the non-zero modes and a contribution $ Z_{\text{zm}}$ from the zero modes of the corresponding kinetic operators, so that we have 
\be
Z_\text{1-loop}(\b,\mu_\a)=  Z_\text{nz}Z_{\text{zm}}\, e^{-\cS_E^\r{class.}}~.
\ee
The one-loop corrected Bekenstein-Hawking entropy, defined in the microcanonical ensemble, takes the form
\be
\label{eqn: BH entropy}
S = {A\/4 G} +  \left( C_{\text{local}} +C_{\text{global}}\right) \log \lambda + \dots~.
\ee
Here $C_\r{local}$ is the local contribution computed using the heat kernel. The global term $C_\r{global}$ is an integer correction due to the zero modes and the change of ensemble from canonical to microcanonical. We now explain how to compute the local contribution. It originates from the non-zero modes
\be
\log Z_\r{nz} = -{1\/2}{\sum_{n}}'\log  \k_n~,
\ee
where $\k_n$ are the eigenvalues of the quadratic operator $\cQ$ and the primed sum runs only over the non-zero eigenvalues $\k_n\neq 0 $. This can be computed by introducing the heat kernel
\be
K(x,s) = \sum_n e^{-\kappa_n s} f_n^\l(x)f_n^{\l'}(x) G_{\l\l'}~,
\ee
where $\{f_n^\l\}$ are the ortho-normalized eigenfunctions of $\cQ$ with eigenvalues $\{\k_n\}$ and $G_{\l\l'}$ is the metric on field space. In particular, we have
\be
\int_\cM d^D x\sqrt{g} \,K(x,s)  = \sum_n e^{-s \k_n}={\sum_n}' e^{-s \k_n} + N_\r{zm}~,
\ee
where  $N_\r{zm}$ is the number of zero modes. We will make use of the relation
\be
\log \k-\log \k^{(0)} = -\lim_{\e\to 0 }\int_{\e}^{\infty} {d s\/s} \le( e^{-s \k }-e^{-s \k^{(0)} }\ri) ~.
\ee
In our scaling regime, the eigenvalues are rescaled according to  \eqref{logs:eigenchange}. This allows us to show that we have
\be\label{Logs:finalintZnz}
\log Z_{\text{nz}}- \log Z_{\text{nz}}^{(0)} = \frac{1}{2} \int_{\e}^{\e \la} \frac{ds}{s} \left(\int_{\cM} d^D x \sqrt{ g}\: K(x,s) - N_{\text{zm}} \right)~.
\ee
The above expression makes it clear that only the range of very small $s$ contributes due to a cancellation between $Z_\r{nz}$ and $ Z_\r{nz}^{(0)}$. We can then use the heat kernel expansion which states the existence of a small $s$ expansion of the form
\be
K(x,s) = \sum_{n\geq 0} s^{n-D/2} a_{2n}(x)
\ee
where $D$ is the dimension of spacetime. The coefficients $a_{2n}(x)$ are known as Seeley-DeWitt coefficients. For smooth manifolds, $a_{2n}(x)$ is a sum of $2n$-derivative terms constructed from the fields appearing in the action \cite{Vassilevich:2003xt}. 

\medskip

We are mainly interested in $D=4$ for which we have
\be
K(x,s) = s^{-2}a_0(x)+ s^{-1}a_2(x)+ s^0 a_4(x) + \cO(s)~.
\ee
We want to compute the $\log \la$ contribution in $\log Z_\r{nz}$. The integral \eqref{Logs:finalintZnz} makes it clear that this comes from the $a_4$ coefficient and we have
\be
\log Z_{\text{nz}}  =C_\r{local} \log \la + \dots~,
\ee
where we have defined
\begin{align}\label{Clocala4}
	C_{\text{local}} \equiv \int d^4 x \sqrt{ g} \, a_4(x)~.
\end{align}
We refer to this as the local contribution as it is given by an integral over spacetime. In general spacetime dimension $D$, $a_4(x)$ should be replaced by $a_D(x)$ in the above formula. Note that this vanishes when $D$ is odd so there is no local contribution in odd-dimensional spacetimes.

The power of the heat kernel expansion lies in the fact that there is a general expression for $a_4(x)$ summarized in \cite{Vassilevich:2003xt}. This allows to compute $C_\r{local}$ without computing the eigenvalues of $\cQ$. 

The other Seeley-DeWitt coefficients $a_0(x)$ and $a_2(x)$ capture one-loop corrections to the cosmological constant, Newton's constant and the other couplings in the Lagrangians. This is discussed, for completeness,  in Appendix \ref{app:renorm}.

\pg{Bosonic fluctuations.} We write the operator of quadratic fluctuations for bosons as
\be\label{Lap}
\cQ^n_m = (\Box) I^n_m + 2 (\w^\mu D_\mu)^n_m + P^n_m~,
\ee
where the Latin indices $m,n$ refer to the different fields and $D_\mu$ is the spacetime covariant derivative. We define $\cD_\mu = D_\mu + \w_\mu$ to complete the square so that
\be
\label{eqn: Lap}
\cQ^n_m = (\cD^\mu \cD_\mu)I^n_m+ E^n_m, \qq
\ee
with
\begin{align}
    \label{eqn: a4E}
    E \equiv  P - \w^\mu\w_\mu - (D^\mu\w_\mu)~.
\end{align}
The Seeley-DeWitt coefficient $a_4(x)$ is then given explicitly by the formula
\bea\label{intro:a4}
(4\pi)^2 a_4(x) \= \Tr\left[{1\over2} E^2 + {1\over6} R E + {1\over 12} \Om_\mn\Om^\mn \ri. \-
&& \hspace{2.5cm}\le.+ {1\over360} I(5 R^2 + 2R_{\mn\rs}R^{\mn\rs} - 2R_\mn R^\mn)\right]~,
\eea
where $\Om_\mn = [D_\mu+\w_\mu,D_\nu+\w_\nu]$ is the curvature associated to the connection $\cD_\mu$. 

\pg{Fermionic fluctuations.} For fermionic fields, the quadratic Lagrangian takes the form  $\cL= \bpsi\cD \psi$ where $\cD = \sD + L$ is a Dirac-type operator and $\psi$ denotes all the fermions of the theory. The prescription is then to use the fact that 
\be
\log \r{det}\,\cD = {1\/2}\log \r{det}\,\cD^\dg\cD,
\ee
so that we can apply the heat kernel method to $\cQ = \cD^\dg\cD$. We have, more explicitly,
\be\label{omegaPforfermions}
    \omega^\mu=\frac{1}{2}\big(\gamma^\mu L-L^\dagger \gamma^\mu\big),\qq  P=\mathcal{R}+\big(\slashed{D}L\big)-L^\dagger L,
\ee
where $\cR =-{1\/4}R$ for spin ${1\/2}$ and $\cR = -{1\/4}g_\mn + {1\/2}\g^\rs R_{\mn \rs}$ for spin ${3\/2}$.

\subsection{Global contribution}

The global contribution consists of an integer correction which is the sum of two contributions
\be \label{Cglobal}
C_\r{global}=C_\r{ens}+C_\r{zm}.
\ee
The first term corresponds to the correction due to changing from the grand canonical to the microcanonical ensemble \cite{Sen:2012dw}.

The zero modes are associated to asymptotic symmetries: gauge transformations with parameters that do not vanish at infinity and are thus, not normalizable. In the path integral, we can treat them by making a change of variable to the parameters of the asymptotic symmetry group. For a field $\Psi$, the Jacobian of this change of variable introduces a factor 
 \be
\lambda^{\b_\Psi}~,
 \ee
 which contributes a logarithmic correction $\b_\Psi \log L$ to the entropy. As a result, the total contribution from the zero modes is
\be
C_{\text{zm}} = \sum_{\Psi} (\b_\Psi -1 ) n_\Psi^0,
\ee
where we are summing over all fields $\Psi$ (including ghosts) and we denote by $n_\Psi^0$  the number of zero modes for $\Psi$. There is a $-1$ because we include here the $-N_{\text{zm}}$ which was in the non-zero mode contribution \eqref{Logs:finalintZnz} (and not included in $C_\r{local}$). The value of $\b_\Psi$ can be computed by normalizing correctly the path integral measure. We refer to \cite{Sen:2012cj} for a more detailed discussion. As an illustration, we report below the values of $\b_\Psi$ for the gauge field, the Rarita-Schwinger field and the graviton in $D$ spacetime dimensions
\be\label{beta}
\b_A = {D\/2}-1 ,\qq\b_{\psi} = D-1,\qq \b_{g} = {D\/2}~. 
\ee

\section{Black hole backgrounds}\label{Sec:Backgrounds}

In this section, we present the background geometries for which we compute the logarithmic corrections. They are solutions of Einstein-Maxwell theory with a negative cosmological constant. 

We give the integrated four-derivative terms as a precursor to the computations of the logarithmic corrections and describe the extremal limit and the near horizon geometry. At this level, we already observe that the local contribution $C_\r{local}$ for extremal black holes is the same in the full geometry and in the near horizon geometry so that the only difference is due to the zero mode contribution.

In the following subsections, we review the metrics of AdS-Schwarzschild, thermal AdS$_4$ and the Reissner-Nordstr\"om AdS$_4$ black hole as simple examples before we consider the general Kerr-Newman AdS$_4$ black hole solution with particular emphasis on its BPS limit. We compute the curvature invariants in both the full solution and the near horizon before giving the general result for the logarithmic corrections to the entropy. The results are written in terms of the theory-dependent coefficients $a_\r{E},c,b_1$ and $b_2$. The computation of these coefficients for the theories of interest will be the subject of subsequent sections.

\subsection{General structure}

The local contribution to the logarithmic correction is given by the Seeley-DeWitt coefficient $a_4(x)$ using \eqref{Clocala4}. For solutions of Einstein-Maxwell-AdS theory, a general four-derivative term can be decomposed as
\be\label{a4general}
(4\pi)^2 a_4(x) =  - a_{\text{E}} \, E_4  + c\, W^2 + b_1 R^2 + b_2 R \FF ~,
\ee
after using the equations of motion for the background fields. 
Here we write the curvature invariants in terms of the Euler density and the Weyl tensor squared given explicitly as
\bea \label{E4 and W2 definition}
     E_4 & \equiv & R_{\mn\rs}R^{\mn\rs}-4 R_\mn R^\mn +R^2~, \\
     W^2 & \equiv &  R_{\mn\rs}R^{\mn\rs} - 2R_\mn R^\mn + {1\/3}R^2 ~.
\eea
Note that the equations of motion implies that $R=4\L = -12/\LFull^2$. The difference with the previous flat space computations lies in the last two terms in \eqref{a4general}, which vanish if $R=0$. These terms are responsible for making the logarithmic correction non-topological. 

To regularize the integral over spacetime, we use the same prescription as in holographic renormalization, which gives an unambiguous finite answer. A consistency check on this procedure is that for the Euler term, the regularized integral gives
\be\label{chiGBC}
\chi  = {1\/32\pi^2} \int d^4 x\sqrt{g} \,E_4,  \qq \text{(regularized)}
\ee
where $\chi$ is the Euler characteristic of spacetime. This is possible because our regularization procedure produces the same boundary as the one appearing in the Gauss-Bonnet-Chern theorem, as we explain in Appendix \ref{app:Euler}. Thus, we see that the logarithmic correction is topological if and only if $a_4(x)$ contains only the Euler term, that is,  $c=b_1=b_2=0$.

\subsection{AdS-Schwarzschild black hole}

\label{subsec: AdSSS}

The Euclidean AdS-Schwarzschild black hole is described by the line element
\begin{equation}
    ds^2=f(r) dt^2+\frac{dr^2}{f(r)}+r^2 d\Om^2,\qq f(r)=1+\frac{r^2}{\LFull^2}-\frac{2 m}{r}~,
\end{equation}
where $m$ is the mass of the black hole and $\LFull$ is the radius of AdS$_4$. Here-forth, Euclidean time is identified with a period proportional to the inverse Hawking temperature $\beta$,
\be
\label{eqn: AdSSST}
t \sim t+\b ,\qq \beta=\frac{4\pi r_+}{1+\frac{3r_+^2}{\LFull^2}}~,
\ee
where $r_+$ is the position of the horizon given by the largest real root of $f(r_+)=0$. The curvature invariants in \eqref{a4general} for this solution are
\begin{equation}
    \label{eqn: SS curvature invariants} 
        E_{4} =\frac{24}{\LFull^4}+\frac{48m^2}{r^6}~,\qq
        W^2 =\frac{48m^2}{r^6}~ ,\qq
        R^2=\frac{144}{\LFull^4}~,\qq R F_{\mu\nu}F^{\mn}= 0~.
\end{equation}
The integrated curvature invariant are divergent due to the infinite volume. To regularize these divergences, we utilize the same prescription as holographic renormalization \cite{Skenderis:2002wp,Natsuume:2014sfa}. Such choice of renormalization is natural given that the logarithmic contributions are corrections to the on-shell action and it allows us to obtain finite and unambiguous results in all cases. A more systematic understanding of this prescription would require a quantum version of holographic renormalization.

The prescription is to impose a cutoff at large $r=r_c$. At the boundary, we add a counter term written in terms of intrinsic data
\be
    a_4^\r{CT} = \int_{\p M} d^3 y \sqrt{h}\, (c_1+ c_2 \cR)~,
\ee
where $\cR$ is the Ricci curvature of the boundary $\p M$. The coefficients $c_1,c_2$ are determined by the requirement that $a_4+ a_4^\r{CT}$ remains finite as we take $r_c\to+\infty$. The regularized integrated invariants take the form
\begin{align}\nt
{1\/(4\pi)^2 }\int d^4 x\sqrt{g}\,E_4 & = 4~, & {1\/(4\pi)^2 }\int d^4 x\sqrt{g} & \,W^2  = {4(\l^2 + r_+^2)^2 \/\l^2(\l^2+3 r_+^2)}~,\\
{1\/(4\pi)^2 }\int d^4 x\sqrt{g}\,R^2 & = {24 r_+^2 (\l^2 - r_+^2) \/\l^2(\l^2+3 r_+^2)}~, &  {1\/(4\pi)^2 }\int d^4 x\sqrt{g} & \,R F_{\mu\nu}F^{\mn}  =0~.
\end{align}
As expected from the Gauss-Bonnet-Chern theorem, the Euler characteristic is
\be
\chi = {1\/32\pi^2}\int d^4x \sqrt{g}\,E_4 = 2.
\ee
In fact, we verify that with the holographic renormalization procedure, the integral of the Euler density is always the Euler characteristic of the spacetime, for all the backgrounds considered in this paper. This suggests that the holographic counterterm reproduces exactly the boundary term comparable to that of the Gauss-Bonnet-Chern theorem. This is evidence that our renormalization procedure is correct and we refer to  Appendix \ref{app:Euler} for details.

The final result for $C_{\text{local}}$ for AdS-Schwarzschild takes the form
\be
C_\r{local} = {4\/\l^2(\l^2+3r_+^2)} \le(  (c-a_\text{E})\l^4 + (2c-3a_\text{E}+6b_1) \l^2 r_+^2 + (c-6b_1)r_+^4\ri)~.
\ee

\subsubsection{Thermal AdS$_4$} \label{subsubsec: thermalads}

We are mainly interested in logarithmic corrections to black hole entropy. However, the dominant saddle-point in the canonical ensemble is not always a black hole in AdS. For temperatures below the Hawking-Page transition \cite{Hawking:1982dh}, it is a thermal AdS. Our computation gives the logarithmic corrections to the free energy of AdS$_4$. The metric of the AdS spacetime with only radiation, {\it thermal AdS}, is given by
\begin{equation} 
    \label{eqn: thermalAdS}
    ds^2=\left(1+\frac{r^2}{\LFull^2}\right)dt^2+\frac{dr^2}{\left(1+\frac{r^2}{\LFull^2}\right)}+r^2 d\Om^2~.
\end{equation}
The curvature invariants for the thermal AdS background read
\be
        E_4 =\frac{24}{\LFull^4},\qq         W^2 = 0,\qq     R^2=\frac{144}{\LFull^4}~,\qq          F_{\mu\nu}F^{\mn}  = 0~.
        \ee
Using the same regularization procedure as above, the integrated invariants all vanish
\begin{align}\nt
{1\/(4\pi)^2 }\int d^4 x\sqrt{g}\,E_4 & = 0~, & {1\/(4\pi)^2 }\int d^4 x\sqrt{g} & \,W^2  = 0~,\\
{1\/(4\pi)^2 }\int d^4 x\sqrt{g}\,R^2 & = 0~, &  {1\/(4\pi)^2 }\int d^4 x\sqrt{g} & \,R  F_{\mu\nu}F^{\mn}  =0~.
\end{align}
This shows that on thermal AdS$_4$, we have $a_4(x) = 0$ so that the local contribution vanishes
\be
C_\r{local} = 0~,
\ee
and the logarithmic correction comes only from the zero mode contribution. Thus we may use $\text{C}_\r{local}$ as an order parameter indicating the Hawking page transition. In the case of Einstein-Maxwell theory, we must include a fixed gauge potential $\Phi$ as thermal AdS \cite{Caldarelli:1999xj,Chamblin:1999hg,Chamblin:1999tk}. Since it is a pure gauge, it does not affect the logarithmic term of the entropy.

\subsection{Reissner-Nordstr\"om black hole}\label{sec: RNBH}

We now turn to the AdS-Reissner-Nordstr\"om black hole and its extremal limit. It is important to note that this black hole is not a BPS solution of minimal gauged supergravity. A non-zero rotation is necessary to solve the BPS equations as we discuss in the next section.

\subsubsection{Non-extremal black hole}

\label{subsubsec: RNAdSFull}

The Euclidean Reissner-Nordstr\"om black hole in AdS is described by
\be
    \label{eqn: RNAdS}
    d s^{2}= f(r) d t^{2}+ {dr^2\/f(r)} +r^{2} d\Om^2~,\qq
     A = \frac{i q_{e}}{r}dt - q_m \cos \, \theta \, d\phi ~.
\ee
with
\be
f(r) = 1+\frac{r^2}{\LFull^2}-{2m\/r}+{q^2_e+q_m^2\/r^2} ~,
\ee
where $m$, $q_e$ and  $q_m$ characterize the mass, the electric charge and the magnetic charge of the black hole, respectively. The horizon $r_+$ is the largest root of $f(r)=0$ and the Hawking temperature is 
\begin{equation}
    T_H =\beta^{-1}= {f'(r_+)\/4\pi} = {1\/4\pi r_+}\le( 1+\frac{3 r_+^2}{\LFull^2}-\frac{(q_e^2+q_m^2)}{r_+^2}\ri)~.
\end{equation}
The curvature invariants are computed to be
 \begin{align}\nt
    R^2 &= \frac{144}{\LFull^4}~, &
    E_4 &=\frac{24}{\LFull^4}+\frac{8 \left(6 m^2 r^2-12 m (q_e^2+q_m^2) r+5 (q_e^2+q_m^2)^2\right)}{r^8}~,   \\
    F_\mn F^\mn &=-\frac{2(q_e^2-q_m^2)}{r^4}~, &
    W^2 &=\frac{48 \left(m r-(q_e^2+q_m^2)\right)^2}{r^8}~.
\end{align}
The integrated invariants can be computed using the same renormalization procedure as described above for the  AdS-Schwarzschild case. The results are
\bea
    \label{eqn: RN curvature invariants}
        {1\/(4\pi)^2} \int d^4 x \sqrt{g}\, E_4 \= 4 ~,\\
       {1\/(4\pi)^2} \int d^4 x \sqrt{g}\,W^2 \= {2\/5}\le(2 - {14 r_+^2\/\l^2} + {16\pi r_+\/\b} + {(\l^4+ \l^2 r_+^2 + 4 r_+^4)\/ \pi \l^4 r_+}\b \ri)~,\\\label{RNR2}
       {1\/(4\pi)^2} \int d^4 x\sqrt{g} \, R^2 \= {12  r_+(r_+^2+\l^2)\/\pi\l^4}\b - {24 r_+^2\/\l^2}~, \\
      {1\/(4\pi)^2}  \int d^4 x \sqrt{g} \,R F_\mn F^\mn \= {6 (3 r_+^4 + \l^2 r_+^2-2 \l^2 q_m^2) \/\pi \l^4 r_+} \b- {24 r_+^2\/\l^2}~.
\eea
The final result for the Reissner-Nordstr\"om black hole takes the following form
\bea
C_\r{local} \= {2\/5}\le( 2( c-5a_\text{E}) - {2 r_+^2\/\l^2}(7c + 30(b_1+b_2)) + {16\pi r_+\/\b}c \ri.\-
&& \hspace{0.4cm} \le.+ {\b\/\pi\l^4 r_+} \Big(c \l^4 + \big(c+30 b_1+15b_2\big) \l^2 r_+^2 + \big(4c + 30 b_1 + 45b_2\big) r_+^4 - 30 b_2 \l^2 q_m^2\Big) \ri)~.
\eea

The appearance of $q_m$ indicates that if the final result has a non-vanishing $b_2$, the logarithmic correction does not preserves the electromagnetic duality. As we will see in section \ref{Sec:N2Sugra}, if we consider $\mathcal{N}=2$ supergravity, we do have a non-trivial $b_2$.

\subsubsection{Extremal limit}

\label{subsubsec: ExtLimit}

The result for the extremal black hole is obtained by taking the $T\to 0$ or $\b\to+\infty$ limit. This limit is naively divergent and we will describe how to implement it in this context. The prescription is as follows. First, the outer horizon is a function of $\beta$, and must be substituted as an explicit expression in terms of $\beta$. We then take the $\beta \to \infty$ limit while keeping the charges fixed and subsequently impose the extremal values of the charges. The low-temperature expansion yields
\be
r_+ = r_0 + {2\pi \l_2^2 \/\b} + O(\b^{-2}),
\ee
where $r_0$ is the position of the extremal horizon and $\l_2$ is the AdS$_2$ radius and can be expressed as
\be\label{r0l2}
r_0^2 = {1\/6}\l(\sqrt{\l^2+12 q^2} - \l)~,\qq \l_2^2 = {r_0^2\/1+ {6 r_0^2\/\l^2}} = {\l^2\/6}\le( 1 -{\l\/\sqrt{\l^2+12 q^2}}\ri)~.
\ee
In the $\b\to+\infty$, we generally have
\be
\int d^4 x\sqrt{g}\,a_4(x) = C_1 \b + C_0 + O(\b^{-1})~.
\ee
The first term, linear in $\b$, is divergent. As this expression is a correction to the effective action, we can interpret this term as a shift of the ground state energy due to one-loop fluctuations. As a result, we ignore this term and define the limit $\b\to+\infty$ to be the constant term $C_0$. The resulting four-derivative terms are
\bea
   \label{RNfourLimit}
    \lim_{\b\to+\infty}    {1\/(4\pi)^2} \int d^4 x \sqrt{g}\, E_4 \= 4 ~,\\
     \lim_{\b\to+\infty}  {1\/(4\pi)^2} \int d^4 x \sqrt{g}\,W^2 \= - {2 ( r_0^2-\l_2^2 )^2 \/ 3r_0^2 \l_2^2 }~,\\\label{RNR2}
   \lim_{\b\to+\infty}    {1\/(4\pi)^2} \int d^4 x\sqrt{g} \, R^2 \=-{2 ( r_0^2-\l_2^2 )^2 \/ r_0^2 \l_2^2}~, \\
   \lim_{\b\to+\infty}   {1\/(4\pi)^2}  \int d^4 x \sqrt{g} \,R F_\mn F^\mn \= -{ (r_0^2-\l_2^2)(r_0^4 +  r_0^2 \l_2^2 - 4q_m^2 \l_2^2) \/r_0^4\l_2^2}~.
\eea
This leads to the final result
\be\label{ClocalRNext}
   \lim_{\b\to+\infty} C_\r{local} = -4\, a_\text{E} -{r_0^2-\l_2^2 \/r_0^2\l_2^2}\le( \le( {2\/3} c + 2 b_1\ri) \Big(r_0^2-\l_2^2\Big) + b_2 \Big(r_0^2+\l_2^2 - {4 \l_2^2 q_m^2\/ r_0^2}\Big) \ri)~.
\ee
Note that in the flat space limit, we have $r_0=\l_2$ and the logarithmic correction is manifestly topological, but such cancellation does not occur for AdS black holes.

\subsubsection{Near horizon geometry}

\label{subsubsec: RNAdSNH}

As we would like to investigate where the quantum degrees of freedom live for asymptotically AdS spacetimes, we compare the basis of curvature invariants of the full solution to that of the near horizon geometry. Let us first consider the extremal black hole. The near horizon geometry can be obtained using the change of coordinates
\begin{align}
    r \to r_0+ \e\, \tilde{r}, \qquad t \to  \l_2^2\,{\tilde{t}\/\e}
\end{align}
and taking the limit $\e\to0$. The result is the AdS$_2\times S^2$ geometry
\be\label{eqn: RNAdSNH}
    ds^2 =\l_2^2 \le(\tilde{r}^2d\tilde{t}^2 + {d\tilde{r}^2\/\tilde{r}^2} \ri) + r_0^2\, d\Omega_2^2 ,\qq
    A = -{i \l_2^2 q_e \/r_0^2} \tilde{r} d\tilde{t} + q_m\, \r{cos}\,\t \, d\phi~,
\ee
where $\l_2$ and $r_0$ are defined in \eqref{r0l2}. For the gauge field, a pure gauge term needs to be added to obtain a smooth $\e\to0$ limit. We can express everything in terms of the two scales $\l_2$ and $r_0$. The AdS$_4$ radius and the extremal charges are given by
\be
{6\/\l^2} = {1\/\l_2^2} -{1\/r_0^2}~,\qq q_e^2+q_m^2 = {r_0^2(r_0^2+\l_2^2)\/2\l_2^2}~.
\ee
In particular we see that we must have $r_0 > \l_2$. Note that in  flat space we obtain $r_0=\l_2$.

The infinite volume of AdS$_2$ is regularized by removing the divergence through a redefinition of the ground state energy in the dual CFT$_1$ \cite{Sen:2008yk,Sen:2008vm}. This leads to a regularized volume of unit AdS$_2$ which is $-2\pi$. The integrated invariants can then be computed  and we find
\be
C_\r{local} = -4 a_\text{E} -{r_0^2-\l_2^2 \/r_0^2\l_2^2}\le( \le( {2\/3} c + 2 b_1\ri) \Big(r_0^2-\l_2^2\Big) + b_2 \left(r_0^2+\l_2^2- {4 \l_2^2 q_m^2\/ r_0^2}\right) \ri)~.
\ee
This expression matches the result \eqref{ClocalRNext} obtained by taking the $\b\to+\infty$ limit of the non-extremal $C_{\text{local}}$. Hence, the computation of $C_\r{local}$ for an extremal black hole can be done either in the full geometry or  in the near horizon region. The difference in logarithmic correction between the full geometry and the near horizon geometry come exclusively from zero modes.

\subsection{Kerr-Newman black hole}\label{sec:KN}

We now turn to the AdS-Kerr-Newman black hole \cite{Plebanski:1976gy,Caldarelli:1999xj}. This solution is particularly interesting because it has a regular BPS limit unlike the Reissner-Nordstr\"om black hole \cite{Romans:1991nq,Kostelecky:1995ei,Hristov:2010ri}. 

\subsubsection{Non-extremal black hole}

As given in \cite{Caldarelli:1999xj}, the line element takes the form , 
\begin{align}
    d s^{2}=-\frac{\Delta_{r}}{\rho^2}\left(d t-\frac{a~\sin ^{2}\theta}{\Xi}   d \phi\right)^{2}+\frac{\rho^2 d r^{2}}{\Delta_{r}}+\frac{\rho^2 d \theta^{2}}{\Delta_{\theta}}+\frac{\Delta_{\theta} \,\sin ^{2} \theta}{\rho^2}\left(a \,d t-\frac{r^{2}+a^{2}}{\Xi} d \phi\right)^{2},
\end{align}
where we have defined
\be
\Delta_{r} = \left(r^{2}+a^{2}\right)\left(1+{r^2\/\l^2}\right)-2 m r+q_{e}^{2}+q_{m}^2, \qq \Delta_{\theta} = 1-{a^2\/\l^2}\,\cos ^{2} \theta~,
\ee
with $ \rho^2 = r^{2}+a^{2} \cos ^{2} \theta$ and $\Xi=  1-{a^2\/\l^2}$. The gauge field is given by
\begin{align}
    A =-\frac{q_e r}{\rho^{2}}\left(d t-\frac{a\, \sin ^{2} \theta}{\Xi} d \phi\right)-\frac{q_m\,\cos\,\theta}{\rho^2}\left(adt-\frac{r^2+a^2}{\Xi}d\phi\right)~,
\end{align}
The parameters satisfy $a^2 <\l^2$ and we take $a\geq 0$ without loss of generality.\footnote{The general result is obtained by replacing $a\ra|a|$ everywhere.} The physical mass $M$, angular momentum $J$, electric charge $Q_e$ and magnetic charge $Q_m$ are given by
\begin{align}
    M = \frac{m}{\Xi^2}, \qquad J = \frac{a m}{\Xi^2}, \qquad Q_e = \frac{q_e}{\Xi},\qquad Q_m=\frac{q_m}{\Xi},
\end{align}
and the inverse temperature is
\begin{align}
\beta=\frac{4 \pi\left(r_{+}^{2}+a^{2}\right)}{r_{+}\left(1+\frac{a^{2}}{\LFull^{2}}+3 \frac{r_{+}^{2}}{\LFull^{2}}-\frac{a^{2}+q^{2}_{e}+q^2_{m}}{r_{+}^{2}}\right)}.
\end{align}
For the non-extremal black hole, the general form is
\be
C_\r{local} = -4 \,a_\r{E}+ (6 A_1+c W_1) \b + (24 A_2+cW_2) + {c\, W_3\/\b}~,
\ee
where the logarithmic corrections depends on five independent parameters $\{r_+,\b,\l,a,q_m\}$. The Euler term simply gives a pure number in agreement with the formula
\be
\chi = {1\/32\pi^2} \int d^4 x\sqrt{g} E_4 = 2~.
\ee
The expressions $A_i$ and $W_i$ are independent of $\b$ and take the form
\begin{align}
    \begin{split}
    A_1 &={  (2b_1+b_2)(a^2+\l^2)r_+^3 +  (2 b_1+3b_2) r_+^5 + ( (2b_1-b_2)a^2 - 2 b_2 q_m^2) r_+ \l^2 \/ \pi\l^2(\l^2-a^2) (a^2+r_+^2)}~,
    \\
    A_2 &=-{b_1 a^2 + (b_1+b_2)r_+^2 \/\l^2-a^2} ~,
    \end{split}
\end{align}
and we have isolated the contribution $W_i$ from the Weyl squared term, explicitly given as
\bea\nt
W_1\= {1\/16\pi a^5 r_+^4\l^2 (\l^2-a^2)(a^2+r_+^2)}\le[ 3 a r_+(a^8(\l^2-r_+^2)^2 + r_+^8 (\l^2+3 r_+^2)^2) \ri. \-
&&  - 4 a^3 r_+^3 (r_+^4(\l^4 - 9r_+^4) + a^4 (\l^4 + 12 \l^2 r_+^2 + 3 r_+^4)  +2 a^5 r_+^5(\l^4 - 14\l^2 r_+^2 + 5 r_+^4) \-
&& \le. -3 (a^2+ r_+^2) (a^2(\l^2-r_+^2) - r_+^2(\l^2+ 3 r_+^2) )^2  (r_+^4-a^4) \,\r{arctan}(a/r_+) \ri]~,\\
W_2 \= {a^2 + r_+^2\/ 2 a^5 r_+^3 (\l^2-a^2)}\le[ 4 a^3\l^2 r_+^3 + 3 a r_+ (a^4(\l^2-r_+^2) - r_+^4(\l^2+3 r_+^2)) \ri. \-
&&\hspace{2.6cm} \le. -3(a^2(\l^2-r_+^2)-r_+^2(\l^2+ 3 r_+^2)) (r_+^4-a^4) \,\r{arctan}(a/r_+) \ri]~,\-
W_3 \= {\pi \l^2 (a^2 + r_+^2)\/  a^5 r_+^2 (\l^2-a^2)}\le[ a r_+ (3 a^4 + 2 a^2 r_+^2 +3 r_+^4 )-3 (r_+^2+a^2) (r_+^4-a^4) \,\r{arctan}(a/r_+)\ri]~.
\eea
We have checked that we reproduce the Reissner-Nordstr\"om results of section \ref{sec: RNBH} in the limit $a=0$.


\sss{Extremal limit}

As was done in the Reissner-Nordstr\"om black hole, the extremal limit can be found by taking the limit $T \to 0$ or $\b\to+\infty$ while keeping the charges fixed. To do this appropriately, we use that for  small temperatures
\be
r_+ = r_0 + {2\pi \l_2^2 \/\b} + O(\b^{-2})~,
\ee
and we take the $\b\to+\infty$ limit while keeping $r_0,\l,a ,q_m$ fixed. The procedure yields a finite piece in $\beta$ as well as a piece linear in $\beta$, which can be removed by a renormalization of the ground state energy. The final result can be written in terms of the four independent parameters $\{r_0,\l,a,q_m\}$. It takes the form
\bea \nt
C_\r{local} \= -4 a_\r{E} + {1\/2 a r_0^5 (\l^2-a^2) (a^2+r_0^2)(a^2+\l^2+6 r_0^2)} \Big[ - 3 a^7 r_0 (16 b_1 r_0^4 + c(\l^2- r_0^2)^2). \-
&& \hspace{1.3cm}+ a^5 r_0^3(c\l^4 + 2( 11c -12 b_2) \l^2 r_0^2 - 3(13 c-8b_2+80 b_1) r_0^4) \-
&&  \hspace{1.3cm}+a^3 r_0^5 (15 c\l^4  + 2 (25 c + 24 b_2) \l^2 r_0^2 -(49 c +336 b_1-48b_2) r_0^4- 48 b_2 \l^2 q_m^2) \-
&&  \hspace{1.3cm}+ 3 a r_0^7 (c\l^4 + 2(3c-4 b_2) \l^2 r_0^2 -(7c+48b_1 + 24b_2) r_0^4 + 16 b_2 \l^2 q_m^2) \\\label{ClocalExtKN}
&& \hspace{1.3cm} - 3 c(a^2 + r_0^2) (a^2 (r_0^2-\l^2)+ r_0^2(\l^2+3 r_0^2))^2 \,\r{arctan}(a/r_0) \Big] ~.
\eea 
We can also compare with the computation performed in the near horizon geometry obtained via
\be\label{toNHGKN}
r \ra r_0 + \e\, \tilde{r},\qq t \ra \l_2^2 \,{\tilde{t}\/\e} ,\qq \phi\ra \phi - { i a \l_2^2(\l^2-a^2)\/\l^2(a^2+r_0^2)}\, {t\/\e}~,
\ee 
while taking $\e\to0$. This leads to
\bea
d\tilde{s}^2 \= {\l_2^2  (r_0^2 + a^2\,\r{cos}^2\t)\/ a^2 + r_0^2} \le(\tilde{r}^2 d\tilde{t}^2 + {d\tilde{r}^2\/\tilde{r}^2} \ri) + {\l^2 (r_0^2 + a^2\,\r{cos}^2\t)\/\l^2 - a^2\,\r{cos}^2\t} d\t^2 \-
&& + {\l^2 (a^2+r_0^2)^2 (\l^2 -a^2 \,\r{cos}^2\t)\,\r{sin}^2\t \/(\l^2-a^2)^2 (r_0^2 + a^2\,\r{cos}^2\t)} \le(d\phi -{2\l_2^2 a r_0(\l^2-a^2)\/ \l^2 (a^2+r_0^2)^2} \,i r dt \ri)^2~,
\eea
where the AdS$_2$ radius is
\be\label{l2KN}
\l_2 = \l \sqrt{a^2+r_0^2\/a^2+\l^2 + 6 r_0^2}~.
\ee
The near horizon geometry is a warped version of AdS$_2$ with a circle fiber, similar to the near horizon of extreme Kerr (NHEK), which we recover in the appropriate limit. The near horizon gauge field takes the form
\bea
\wt{A} \= {1\/r_0^2 + a^2\,\r{cos}^2\t} \le[ -{i\l_2^2\/a^2 +r_0^2} (q_e (r_0^2 -a^2\,\r{cos}^2\t) + 2 q_m \,a r_0\,\r{cos}\,\t) \tilde{r} d\tilde{t} \ri.\-
&& \hspace{5cm}\le. + {\l^2\/\l^2-a^2}( q_e\, a r_0 \,\r{sin}^2\t + q_m (a^2+r_0^2)\,\r{cos}\,\t)d\tilde\phi \ri]~.
\eea
We can perform more general near horizon limits by taking at the same time a near-extremal limit. Instead of setting $q_e = q_e^\ast$, we can consider a deformation $q_e = q_e^\ast + \d q_e \e^2$ parametrized by the same $\e$ as in \eqref{toNHGKN}. Moreover, keeping subleading corrections in $\beta^{-1}$ would yield corrections to the entropy in the near-extremal regime. The non-zero energy associated to this large diffeomorphism can be understood in terms of the Schwarzian action of Jackiw-Teitelboim gravity \cite{Maldacena:2016upp}.

We are now in a position to compute the logarithmic corrections in the near horizon geometry and we find that the result is equal to \eqref{ClocalExtKN} obtained by taking the extremal limit appropriately, {\it i.e.}, fixing the charges while taking $\beta \to +\infty$. Thus, the local contribution is the same in the near horizon region and the full geometry.

\sss{BPS limit}
\label{subsubsec: AdSKNBPS}

The BPS limit can be obtained by imposing the additional conditions to the extremal black hole 
\be
r_0 = \sqrt{a \l},\qq q_m=0~.
\ee
The resulting black hole preserves half of the supersymmetries \cite{Kostelecky:1995ei}. Its charges are given by
\be
M =\frac{\sqrt{a\LFull }}{\left(1-\frac{a}{\LFull}\right)^{2}}, \qquad Q_e=\frac{\sqrt{a\LFull }}{1-\frac{a}{\LFull}}, \qq Q_{m}=0, \qq J=\frac{a\sqrt{a\LFull}}{\left(1-\frac{a}{\LFull}\right)^{2}}
\ee 
and it satisfies a BPS bound:
\begin{align}
M=Q_e+\frac{J}{\LFull}~.
\end{align}
The BPS result can be written in terms of the two independent parameters $\l$ and $a$
\bea\nt
C_\r{local} \= - 4 a_\r{E} + {3\l_2^2\/2 a\l^2(\l^2-a^2)} \le[ (9c-8b_2) a \l^2 - (9c +48 b_1- 8b_2)a^2 \l - (c+16 b_1)a^3\vphantom{c (a+\l)^4\/\sqrt{al}} \ri.\\\label{ClocalBPS}
&&\hspace{6cm}\le.+ c\l^3 - {c (a+\l)^4\/\sqrt{al}} \,\r{arctan}(\sqrt{a/\l})\ri],
\eea
where the AdS$_2$ radius given in \eqref{l2KN} is
\be
\l_2= \l\sqrt{a(a+\l)\/a^2+ 6 a\l+\l^2} \qq \text{(BPS case)}~.
\ee
It is clear from this formula that there is no non-rotating BPS solution as the limit $a\to 0$ is singular.

\subsection{AdS-Rindler geometry}\label{sec:hypBH}

Our computation of the logarithmic correction can also be applied to the so-called hyperbolic black hole of \cite{Casini:2011kv}, \emph{i.e.}, the AdS$_4$-Rindler geometry. The entropy of this black hole is the entanglement entropy 
\be
S_\r{EE} = -\Tr\,\rho_A \log\rho_A,
\ee
associated to a ball-shaped boundary subregion $A$. Here $\rho_A$ is the reduce density matrix defined by tracing over the complement $\bar{A}$ 
\be
\rho_A = \Tr_{\bar{A}} |0\rangle\langle 0|,
\ee
where $|0\rangle$ is the global vacuum. Here the only length scale is the AdS$_4$ radius $\l$ so we are considering the regime of large $\l$ and computing
\be
S_\r{EE} = {\r{Area}\/4 G}+ (C_\r{local}+C_\r{zm}) \log \l+ \dots .
\ee
The geometry of the hyperbolic black hole is given by
\be
ds^2 = \le( {\rho^2\/\l^2} - 1\ri)dt^2 + {d\rho^2\/{\rho^2\/\l^2}-1 }+\rho^2 ds^2_{H_2}~,\qq ds^2_{H_2}=du^2 + \r{sinh}^2 u\, d\phi^2,
\ee
where $\rho \geq \l$, $u\geq 0$ and $t\sim t+\b$. The inverse temperature is given by
\be
\b=2\pi \l~.
\ee
We regularize the integral over spacetime using holographic renormalization. In this case, there is also a divergence coming from the volume of $H_2$ and we take a regulator such that $\r{vol}(H_2) = -2\pi$. The integrated four-derivative invariants are given by
\begin{align}\nt
{1\/(4\pi)^2 }\int d^4 x\sqrt{g}\,E_4 & = 2~, & {1\/(4\pi)^2 }\int d^4 x\sqrt{g}\,W^2 & = 0~,\\
{1\/(4\pi)^2 }\int d^4 x\sqrt{g}\,R^2 & = 12~, &  {1\/(4\pi)^2 }\int d^4 x\sqrt{g}\,R F_\mn F^\mn & =0~.
\end{align}
This implies that we have
\be
C_\r{local} = -2 a_\text{E}+ 12 b_1~.
\ee
Note that the Gauss-Bonnet-Chern theorem gives
\be
\chi = {1\/32\pi^2 }\int d^4 x\sqrt{g}\,E_4  = 1~,
\ee
as expected since $M$ is topologically $D_2\times H_2$ where $D_2$ is a disk and we have $\chi(M) = \chi(D_2)\chi(H_2) = 1$ since $\chi(D_2)=\chi(H_2)=1$. This is a non-trivial consistency check for our regularization procedure.

\subsection{Global contribution}

We now compute the global contribution \eqref{Cglobal} which comes from the zero modes and the change of ensemble.  The results are  summarized in Table \ref{tab:global}. In the full geometry, the contribution from the bosonic zero modes in the full  asymptotically AdS$_4$ geometry vanishes  \cite{Sen:2012dw}. Indeed, the fact that AdS$_4$ admits a 2-form zero mode follows from the general result of Camporesi and Higuchi who established that AdS$_{2M}$ admits a M-form zero mode \cite{CAMPORESI199457}. This 2-form zero mode is central in generating the logarithmic correction in asymptotically AdS$_4$ backgrounds embedded in eleven-dimensional supergravity \cite{Bhattacharyya:2012ye,Liu:2017vbl}. However, in the four-dimensional theories we consider in this manuscript, there is no contribution from such a 2-form zero mode. 

Hence we have
\be
C_\r{zm}=0\qq \text{(full geometry)}~.
\ee
In the near horizon geometry, additional zero modes come from the AdS$_2$ factor. The metric contributes $-3$ zero modes. In the near horizon geometry of BPS black holes, we  also have $8$ fermionic zero modes. The zero mode contribution for extremal black holes in the near horizon geometry is then given by
\be
C_\r{zm}= - 3 + 8\, \d_\r{BPS}\qq \text{(near horizon geometry)},
\ee
where $\d_\r{BPS}=1$ in the BPS case and $0$ otherwise. It is interesting to observe that this contribution can be interpreted in the context of nearly AdS$_2$ holography \cite{Maldacena:2016upp}. The asymptotic symmetry group of AdS$_2$ is $\r{Diff}(S^1)/\r{SL}(2,\R)$. Upon a choice of configuration, the number of broken symmetries is $n_0 = +\infty - 3$, the infinite piece being absorbed in a renormalization of the energy. So the $-3$ zero modes come from the unbroken $\r{SL}(2,\R)$ symmetry of AdS$_2$. A similar argument for BPS black holes explains the $8$ fermionic zero modes as arising from the eight fermionic generators of the $\r{PSU}(1,1|2)$ near horizon symmetry. These patterns of symmetry breaking can be studied using Jackiw-Teitelboim gravity \cite{Almheiri:2014cka,Maldacena:2016upp, Nayak:2018qej,Heydeman:2020hhw,Iliesiu:2020qvm,Castro:2019crn,Castro:2021csm,Castro:2021wzn}.

We also include in $C_\r{global}$ the correction that comes from the change of ensemble from canonical to microcanonical. Following \cite{Sen:2012dw}, the change of ensemble gives a contribution
\be
C_\r{ens}= -K  ~,
\ee
where $K$ is the number of rotational symmetries of the black hole.

\begin{table}
	\centering\arraycolsep=4pt\def\arraystretch{1.2}
\begin{tabular}{|l|c|c||c|}
	\hline
Background spacetime &	$C_\r{zm}$ 	& $C_\r{ens}$ & $C_\r{global}$  \\\hline
	Schwarzschild & 0 & $-3$ & $-3$\\
	Reissner-Nordstr\"om  & 0 & $-3$ & $-3$\\
	Kerr  & 0 & $-1$ & $-1$\\
	Kerr-Newman  & $0$ & $-1$ & $-1$\\
	BPS Kerr-Newman & 0& $-1$& $-1$\\
	Reissner-Nordstr\"om near horizon  & $-3$ & $-3$ & $-6$\\
	Kerr-Newman near horizon & $-3$ & $-1$ & $-4$ \\
	BPS Kerr-Newman near horizon & $5$ & $-1$ & $4$ \\\hline
	Thermal AdS$_4$  & $0$ & $-3$ & $-3$ \\
	AdS$_4$-Rindler & $0$ & $-3$ & $-3$ \\
	\hline

\end{tabular}\caption{Global contribution to the logarithmic correction.}\label{tab:global}
\end{table}

\section{Minimally coupled matter}

To obtain the logarithmic corrections, we need to compute the coefficients $a_\r{E},c,b_1,b_2$ that appear in the general expression \eqref{a4general}. Our ultimate aim is to evaluate logarithmic corrections in theories that can arise as consistent low-energy truncations from string and M-theory. However, in the next sections, we compute these logarithmic corrections in Einstein-Maxwell theory with a negative cosmological constant and in minimal $\cN=2$  gauged supergravity. As a warm-up, we also present the logarithmic corrections to AdS black holes due to minimally coupled fields, as was done for flat space black holes in \cite{Sen:2012dw}.

\subsection{Minimal theories}

In this subsection, we compute $C_\r{local}$ for minimal scalars, fermions, vectors and gravitini.

\pg{Free scalar.}

We consider a scalar field of mass $m$ described by the action
\be
S = -{1\/2}\int d^4 x\sqrt{g}
\le((\p\phi)^2+m^2\phi^2 \ri).
\ee
The result for a scalar field is obtained by setting 
\be
P=E=m^2, \qq\Om=0~,
\ee
in equation \eqref{intro:a4}. As explained in section \ref{sec:scaling}, we consider a regime where every length scales with a factor $\la$. So here $m$ scales as $\la^{-1}$ and what is fixed is the  conformal dimension
\be
\D = {1\/2}\le(3+\sqrt{9+4m^2\l^2}\ri)~.
\ee
This is to be contrasted with flat space where massive fields do not contribute to the logarithmic correction as explained in \cite{Sen:2012kpz,Sen:2012cj,Sen:2012dw}. 

The heat kernel takes the form
\be
(4\pi)^2 a_4(x) = -{1\/360}E_4 +{1\/120}W^2+{1\/288}(\D(\D-3)-2)^2 R^2~.
\ee
The explicit result for $C_\r{local}$ can be obtained using \eqref{Clocala4} and \eqref{a4general}. We report the result for the extremal black hole 
\be
C_\r{local} = -{1\/90} - { \LS^2\/20 \l^4} (24 + 5(\D+1)\D(\D-3)(\D-4))~.
\ee

\pg{Free fermion.}

We consider a free Dirac fermion with Euclidean  action
\be
S =\int d^{d+1}x \sqrt{g}\,\bar\psi\le(   \g^\mu\n_\mu -m\ri) \psi
\ee
This is dual to an operator with scaling dimension \cite{Henningson:1998cd}
\be
\D = {3\/2} + m\l.
\ee
The result is
\be
(4\pi)^2a_4(x) = {11\/360} E_4 + {1\/20} W^2 +{1\/72} \le( \D-\tfrac32\ri)^2\le(\le( \D-\tfrac32\ri)^2-2 \ri)R^2.
\ee

\pg{Free vector.}

We now consider a free Maxwell field $a^\mu$ with the Lagrangian
\be
\cL = -{1\/4}f_\mn f^\mn,
\ee
where $f_\mn = \n_\mu a_\nu - \n_\nu a_\mu$. We add the gauge fixing term
\be
\cL_\r{g.f.} = {1\/2} (\n_\mu a^\mu)^2 ,
\ee
so that the total Lagrangian becomes
\be
\cL + \cL_\r{g.f.} = a^\mu \Box a_\mu - a^\mu R_\mn a^\nu~.
\ee
The gauge-fixing induces two massless scalar fields with fermionic statistics. We obtain the result
\be
(4\pi)^2 a_4(x) = -{31\/180} E_4 + {1\/10}W^2~.
\ee

\pg{Free Rarita-Schwinger field.}

We consider here a Majorana spin-${3\/2}$ field described by the Lagrangian
\be
\cL_{3/2} = -\bar\psi_\mu \g^{\mu\rho\nu} \n_\rho \psi_\nu.
\ee
We use the gauge-fixing condition $\g^\mu\psi_\mu=0$. This is implemented with the gauge-fixing term
\be
\cL_\text{g.f} = -{1\/2}(\bar\psi_\mu \g^\mu) \g^\rho \n_\rho (\g^\nu \psi_\nu)~,
\ee
so that the total Lagrangian is
\be
\cL_{3/2} + \cL_\text{g.f}=  \bchi_\mu \g^\nu D_\nu \chi^\mu,
\ee
after using the field redefinition $\psi_\mu = \chi_\mu - {1\/2} \g_\mu \g^\nu\chi_\nu$. The gauge-fixing leads to three Majorana ghosts which are free massless fermions. We refer the reader to section \ref{sec:FPfermions} for details on the gauge-fixing procedure. Hence, we find
\be
(4\pi)^2 a_4(x) = {229\/720} E_4- {77\/120}W^2 - {1\/9}R^2.
\ee

\subsection{Logarithmic corrections}

The results for minimally coupled scalars are summarized in Table \ref{tab:results}.

\subsubsection{Massless fields}

For massless fields, we can present the result as
\bea
a_\text{E} \= {1\/720}( 2 n_S + 22 n_F+124 n_V -229 n_\psi)~,\\
c\= {1\/120}( n_S+6n_F + 12 n_V- 77 n_\psi)~,\\
b_1 \= {1\/72}(n_S - 8 n_\psi)~.
\eea
where $n_S,n_F,n_V,n_\psi$ the number of scalars, spin-${1\/2}$ Majorana fermion, vector and gravitini. The result for AdS-Schwarzschild takes the form
\bea
C_\r{local}\= {1\/180\l^2(\l^2 + 3r_+^2)} \le[ \l^4 (4 n_S + 14 n_F-52 n_V -233 n_\psi)\ri. \-
&& \le.+ 3 r_+^2\l^2 (22 n_S+2 n_F- 76 n_V-239 n_\psi) + 18  r_+^4 (-3 n_S +2 n_F+4n_V +n_\psi) \ri].
\eea
It is easily seen that in the flat space limit, we have
\be
\lim_{\l\to+\infty}C_\r{local} = {1\/180} (4 n_S + 14 n_F-52 n_V -233  n_\psi).
\ee
which reproduces the results of \cite{Sen:2012dw}.

\subsubsection{Corrections to entanglement entropy}
Our result can also be applied to compute logarithmic correction to entanglement entropy. We consider a ball-shaped region $A$ in the boundary. The entanglement entropy of $A$ is given by the area of the hyperbolic black hole discussed in section \ref{sec:hypBH}. The logarithmic corrections to entanglement entropy are given by
\be
S_\r{EE} = {\r{Area}\/4G} + C\, \log \b+\dots .
\ee
The contribution of a minimal scalar field of conformal dimension $\D$ gives
\be\label{CEEscalar}
C = {29\/180} + {1\/24}(\D+1)\D(\D-3)(\D-4)~.
\ee
We have here $C=C_\r{local}$ since there is no zero mode for the scalar field. Quantum corrections to entanglement entropy can also be interpreted in terms of bulk entanglement entropy \cite{Faulkner:2013ana}. It would be interesting to see if we can understand \eqref{CEEscalar} as the logarithmic piece of the bulk entanglement entropy of a scalar field in the Rindler wedge.

\section{Einstein-Maxwell-AdS theory}
\label{Sec:EMLambda}

We now consider Einstein-Maxwell theory with a negative cosmological constant. This is the minimal theory that contains the AdS-Kerr-Newman black hole and is the bosonic part of  minimal $\cN=2$  gauged supergravity studied in section \ref{Sec:N2Sugra}. The action is given by
\begin{align}
    \label{EM action}
    S=\int d^{4} x \sqrt{ g}\left(R-2\Lambda-F_{\mu \nu} F^{\mu \nu}\right),
\end{align}
where $F_{\mu\nu}=\partial_\mu A_\nu-\partial_\nu A_\mu$ is the field strength with $A_\mu$ the gauge potential. Note that we find it convenient to use a convention $4\pi G=1$.

The computation is easily performed using the algorithm described in Appendix \ref{app:algo}. We have also performed an independent computation by hand, as detailed in Appendix \ref{app:bosons}.

\subsection{Bosonic fluctuations}
\label{subsec: bosonquadexpansion}

We consider variations of the metric and gauge field
\be
\d g_\mn =\sqrt{2} h_\mn~,\qq \d A_\mu = \frac{1}{2} a_\mu~,
\ee
where $h_{\mu\nu}$ and $a_\alpha$ are the graviton and graviphoton,  respectively. We impose a particular gauge to the theory by adding a suitable gauge-fixing Lagrangian 
\begin{align}
\label{eqn: EMgf}
S=    -\int d^{4} x \sqrt{\operatorname{det} g}\left\{\left(D^{\mu} h_{\mu \rho}-\frac{1}{2} D_{\rho} h\right)\left(D^{\nu} h_{\nu}^{\rho}-\frac{1}{2} D^{\rho} h\right)+\frac{1}{2}\left(D^{\mu} a_{\mu}\right)\left(D^{\nu} a_{\nu}\right)\right\},
\end{align}
and the corresponding ghost action to the action \eqref{EM action}. We then expand the action up to quadratic order. The linear order variation yields the equation of motion for the background fields 
\begin{align}
    \label{eqn: gravityEoM}
    R_{\mu\nu}-\frac{1}{2}g_{\mu\nu}R+g_{\mu\nu}\Lambda=&2F_{\mu\rho}F_\nu^{~\rho}-\frac{1}{2}g_{\mu\nu}F_{\alpha\beta}F^{\alpha\beta}~,\\
    \label{eqn: gaugeEoM}
    D^\mu F_{\mu\nu}=&0~.
\end{align}
Note that the equations of motion implies that $R=4\L = -12/\LFull^2$. It is also worth mentioning the Bianchi identity for the gravitational field and gauge field
\begin{align}
    \label{eqn: gravitybianchi}
    D_{[\mu}F_{\nu\rho]}=&0~,\\
    \label{eqn: gaugebianchi}
    R_{\mu[\nu\rho\sigma]}=&0~,
\end{align}
as they serve as handy tools for simplifying our calculations. The quadratic action can be put in the canonical form \eqref{eqn: Lap}. The details can be found in appendix \ref{appendix:fluctuations} where we present the explicit form of the quadratic fluctuations. This allows us to extract the matrices $I$, $E$ and $\Om$:
\begin{align}
        \label{eqn: a4BI}
        \phi_{m}I^{mn}\phi_{n}=& h_{\mu \nu}\left(\frac{1}{2}g^{\mu\alpha}g^{\nu\beta}+\frac{1}{2}g^{\mu\beta}g^{\nu\alpha}-\frac{1}{2}g^{\mu\nu}g^{\alpha\beta}\right)h_{\alpha\beta}+a_\alpha g^{\alpha\beta}a_\beta,\\
    \begin{split}
        \label{eqn: a4BE}
        \phi_{m} E^{m n} \phi_{n}=& h_{\mu \nu}\left(R^{\mu \alpha \nu \beta}+R^{\mu \beta \nu \alpha}-g^{\mu \nu} R^{\alpha \beta}-g^{\alpha \beta} R^{\mu \nu} + \Lambda g^{\mu\nu}g^{\alpha\beta} \right) h_{\alpha \beta} \\
        &+a_{\alpha}\left(\frac{3}{2} g^{\alpha \beta} F_{\mu \nu} F^{\mu \nu} - \Lambda g^{\alpha\beta}\right) a_{\beta}+\frac{\sqrt{2}}{2} h_{\mu \nu}\left(D^{\mu} F^{\alpha \nu}+D^{\nu} F^{\alpha \mu}\right) a_{\alpha} \\
        &+\frac{\sqrt{2}}{2} a_{\alpha}\left(D^{\mu} F^{\alpha \nu}+D^{\nu} F^{\alpha \mu}\right) h_{\mu \nu}~, 
    \end{split}\\
    \begin{split}
        \label{eqn: a4BOmega}
        \phi_{m}\left(\Omega^{\rho \sigma}\right)^{m n} \phi_{n}=& h_{\mu \nu}\left\{\frac{1}{2}\left(g^{\nu \beta} R^{\mu \alpha \rho \sigma}+g^{\nu \alpha} R^{\mu \beta \rho \sigma}+g^{\mu \beta} R^{\nu \alpha \rho \sigma}+g^{\mu \alpha} R^{\nu \beta \rho \sigma}\right)\right.\\
        &\left.+\left[\omega^{\rho}, \omega^{\sigma}\right]^{\mu \nu \alpha \beta}\right\} h_{\alpha \beta}+a_{\alpha}\left\{R^{\alpha \beta \rho \sigma}+\left[\omega^{\rho}, \omega^{\sigma}\right]^{\alpha \beta}\right\} a_{\beta} \\
        &+h_{\mu \nu} \left(D^{[\rho} \omega^{\sigma]}\right)^{\mu \nu \alpha} a_{\alpha}+a_{\alpha}\left( D^{[\rho}\omega^{\sigma]}\right)^{\alpha \mu \nu} h_{\mu \nu}~,
    \end{split}
\end{align}
where $\omega^\rho$ is the spin-connection given by
\begin{equation}
    \begin{split}
        \phi_{m}\left(\omega^{\rho}\right)^{m n} \phi_{n}=&\frac{\sqrt{2}}{2} h_{\mu \nu}\Big(g^{\alpha\mu}F^{\rho\nu}+g^{\alpha\nu}F^{\rho\mu}-g^{\mu\rho}F^{\alpha\nu}-g^{\nu\rho}F^{\alpha\mu}-g^{\mu\nu}F^{\rho\alpha} \Big)a_{\alpha}\\
       &-\frac{\sqrt{2}}{2}  a_{\alpha}\Big(g^{\alpha\mu}F^{\rho\nu}+g^{\alpha\nu}F^{\rho\mu}-g^{\mu\rho}F^{\alpha\nu}-g^{\nu\rho}F^{\alpha\mu}-g^{\mu\nu}F^{\rho\alpha} \Big) h_{\mu \nu}~.
    \end{split}
\end{equation}

We then find the trace of \eqref{eqn: a4BI}-\eqref{eqn: a4BOmega}. The computation is tedious, but it may also be illuminating for some readers. We present the intermediate steps in appendix \ref{appendix: bosontrace}. The final contribution to the heat kernel coefficient is
\begin{align}
\label{Eq:a4EM}
    (4\pi)^2 a_{4}^{\text{EM}}(x) &= -{277\/180}E_4 + {38\/15}W^2 + {7\/18}R^2~.
\end{align}
The reader familiar with the literature might notice that we have not treated the {\it trace mode} in the graviton. Traditionally, as in the literature \cite{Charles:2015eha, Christensen:1979iy, Gibbons:1976ue}, one decomposes the fields appearing in the Lagrangian into the irreducible fields $\phi(A,B)$ which transform according to the irreducible $(A,B)$ representation of SO(4). For example, in \cite{Christensen:1979iy}, the authors considered the decomposition of fluctuation of geometry $h_{\mu\nu}$ into a  $(1,1)$ symmetric traceless tensor, a scalar characterizing the trace part transforming in $(0,0)$ and the corresponding vector ghost field in $(\frac{1}{2},\frac{1}{2})$. Here, we choose the operator $I^{mn}$ as the effective metric, which is equivalent to making this decomposition.

\subsubsection{Ghost contribution}

\label{subsec:bosonghost}

The addition of the gauge-fixing Lagrangian \eqref{eqn: EMgf} gives an action for the ghosts
\begin{align}
\label{eqn: EMgh}
    \mathcal{S}_{\mathrm{ghost}, b}=\frac{1}{2} \int d^{4} x \sqrt{g}\Big\{2 b_{\mu}\big(g^{\mu \nu} \Box+R^{\mu \nu}\big) c_{\nu}+2 b \, \Box \, c-4 b F^{\rho \nu} D_{\rho} c_{\nu}\Big\},
\end{align}
where $b_{\mu}$ and $c_{\mu}$ are vector fields and $b$ and $c$ are scalar fields. From these expression, we can extract the matrices $E$ and $\Omega$ as
\begin{align}
    \begin{aligned}
        \phi_{n} E_{m}^{n} \phi^{m}=& b_{\mu}\left(R_{~\nu}^{\mu}\right) b^{\nu}+c_{\mu}\left(R_{~\nu}^{\mu}\right) c^{\nu} ,
        \\
        \phi_{n}\left(\Omega_{\alpha \beta}\right)_{m}^{n} \phi^{m}=& b_{\mu}\left(R_{~\nu \alpha \beta}^{\mu}\right) b^{\nu}+c_{\mu}\left(R_{~\nu \alpha \beta}^{\mu}\right) c^{\nu}-\frac{1}{2}\left(b_{\mu}-i c_{\mu}\right)\left(D^{\mu} F_{\alpha \beta}\right)(b+i c) 
        \\
        &+\frac{1}{2}(b+i c)\left(D_{\nu} F_{\alpha \beta}\right)\left(b^{\nu}-i c^{\nu}\right),
    \end{aligned}
    \label{eqn: bosonghost}
\end{align}
The result for the Seeley-DeWitt coefficient is
\be
a_4^\text{ghosts}(x) = {13\/36} E_4 - {1\/4} W^2-{3\/4}R^2~,
\ee
where we have already included here the minus sign due to the opposite statistics.

\subsection{Logarithmic correction}

Adding the above results, the heat kernel for Einstein-Maxwell theory takes the form,
\be\label{a4B}
(4\pi)^2 a_4^{\text{B}}(x) = -{53\/45}E_4 + {137\/60} W^2 - {13\/36} R^2 ~.
\ee
We can read off the coefficients from \eqref{a4general} to be
\be
a_\text{E} = {53\/45},\qq c = {137\/60},\qq b_1 = -{13\/36}~,\qq b_2= 0.
\ee
We note that in the flat limit $\l\to+\infty$, the coefficients $a$ and $c$ match the known flat space computations in \cite{Bhattacharyya:2012wz, Charles:2015eha, Karan:2019gyn} while the coefficients $b_1$ and $b_2$ are unique to AdS. We can also note that the result does not explicitly depend on $F^{\mu\nu}$ as $b_2=0$. This implies that the final result is invariant under electric-magnetic duality. This property has also been observed in the asymptotically flat case in \cite{Bhattacharyya:2012wz,Karan:2019gyn}. Another sanity check is to consider the truncation of the terms involving $F_{\mu\nu}$ in the fluctuations. Then \eqref{a4B} reduces to the neutral limit which was first obtained in \cite{Christensen:1979iy}; we show this in detail in Appendix \ref{sec:neutrallimit}.

We can evaluate this result for the BPS Kerr-Newman solution described in section \ref{subsubsec: AdSKNBPS}. The result is
\bea
C_\r{local} \= -{212\/45}+ {\l_2^2\/120 a\l^2(\l^2-a^2)} \le(629 a^3 - 579 a^2\l + 3699 a\l^2 \vphantom{{(a+\l)^4\/\sqrt{a\l}}}\ri. \-
&&\hspace{6cm} \le. + 411 \l^3 - 411 {(a+\l)^4\/\sqrt{a\l}}\,\r{arctan}(\sqrt{a/\l}) \ri).
\eea

\section{Minimal ${\cal N}=2$ gauged supergravity}
\label{Sec:N2Sugra}

We now consider the simplest supersymmetric theory with a consistent truncation to Einstein-Maxwell theory with a negative cosmological constant. This is minimal ${\cal N}=2$  gauged supergravity \cite{Freedman:1976aw,Fradkin:1976xz,Romans:1991nq,Caldarelli:2003pb}. In this section, we compute the logarithmic corrections in this theory. We find that in contrast to flat space, the logarithmic correction for BPS black holes is not topological. The results of this section were obtained using a Mathematica algorithm described in Appendix \ref{app:algo} which we have made publicly available \cite{github}.

Ultimately, we would like to compute the logarithmic correction for AdS black holes where a microscopic counting is available. Although the techniques of this paper are applicable in those cases, the computations are more involved due to additional matter multiplets.

\subsection{Fermionic fluctuations}

The bosonic Lagrangian of minimal $\cN=2$ supergravity is the same as \eqref{EM action}. Hence, the result of the previous section can be applied and gives \eqref{a4B}. In this section, we will compute the contribution from the fermions. In the conventions of \cite{Caldarelli:2003pb}, the fermionic Lagrangian takes the form
\bea
\cL_f \=  {1\/2} \bpsi_\mu\g^{\mu\nu\rho} D_\nu\psi_\rho + {i\/4}  F^\mn \bpsi_\rho \g_{\mu} \g^\rs \g_{\nu} \psi_\s -{1\/2\l} \bpsi_\mu \g^\mn \psi_\nu~,
\eea
where the gravitino $\psi_\mu$ is a Dirac spin-${3\/2}$ field with charge one, in units of the AdS$_4$ length, under the $\r{U}(1)$ gauge symmetry. The action of the covariant derivative is
\be
D_\mu \psi_\nu = \n_\mu \psi_\nu - {i\/\l}  A_\mu \psi_\nu~.
\ee
We now put the fermionic Lagrangian in a form  suitable for the heat kernel computation. Firstly, we fix the gauge by adding the following gauge-fixing Lagrangian
\be
\label{eqn: gravitinigf}
\cL_\r{g.f.}=- {1\/4} (\bpsi_{\mu} \g^\mu) \g^\nu D_\nu (\g^\rho \psi_\rho)~.
\ee
This choice is convenient because after we perform the field redefinition 
\be
\psi_\mu = \sqrt{2}\le( \chi_\mu - {1\/2}\g_\mu \g^\nu \chi_\nu\ri)~,
\ee
we obtain a simpler kinetic term. The resulting Lagrangian takes the form
\bea
\cL_f \=  g^{\mn}\bchi_\mu\g^{\rho} D_\rho\chi_\nu + {i\/2}  F^\mn \bchi_\rho \g_{\mu} \g^\rs \g_{\nu} \chi_\s -{1\/\l} \bchi_\mu \g^\mn \chi_\nu~.
\eea
More details on this computation are given in Appendix \ref{app:fermions}. We then write the Dirac spinor as
\be
\chi^\mu = \chi^\mu_1+i \chi^\mu_2~,
\ee
where $\chi_1$ and $\chi_2$ are Majorana spin-${3\/2}$ spinors \footnote{For definiteness, we can use here the ``really real'' representation of the Clifford algebra in which the Majorana condition is just the reality condition \cite{Freedman:2012zz}.}. We use the label $A=1,2$ for the two Majorana spinors. The covariant derivative acting on $\chi_A^\mu$ takes the form
\be
D_\mu \chi^\nu_A = \le(\d_{AB} \n^\mu +{1\/\l}\,\ve_{AB} A_\mu\ri) \chi^\nu_B~,
\ee
where $\ve_{AB}$ is the antisymmetric symbol with $\ve_{12}=1$. This is necessary if we want to preserve the reality condition. It is useful to use the Majorana flip identities \eqref{MajoranaFlip} reviewed in Appendix \ref{app:fermions}. The computation detailed there leads to the  Lagrangian in terms of Majorana spinors
\bea\label{LagMajoranas}
\cL_f \=  \d_{AB} g_{\mn}\bchi^\mu_A\g^{\rho} D_\rho\chi^\nu_B -{1\/2}\ve_{AB}  F^\mn \bchi^\rho_A \g^{\mu} \g_\rs \g^{\nu} \chi^\s_B -{1\/\l}\d_{AB} \bchi^\mu_A \g_\mn \chi^\nu_B~.
\eea
Finally, we reinterpret this Lagrangian as being a Euclidean Lagrangian in which $\chi_\mu^A$ are Euclidean spinors satisfying $\bar\chi_\mu^A= (\chi_\mu^A)^\dg$. This Lagrangian can then be used in the algorithm to obtain the result for the heat kernel. Note that we can question the validity of the Wick rotation here because Majorana spinors actually do not exist in four Euclidean dimensions. This can be addressed by using symplectic Majorana spinors. We find, however, that this procedure actually gives  the same result as the naive Wick rotation.

\subsubsection{Symplectic Majoranas}

The Lagrangian \eqref{LagMajoranas} is written in terms of Majorana spinors in $(1,3)$ signature. We would like to Wick rotate this Lagrangian to $(0,4)$ signature. As mentioned above, this appears problematic because Majorana spinors do not exist in $(0,4)$ signature.  A better approach is to use symplectic Majorana spinors which exist in both $(1,3)$ and $(0,4)$ signature \cite{deWit:2017cle}. 

 It is shown in \cite{Cortes:2003zd} that one can map Majorana spinors $\chi_A^\mu$ to symplectic Majorana spinors $\la_A^\mu$ using
\bea
\chi_1^\mu \= {1\/2}(\la_1^\mu +\g^5 \la_2^\mu)~,\-
\chi_2^\mu \= {i\/2}(\la_1^\mu -\g^5 \la_2^\mu)~.
\eea
This allows us to write the Lagrangian in terms of $\la_A^\mu$. We find that the two flavors actually decouple as
\be
\cL_f = \cL_1+\cL_2,
\ee
with
\bea
\cL_1 \= g_\mn \bla_1^\mu\g^\rho(\n_\rho + i\l^{-1} A_\rho)\la_1^\nu - {i\/2}   F^\mn  \bla_1^\rho  \g_{\mu} \g_\rs \g_{\nu} \la_1^\s  -{1\/\l} \bla_1^\mu\g_\mn \la_1^\nu~,\\
\cL_2 \= g_\mn \bla_2^\mu\g^\rho(\n_\rho - i\l^{-1} A_\rho)\la_2^\nu - {i\/2}   F^\mn  \bla_2^\rho  \g_{\mu} \g_\rs \g_{\nu} \la_2^\s  +{1\/\l} \bla_2^\mu\g_\mn \la_2^\nu~.
\eea
The Wick rotation is done by reinterpreting $\la_A^\mu$ as symplectic Majorana spinors in $(0,4)$ signature with $\bla_A^\mu = (\la_A^\mu)^\dg$. This can then be used in the algorithm, described in Appendix \eqref{app:algo}, to compute the heat kernel\footnote{The contribution to the heat kernel of $\cL_1$ and $\cL_2$ are equal because the two Lagrangian differs by $\l\ra-\l$ and the four-derivative terms are invariant under that change.}. We obtain the gravitini contribution
\be
(4\pi)^2 a_4^\text{F,gravitini}(x) = {139\/90} E_4-{32\/15}W^2 -{2\/9}R^2 + {8\/9} R\FF~.
\ee
Note that the result is ultimately the same as what we obtain naively by using directly the Majorana Lagrangian \eqref{LagMajoranas} in the algorithm.

\subsubsection{Ghost contribution}

The gauge-fixing of the gravitini leads to three pairs of ghosts. The gauge condition $\g^\mu\psi_\mu^A=0$ leads to a $bc$ ghost Lagrangian given as
\be
\cL_{bc}= \d_{AB}\bar{b}_A \g_\mu\d_c\psi^\mu_B,
\ee
where $\d_c\psi^\mu$ is the supersymmetry transformation with parameter $c$. This gives
\be
\cL_{bc} =\d_{AB} \bar{b}_A\le(\g^\mu D_\mu + {2\/\l}\ri) c_B~.
\ee
We can get a diagonal kinetic term by a suitable redefinition. This leads to two pairs of ghosts which are charged spin-${1\/2}$ fermions with mass $m={2\/\l}$. In addition, implementing the gauge-fixing term in the path integral leads to an additional pair of massless charged ghosts, giving us a ghost for ghost phenomena \cite{Nielsen:1978mp,Siegel:1980jj}. Details are given in Appendix \ref{app:fermions}.

The total heat kernel of the fermionic ghosts is
\be
(4\pi)^2a_4^\r{F,ghosts}(x) = {11\/120}E_4 -{3\/20}W^2+{2\/9}R^2  - {1\/6} R \FF ~,
\ee
where we have already included the minus sign due to the opposite statistics of ghosts.

\subsection{Logarithmic correction}

The total fermionic contribution  is
\be\label{ResGravitini}
(4\pi)^2a_4^\r{F}(x) = {589\/360}E_4 -{137\/60}W^2 + {13\/18} R F_\mn F^\mn~.
\ee
Adding this to \eqref{a4B} from the bosonic computation, we obtain for minimal $\cN=2$ supergravity
\be 
(4\pi)^2a_4(x) = {11\/24} E_4 -{13\/36} R^2 + {13\/18} R \FF~.
\ee
We find that the full result is not only given by the Euler term as other four-derivative invariants are present. This indicates that the logarithmic correction is non-universal. We note that the $ W^2$ term, which would give another non-universal contribution, does cancel between bosons and fermions. This is expected from the flat space result \cite{Charles:2015eha}, which we recover in the flat limit.

\subsubsection{Evaluation}

We can evaluate the heat kernel coefficient on the backgrounds summarized in section \ref{Sec:Backgrounds}.  For the non-extremal Kerr-Newman black hole, we get
\be\label{ClocalKNres}
C_\r{local} = {11\/6} + {26 \le[r_+\b (r_+^4-\l^2(a^2+q_m^2) ) - \pi\l^2(r_+^4-a^4) \ri]\/3\pi\l^2 (\l^2-a^2)(a^2+r_+^2)}~.
\ee
We note that the fermionic contribution breaks electromagnetic duality as it generates a non-zero $b_2$. This is reflected by the dependence in the magnetic charge $q_m$ in the above expression.

The result for extremal Kerr-Newman takes the form
\be\label{evalClocalextKN}
C_\r{local} = {11\/6} + {26  \l_2^2 \le[ (a^2(a+\l)+r_0^2 (3a-\l))(a^2(a-\l)+r_0^2 (3a+\l))+ 2 \l^2 q_m^2(r_0^2-a^2) \ri]\/3\l^2 (\l^2-a^2)(a^2+r_0^2)^2},
\ee
where here the AdS$_2$ radius is $\l_2 = \l\sqrt{a^2+r_0^2\/a^2+\l^2+6r_0^2}$. As explained in section \ref{sec:KN}, this is obtained by either taking the extremal limit of \eqref{ClocalKNres} or by doing the computation in the near horizon geometry. 

We are particularly interested in evaluating the logarithmic corrections on BPS black holes. Rotation is necessary to have a regular BPS background in minimal gauged supergravity as the extremal AdS-Reissner-Nordstr\"om is singular in the BPS limit \cite{Caldarelli:1999xj, Caldarelli:1998hg, Kostelecky:1995ei}. We obtain the BPS result by imposing the BPS constraints $r_0=\sqrt{a\l}$ and $q_m=0$ on \eqref{evalClocalextKN}. The contribution is 
\be\label{ClocalBPSres}
    C_\r{local} = {11\/6} -{26\/3}{a(\l^2-4a\l-a^2)\/(\l-a)(a^2+ 6 a\l + \l^2)},
\ee
where the first term comes from the topological Euler term and the second term comes $R^2$ and $RF^2$ and constitute the non-topological piece. We discuss the significance of this non-topological term in the next subsection.

\ss{Implications}

We shall now comment on the non-topological nature of the logarithmic correction.  For the flat space Kerr-Newman black hole, the heat kernel $a_4(x)$ is the sum of only two terms: the Euler term and the Weyl squared term. Although $W^2=0$ for extremal non-rotating black holes in flat space, it is non-zero for extremal rotating black holes. It was shown in \cite{Charles:2015eha,Larsen:2018cts} that supersymmetry ensures that the coefficient $c$ multiplying $W^2$ actually cancels. This shows that supersymmetry makes the logarithmic correction topological in ungauged supergravity.

In AdS$_4$, the $ W^2$ term never vanishes in the near horizon geometry (even without rotation) and there are two additional terms. We also obtain that supersymmetry ensures $c=0$ due to cancellations between bosons and fermions. This could be expected from the flat space results of \cite{Charles:2015eha} which we reproduce in the flat limit $\l\to+\infty$. Hence, even with a negative cosmological constant, supersymmetry makes the logarithmic correction less complicated as it removes the non-topological term $W^2$. This term is a complicated function  of the black hole parameters. For the BPS Kerr-Newman AdS black hole, it takes the form
\be\label{Weyl2expr}
{1\/(4\pi)^2}\int d^2 x\sqrt{g}\,W^2 ={3\l_2^2\/2 a\l^2(\l^2-a^2)} \le( (\l-a)(\l^2+10 a \l + a^2) - {(a+\l)^4 \/\sqrt{a\l}} \r{arctan}(\sqrt{a/\l}) \ri) ~.
\ee
The two other four-derivative, $R^2$ and $RF^2$, terms do not cancel so supersymmetry does not imply that the logarithmic corrections are topological. However, we see that the BPS result \eqref{ClocalBPSres} is still a simpler function of $a$ and $\l$ as it has $c=0$. It is a rational function rather than a transcendental one.

It is natural to expect topological logarithmic corrections in the UV given the known examples of microscopic counting of black hole entropy \cite{Mandal:2010cj,Sen:2014aja,Belin:2016knb,Liu:2017vll,Benini:2019dyp,Gang:2019uay,PandoZayas:2020iqr,Liu:2018bac}. This is also automatic if the 4d theory comes from an odd-dimensional theory by Kaluza-Klein reduction because $C_\r{local}=0$ in odd dimensions. Hence, the logarithmic correction can be a useful probe of whether a low-energy effective theory can have a UV completion. The idea is that from a bottom-up perspective, we should prefer low-energy theories which have topological logarithmic correction. This is only possible if $c=b_1=b_2=0$ which is a rather strong constraint on the low-energy Lagrangian, analogous to anomaly cancellation.

\section{Discussion}\label{Sec:Discussion}

In this paper we have computed the logarithmic corrections to the entropy of black holes in minimal gauged supergravity using the four dimensional heat kernel expansion. The inclusion of a negative cosmological constant leads to new features compared to the case of asymptotically flat black holes. In the especially interesting case of BPS black holes, the logarithmic corrections present a richer structure and can be non-topological.

The original explicit logarithmic corrections performed for asymptotically AdS$_4\times S^7$ black holes based on Sen's entropy function formalism, using the near horizon geometry, did not agree with the field theory computations \cite{Liu:2017vll,Jeon:2017aif}. It was only in \cite{Liu:2017vbl} that agreement was found by considering the full geometry. The results of this manuscript clarify that the difference between the two approaches comes from the contributions of the zero modes which are indeed different in the two geometries. Namely, we have shown that for extremal black holes the {\it local} contribution to the logarithmic correction, $C_\r{local}$, is the same when computed either from the full AdS$_4$ asymptotic region or for the near horizon  geometry. This result elucidates the question of where the degrees of freedom responsible for the quantum entropy live.

For the BPS Kerr-Newman black hole in $\mathcal{N}=2$ minimally gauged supergravity, we have found that the logarithmic correction, given in \eqref{ClocalBPSres}, is non-topological. To obtain this result, we have used holographic renormalization to regularize the divergent volume integrals. This appears to be the right prescription as, for example, it gives the correct counterterm to obtain the Euler characteristic when integrating the Euler density, see Appendix \ref{app:Euler} for more details.

The non-topological nature of the logarithmic corrections suggests that they might contain more information than the flat space counterpart, providing a wider ``infrared window into the microstates''. Moreover, this non-topological nature is interesting because for all the available examples of microscopic counting, see for example \cite{GonzalezLezcano:2020yeb, Benini:2019dyp, Hanada:2012si, Gang:2019uay}, and for BPS black holes in flat space \cite{Charles:2015eha,Charles:2017dbr}, the logarithmic correction is always topological, {\it i.e.}, the coefficient of the logarithm is a pure number. In $\mathcal{N}=2$ minimal gauged supergravity, we find that it is instead a rather non-trivial function of the black hole parameters.

In addition, we have shown that supersymmetry makes the logarithmic correction simpler: the bosonic and fermionic contributions to the coefficient, $c$, of the Weyl squared term \eqref{Weyl2expr} cancel each other. With supersymmetry, the dependence of $a$ and $\ell$ in $C_{\text{local}}$ is now rational instead of transcendental. Nonetheless, we have seen that supersymmetry is not enough to make the other terms cancel which shows the logarithmic corrections can be non-topological for BPS black holes.  

It is illuminating to compare this result to recent investigations using supergravity localization \cite{Hristov:2018lod,Hristov:2019xku,Hristov:2021zai}. In \cite{Hristov:2021zai}, the general structure of the logarithmic correction of 4d $\cN=2$ gauged supergravity on BPS backgrounds was studied using index theory. It was shown that the universal piece coming from the Euler term arises from the application of the Atiyah-Singer theorem to an appropriate supercharge. We surmise that the non-universal piece that we obtained should be interpreted as the contribution from the $\eta$ invariant, not considered in \cite{Hristov:2021zai}, which is a non-topological correction due to the presence of a boundary \cite{atiyah_patodi_singer_1975}. It is also possible that the contributions of auxiliary fields in off-shell supergravity have to be included to obtain topological logarithmic corrections. Supergravity localization has the potential of ultimately providing the full quantum entropy of the black holes and it would be fruitful to test it against one-loop supergravity results such as ours. 

We might hope to use the topological nature of logarithmic corrections as a criterion for a low-energy theory to admit a UV completion. In the available examples of microscopic counting, the logarithmic correction is indeed topological \cite{Mandal:2010cj,Sen:2014aja,Belin:2016knb,Liu:2017vll,Benini:2019dyp,Gang:2019uay,PandoZayas:2020iqr,Liu:2018bac}. Such a criteria would greatly constrain effective supergravity theories as it gives rather stringent conditions similar to anomaly cancellation. Note that in odd dimensions, the logarithmic correction is automatically topological because the local contribution is trivially zero. This suggests that the topological nature of the local contribution should remain when computing the logarithmic corrections in 4d. The addition of the full KK tower of modes could remedy the logarithmic corrections to be topological. On the other hand, the situation is quite different in ten dimensional theories, as there is a local contribution coming from the tenth Seeley-deWitt coefficient and as a result, the topological criterion should be much more constraining. 

We have obtained the logarithmic correction for the simplest gauged supergravity in four dimensions. Our goal is to grow this direction towards more interesting theories and to relate our results to other approaches such as the computations performed in eleven-dimensional supergravity \cite{Liu:2017vbl,Gang:2019uay,PandoZayas:2020iqr,Benini:2019dyp}.  It should be possible to perform the same computation in the gauged $\r{U}(1)^4$ supergravity which comes from eleven dimensional supergravity on AdS$_4\times S^7$. Similarly, the logarithmic correction to the entropy of black holes in AdS$_4\times SE_7$ has been computed both in field theory and supergravity for a large class of Sasaki-Einstein seven-dimensional manifolds \cite{PandoZayas:2020iqr}. In both these cases, the topological nature follows from the fact that the parent theory is odd-dimensional. It would be interesting to see explicitly how this is realized from a four-dimensional perspective.

More challenging would be the cases where the AdS$_4$ black holes are embedded in ten-dimensional theories such as massive IIA supergravity. A matching of the Bekenstein-Hawking entropy at leading order  was presented in \cite{Azzurli:2017kxo,Hosseini:2017fjo,Benini:2017oxt}. The available sub-leading, microscopic analysis confirms the topological nature of the logarithmic term \cite{Liu:2018bac}. However, the supergravity computations need to be in agreement with the nontrivial nature of $C_{\rm local}$. We hope to address some of these issues in the future. 

\section*{Acknowledgements}

We would like to thank Rodrigo de Le\'on Ard\'on and  Alberto Faraggi for discussions. We are particularly grateful to Kiril Hristov and Valentin Reys for engaging and clarifying correspondence and to Alejandra Castro and Ashoke Sen for insightful comments on the draft.
This  work was supported in part by the U.S. Department of Energy under grant DE-SC0007859. M.D. was supported by the NSF Graduate Research Fellowship Program under NSF Grant Number: DGE 1256260. 
VG acknowledges the postdoctoral program at ICTS for funding support through the Department of Atomic Energy, Government of India, under project no. RTI4001.

\appendix
\addtocontents{toc}{\protect\setcounter{tocdepth}{1}}

\section{Mathematica algorithm}\label{app:algo}

We describe the Mathematica algorithm written with xAct \cite{xAct} and xPert \cite{Brizuela:2008ra} to compute the Seeley-DeWitt coefficients $a_4(x)$ presented in this paper. An executable code reproducing the results of this paper is available at \cite{github}. The purpose of the algorithm is to compute $a_4(x) $ via the expression
\bea\nt
(4\pi)^2 a_4(x) \= \Tr\left[{1\over2} E^2 + {1\over6} R E + {1\over 12} \Om_\mn\Om^\mn+ {1\over360} (5 R^2 + 2R_{\mn\rs}R^{\mn\rs} - 2R_\mn R^\mn)\right]~,\\\label{app:a4}
\eea
where $E$ and $\Om$ are determined by the two-derivative action as defined in section \eqref{sec:HKexpansion}. This computation is straightforward but tedious to do by hand, especially for fermions. The algorithm uses xTensor and our own implementation of Euclidean spinors (as xSpinor can only treat Lorentzian spinors). The resulting expression is then reduced using various spinorial and geometrical identities, as well as the equations of motion. 

For bosons, the algorithm uses xPert \cite{Brizuela:2008ra} to expand any Lagrangian to quadratic order. It then extracts the matrices $P$ and $\w$ which allows us to compute $E$ and $\Om$, evaluates and simplifies $a_4(x)$. For fermions, the input is the matrix $L$ which defines the quadratic Lagrangian as $\cL = \bpsi (\sD + L) \psi$ where $\psi$ refers to all the fermionic fields of the theory. The heat kernel method is then applied to  the operator $\cQ = \cD^\dg \cD$ using the formula \eqref{omegaPforfermions} to obtain $P$ and $\w$. The algorithmic approach is useful because we can automatize simplification using gamma matrix identities.

The algorithm was used in this paper to verify the bosonic result, also computed by hand, and to obtain the fermionic result, which appeared too tedious to compute by hand. It has also been used to obtain the results for minimal couplings, which can also be easily obtained by hand.

Let us mention various checks that have been performed on this algorithm.  It gives the correct logarithmic contribution for various results in the literature such as minimally coupled fields \cite{Vassilevich:2003xt} and ungauged $\cN\geq 2$ supergravity \cite{Charles:2015eha}. The same algorithm was used in \cite{Castro:2018hsc} to compute the logarithmic correction in the non-BPS branch of ungauged $\cN\geq2$ supergravity. The results of \cite{Castro:2018hsc} were subsequently checked by a completely independent approach \cite{Charles:2015eha} computing directly $a_4(x)$ from eigenvalues, with agreement in all  cases. Given that the Lagrangians involved in these computations were fairly complicated, this gives us confidence that the  algorithm performs correctly.

\section{Bosonic computation}\label{app:bosons}

For the interested reader, we present a self-contained computation of the heat kernel coefficient $a_4(x)$ for the Einstein-Maxwell-AdS theory.

\subsection{Quadratic fluctuations in Einstein-Maxwell AdS theory} \label{appendix:fluctuations}

The action is given by
\begin{align}
    \label{eqn: EM action app}
    S=\int d^{4} x \sqrt{ g}\left(R-2\Lambda-F_{\mu \nu} F^{\mu \nu}\right),
\end{align}
where $F_{\mu\nu}=\partial_\mu A_\nu-\partial_\nu A_\mu$ is the field strength with $A_\mu$ the gauge potential. Note that we find it convenient to use the convention $4\pi G=1$. We consider variations of the metric and gauge field
\be
\d g_\mn =\sqrt{2} h_\mn,\qq \d A_\mu = \frac{1}{2} a_\mu~,
\ee
where $h_{\mu\nu}$ and $a_\alpha$ are the graviton and graviphoton respectively. We impose a particular gauge to the theory by adding a suitable gauge-fixing Lagrangian 
\begin{align}
\label{eqn: EMgf app}
S=    -\int d^{4} x \sqrt{\operatorname{det} g}\left\{\left(D^{\mu} h_{\mu \rho}-\frac{1}{2} D_{\rho} h\right)\left(D^{\nu} h_{\nu}^{\rho}-\frac{1}{2} D^{\rho} h\right)+\frac{1}{2}\left(D^{\mu} a_{\mu}\right)\left(D^{\nu} a_{\nu}\right)\right\},
\end{align}
and the corresponding ghost action to the action \eqref{eqn: EM action app}. We then expand the action up to quadratic order. The linear order variation yields the equation of motion for the background fields 
\begin{align}
    \label{eqn: gravityEoM app}
    R_{\mu\nu}-\frac{1}{2}g_{\mu\nu}R+g_{\mu\nu}\Lambda=&2F_{\mu\rho}F_\nu^{~\rho}-\frac{1}{2}g_{\mu\nu}F_{\alpha\beta}F^{\alpha\beta}~,\\
    \label{eqn: gaugeEoM app}
    D^\mu F_{\mu\nu}=&0~.
\end{align}
Note that the equations of motion implies that $R=4\L = -12/\LFull^2$. It is also worth mentioning the Bianchi identity for the gravitational and gauge fields
\begin{align}
    \label{eqn: gravitybianchi app}
    D_{[\mu}F_{\nu\rho]}=&0~,\\
    \label{eqn: gaugebianchi app}
    R_{\mu[\nu\rho\sigma]}=&0~.
\end{align}
Writing the quadratic action in the standard form \eqref{Lap}, we find
\begin{align} \label{quadratic action with cosmological constant}
    \begin{aligned}
        \phi_{m} \mathcal{Q}^{m n} \phi_{n}=& G^{\mu \nu \alpha \beta} h_{\mu \nu} \Box h_{\alpha \beta}+g^{\alpha \beta} a_{\alpha} \Box a_{\beta}-a_{\alpha} R^{\alpha \beta} a_{\beta}
        \\&+
        h_{\mu \nu}\left\{R^{\mu \alpha \nu \beta}+R^{\mu \beta \nu \alpha}-\frac{1}{2}\left(g^{\mu \alpha} R^{\nu \beta}+g^{\mu \beta} R^{\nu \alpha}\right.\right.
        \\&
        \left.+g^{\nu \alpha} R^{\mu \beta}+g^{\nu \beta} R^{\mu \alpha}\right)-2\left(F^{\mu \alpha} F^{\nu \beta}+F^{\mu \beta} F^{\nu \alpha}\right)
        \\&
        \left.+\frac{1}{2}\left(g^{\mu \nu} g^{\alpha \beta}-g^{\mu \alpha} g^{\nu \beta}-g^{\mu \beta} g^{\nu \alpha}\right)\left(F_{\theta \varphi} F^{\theta \varphi}\right)  + 2G^{\alpha\beta\mu\nu}\Lambda \right\} h_{\alpha \beta} 
        \\&
        -h_{\mu \nu}\left\{\frac{1}{4}\left(D_{\rho} K^{\rho}\right)^{\mu \nu \alpha}-\frac{1}{2}\left(K^{\rho}\right)^{\mu \nu \alpha} D_{\rho}\right\} a_{\alpha}\\
        &-a_{\alpha}\left\{\frac{1}{4}\left(D_{\rho} K^{\rho}\right)^{\mu \nu \alpha}+\frac{1}{2}\left(K^{\rho}\right)^{\mu \nu \alpha} D_{\rho}\right\} h_{\mu \nu},
    \end{aligned}
\end{align}
where 
\begin{equation}
\label{eqn: bosonK}
    \left(K^\rho\right)^{\mu\nu\alpha}=2\sqrt{2}\Big(g^{\alpha\mu}F^{\rho\nu}+g^{\alpha\nu}F^{\rho\mu}-g^{\mu\rho}F^{\alpha\nu}-g^{\nu\rho}F^{\alpha\mu}-g^{\mu\nu}F^{\rho\alpha} \Big).
\end{equation}
Note that we have used the symmetry properties of the graviton to write the term proportional to $\Lambda$ using the DeWitt metric
\begin{align}
    G^{\mu \nu \alpha \beta}=\frac{1}{2}\left(g^{\mu \alpha} g^{\nu \beta}+g^{\mu \beta} g^{\nu \alpha}-g^{\mu \nu} g^{\alpha \beta}\right),
\end{align}
as $ h_{\mu\nu}( 2g^{\mu\alpha}g^{\nu\beta} \Lambda -g^{\mu\nu}g^{\alpha\beta} \Lambda)h_{\alpha\beta} = h_{\mu\nu}(2G^{\alpha\beta\mu\nu}\Lambda)h_{\alpha\beta}$. We note that the results pertaining to the cosmological constant terms agree with what we expect from \cite{Christensen:1979iy}. With \eqref{quadratic action with cosmological constant}, we can explicitly read out the matrices $I^{mn}$, $\omega^{\rho}$ and $P^{mn}$ in \eqref{Lap}:
\begin{align}
    \phi_{m} I^{m n} \phi_{n}=&h_{\mu \nu} G^{\mu \nu \alpha \beta} h_{\alpha \beta}+a_{\alpha} g^{\alpha \beta} a_{\beta}~.\\
    \phi_{m}\left(\omega^{\rho}\right)^{m n} \phi_{n}=&\frac{1}{4} h_{\mu \nu}\left(K^{\rho}\right)^{\mu \nu \alpha} a_{\alpha}-\frac{1}{4} a_{\alpha}\left(K^{\rho}\right)^{\mu \nu \alpha} h_{\mu \nu}~.\\
    \begin{split}
    \phi_{m} P^{m n} \phi_{n}=& h_{\mu \nu}\left\{R^{\mu \alpha \nu \beta}+R^{\mu \beta \nu \alpha}-\frac{1}{2}\left(g^{\mu \alpha} R^{\nu \beta}+g^{\mu \beta} R^{\nu \alpha}\right.\right.\\
    &\left.+g^{\nu \alpha} R^{\mu \beta}+g^{\nu \beta} R^{\mu \alpha}\right)-2\left(F^{\mu \alpha} F^{\nu \beta}+F^{\mu \beta} F^{\nu \alpha}\right) \\
    &\left.+\frac{1}{2}\left(g^{\mu \nu} g^{\alpha \beta}-g^{\mu \alpha} g^{\nu \beta}-g^{\mu \beta} g^{\nu \alpha}\right)\left(F_{\theta \varphi} F^{\theta \varphi}\right) \right.\\
    &\left. + 2g^{\mu\alpha}g^{\nu\beta} \Lambda -g^{\mu\nu}g^{\alpha\beta} \Lambda \right\} h_{\alpha \beta}~, \\
    &-a_{\alpha} R^{\alpha \beta} a_{\beta}+\frac{\sqrt{2}}{2} h_{\mu \nu}\left\{D^{\mu} F^{\alpha \nu}+D^{\nu} F^{\alpha \mu}\right\} a_{\alpha} \\
    &+\frac{\sqrt{2}}{2} a_{\alpha}\left\{D^{\mu} F^{\alpha \nu}+D^{\nu} F^{\alpha \mu}\right\} h_{\mu \nu}~.
    \end{split}
\end{align}

\subsection{Trace computation}
\label{appendix: bosontrace}
Here, we present various trace computations. The field strength $\Omega$ is given by
\begin{align}
     \label{eqn: a4Omega}
     \phi_m\Om_\mn^{mn}\phi_n =&\phi_m [D_\mu+\w_\mu,D_\nu+\w_\nu]\phi_n
     =\phi_m\Big\{[D_\mu, D_\nu]^{mn}+D_{[\mu}\omega_{\nu]}^{~~mn}+[\omega_\mu,\omega_\nu]^{mn}\Big\}\phi_n~.
 \end{align}
To compute the matrix $E$ and $\Omega$ using \eqref{eqn: a4E} and \eqref{eqn: a4Omega}, we need $[D_\mu. D_\nu]$, $\left(D^{\rho} \omega_{\rho}\right)^{m n}$ and $\left(\omega^{\rho}\right)^{m p}\tensor{\left(\omega_{\rho}\right)}{_{p}^{n}}$:
\begin{align}
    \begin{split}
        \label{eqn: bosoncommutator}
        \phi_{m}\left[D^{\rho}, D^{\sigma}\right]^{m n} \phi_{n} =&h_{\mu \nu}\left[D^{\rho}, D^{\sigma}\right] h^{\mu \nu}+a_{\alpha}\left[D^{\rho}, D^{\sigma}\right] a^{\alpha} \\
        =&h_{\mu \nu}\left\{g^{\nu \beta} R^{\mu \alpha \rho \sigma}+g^{\mu \beta} R^{\nu \alpha \rho \sigma}\right\} h_{\alpha \beta}+a_{\alpha} R^{\alpha \beta \rho \sigma} a_{\beta}
        \\=&
        \frac{1}{2}h_{\mu \nu}\left\{g^{\nu \beta} R^{\mu \alpha \rho \sigma}+g^{\nu \alpha} R^{\mu \beta \rho \sigma}+g^{\mu \beta} R^{\nu \alpha \rho \sigma}+g^{\mu \alpha} R^{\nu \beta \rho \sigma}\right\} h_{\alpha \beta}
        \\&+a_{\alpha} R^{\alpha \beta \rho \sigma} a_{\beta},
    \end{split}\\
    \begin{split}
        \phi_{m} \left(D^{\rho} \omega_{\rho}\right)^{m n} \phi_{n}=& - \frac{\sqrt{2}}{2}h_{\mu\nu} \left(D^{\nu}F^{\alpha \mu}+D^{\mu} F^{\alpha \nu}\right)a_{\alpha}+\frac{\sqrt{2}}{2}a_{\alpha} \left(D^{\nu}F^{\alpha \mu}+D^{\mu} F^{\alpha \nu}\right)h_{\mu\nu},
    \end{split}\\
    \begin{split}
    \label{eqn: bosonomega2}
        \phi_{m} \left(\omega^{\rho}\right)^{m p}\tensor{\left(\omega_{\rho}\right)}{_{p}^{n}}\phi_{n}
        =&
        \frac{1}{16}h_{\mu\nu} \left(K^{\rho}\right)^{\mu \nu \alpha} \left(-K_{\rho}\right)^{\delta \gamma\beta}g_{\alpha\beta} h_{\delta \gamma} 
        \\& \quad + \frac{1}{16} a_{\alpha} \left(K^{\rho}\right)^{\mu \nu \alpha} \left(-K_{\rho}\right)^{\delta\gamma}\tensor{}{^{\beta}}G_{\mu\nu\delta\gamma} a_{\beta}.
    \end{split}
\end{align}
For \eqref{eqn: bosonomega2}, using the definition of $\left(K^\rho\right)^{\mu\nu\alpha}$ \eqref{eqn: bosonK}, we find
\begin{align}
\label{eqn: bosonomega2-1}
    \begin{split}
    \phi_{m} \left(\omega^{\rho}\right)^{m p}\tensor{\left(\omega_{\rho}\right)}{_{p}^{n}}\phi_{n}
    =&
     h_{\mu\nu}\left(-2 F^{\mu \delta} F^{\nu \gamma} - 2 F^{\mu \gamma} F^{\nu \delta} + 2 F^{ \gamma \rho} F^{\delta}{}_{\rho} g^{\mu\nu } -  F^{\nu \rho} F^{\delta}{}_{\rho} g^{\mu \gamma} \right.
    \\&
    \left. -  F^{\nu \rho} F^{\gamma}{}_{\rho} g^{\mu \delta} -  F^{\mu \rho} F^{\delta}{}_{\rho} g^{\nu \gamma} -  F^{\mu \rho} F^{\gamma}{}_{\rho} g^{\nu \delta} + 2 F^{\mu \rho} F^{\nu }{}_{\rho} g^{\gamma\delta} \right.
    \\& \left. -  \tfrac{1}{2} F_{\rho \sigma} F^{\rho \sigma} g^{\mu \nu } g^{\gamma\delta} \right)h_{\gamma\delta} +
    a_{\alpha}\left(-2 F^{\alpha \gamma} F^{\beta}{}_{\gamma} -  F_{\theta\phi} F^{\theta\phi} g^{\alpha\beta} \right)a_{\beta}.
    \end{split}
\end{align}
Note that \eqref{eqn: bosonomega2-1} is the same expression as in the asymptotically flat case \cite{Bhattacharyya:2012wz} and changes only when we plug in the equations of motion \eqref{eqn: gravityEoM app},
\begin{align}
    \begin{split}
    \phi_{m} \left(\omega^{\rho}\right)^{m p}\tensor{\left(\omega_{\rho}\right)}{_{p}^{n}}\phi_{n}
    =&
     h_{\mu \nu }\left(-2 F^{\mu \alpha} F^{\nu \beta} - 2 F^{\mu \beta} F^{\nu \alpha} -  \tfrac{1}{2} F_{ad} F^{ad} g^{\mu \alpha} g^{\nu \beta} -  \tfrac{1}{2} F_{ad} F^{ad} g^{\mu \beta} g^{\nu \alpha} \right.
    \\& \left. + \tfrac{1}{2} F_{ad} F^{ad} g^{\mu \nu } g^{\beta\alpha} + g^{\beta\alpha} R^{\mu \nu } -  \tfrac{1}{2} g^{\nu \alpha} R^{\mu \beta} -  \tfrac{1}{2} g^{\nu \beta} R^{\mu \alpha} -  \tfrac{1}{2} g^{\mu \alpha} R^{\nu \beta} \right.
    \\& \left.  - \tfrac{1}{2} g^{\mu \beta} R^{\nu \alpha} + g^{\mu \nu } R^{\beta\alpha} + \Lambda g^{\mu \alpha} g^{\nu \beta} + \Lambda g^{\mu \beta} g^{\nu \alpha} - 2 \Lambda g^{\mu \nu } g^{\beta\alpha}\right)h_{\alpha\beta}
    \\&
    + \frac{1}{2}a_{\alpha}\left(- \tfrac{3}{2} F_{\theta\phi} F^{\theta\phi} g^{\alpha\beta} -  R^{\alpha\beta}+ \Lambda g^{\alpha\beta}\right)a_{\beta}.
    \end{split}
\end{align}
Extracting the information from the quadratic action, we find
\begin{align}
    \begin{split}
    \label{appeqn: bosonbulkE}
        \phi_{m} E^{m n} \phi_{n}=& h_{\mu \nu}\left(R^{\mu \alpha \nu \beta}+R^{\mu \beta \nu \alpha}-g^{\mu \nu} R^{\alpha \beta}-g^{\alpha \beta} R^{\mu \nu} + \Lambda g^{\mu\nu}g^{\alpha\beta} \right) h_{\alpha \beta} \\
        &+a_{\alpha}\left(\frac{3}{2} g^{\alpha \beta} F_{\mu \nu} F^{\mu \nu} - \Lambda g^{\alpha\beta}\right) a_{\beta}+\frac{\sqrt{2}}{2} h_{\mu \nu}\left(D^{\mu} F^{\alpha \nu}+D^{\nu} F^{\alpha \mu}\right) a_{\alpha} \\
        &+\frac{\sqrt{2}}{2} a_{\alpha}\left(D^{\mu} F^{\alpha \nu}+D^{\nu} F^{\alpha \mu}\right) h_{\mu \nu},
    \end{split}
\end{align}
\begin{align}
\label{appeqn: bosonbulkOmega}
    \begin{aligned}
        \phi_{m}\left(\Omega^{\rho \sigma}\right)^{m n} \phi_{n}=& h_{\mu \nu}\left\{\frac{1}{2}\left(g^{\nu \beta} R^{\mu \alpha \rho \sigma}+g^{\nu \alpha} R^{\mu \beta \rho \sigma}+g^{\mu \beta} R^{\nu \alpha \rho \sigma}+g^{\mu \alpha} R^{\nu \beta \rho \sigma}\right)\right.\\
        &\left.+\left[\omega^{\rho}, \omega^{\sigma}\right]^{\mu \nu \alpha \beta}\right\} h_{\alpha \beta}+a_{\alpha}\left\{R^{\alpha \beta \rho \sigma}+\left[\omega^{\rho}, \omega^{\sigma}\right]^{\alpha \beta}\right\} a_{\beta} \\
        &+h_{\mu \nu} \left(D^{[\rho} \omega^{\sigma]}\right)^{\mu \nu \alpha} a_{\alpha}+a_{\alpha}\left( D^{[\rho}\omega^{\sigma]}\right)^{\alpha \mu \nu} h_{\mu \nu}.
    \end{aligned}
\end{align}
We explicitly compute the traces involving the endomorphism $E$
\begin{align}
    \begin{split}
    \Tr (R E) &= \Tr R\left[R^{\mu \alpha \nu \beta}+R^{\mu \beta \nu \alpha}-g^{\mu \nu} R^{\alpha \beta}-g^{\alpha \beta} R^{\mu \nu}  + \Lambda g^{\mu\nu}g^{\alpha\beta}  \right] 
    \\& \quad + \Tr \left(\frac{3}{2}g^{\alpha \beta} RF_{\mu \nu} F^{\mu \nu} - R\Lambda g^{\alpha\beta} \right)
    \\&=
    R \left(R^{\mu \alpha \nu \beta}+R^{\mu \beta \nu \alpha}-g^{\mu \nu} R^{\alpha \beta}-g^{\alpha \beta} R^{\mu \nu} +g^{\mu\nu}g^{\alpha\beta} \Lambda \right)G_{\mu\nu\alpha\beta} 
    \\& \quad + R \left(\frac{3}{2}g^{\alpha \beta} F_{\mu \nu} F^{\mu \nu} - \Lambda g^{\alpha\beta}\right)g_{\alpha\beta},
    \end{split}
\end{align}
which after expanding, we find
\begin{align}
\begin{split}
    \Tr (RE) &= R \left(-2R+(D-2)R+\left(D-\frac{D^2}{2}\right)\Lambda \right) + R\left(\frac{3D}{2} F_{\mu \nu} F^{\mu \nu}-d\Lambda\right)
    \\&=
    -32 \Lambda^2 + 6 R  F_{\mu \nu} F^{\mu \nu},
\end{split}
\end{align}
where $D=4$ is the dimension of the space and we have imposed $R=4\Lambda$. Next, we consider
\begin{align} \label{E squared}
    \begin{split}
    \Tr (E^2)
    =& 
    \Tr \left(R^{\mu \alpha \nu \beta}+R^{\mu \beta \nu \alpha}-g^{\mu \nu} R^{\alpha \beta}-g^{\alpha \beta} R^{\mu \nu}  + g^{\mu\nu}g^{\alpha\beta}\Lambda\right)
    \\& 
    \times \left(R^{\rho \tau \sigma \delta}+R^{\rho \delta \sigma \tau}-g^{\rho\sigma} R^{\tau \delta}-g^{\tau\delta} R^{\rho\sigma} + g^{\rho\sigma}g^{\tau\delta}\Lambda \right) 
    \\&+ 
    \Tr \left(\left(\frac{3}{2}g^{\alpha \beta} F_{\mu \nu} F^{\mu \nu} - \Lambda g^{\alpha\beta} \right)\left(\frac{3}{2}g^{\tau \delta} F_{\theta\phi} F^{\theta\phi} - \Lambda g^{\tau\delta} \right)\right)
    \\&+ 
    \frac{1}{2}\Tr \left(\left(D^{\mu} F^{\alpha \nu}+D^{\nu} F^{\alpha \mu}\right) \left(D^{\rho} F^{\beta \sigma}+D^{\sigma} F^{\beta \rho}\right)\right)
    \\&+
    \frac{1}{2}\Tr \left(\left(D^{\mu} F^{\alpha \nu}+D^{\nu} F^{\alpha \mu}\right)\left(D^{\rho} F^{\beta \sigma}+D^{\sigma} F^{\beta \rho}\right)\right).
    \end{split}
\end{align}
For the first term of \eqref{E squared}, we have
\begin{align}
    \begin{split}
    &\left(R^{\mu \alpha \nu \beta}+R^{\mu \beta \nu \alpha}-g^{\mu \nu} R^{\alpha \beta}-g^{\alpha \beta} R^{\mu \nu}  + g^{\mu\nu}g^{\alpha\beta}\Lambda \right)
    \\& 
    \times \left(R^{\rho \tau \sigma \delta}+R^{\rho \delta \sigma \tau}-g^{\rho\sigma} R^{\tau \delta}-g^{\tau\delta} R^{\rho\sigma} + g^{\rho\sigma}g^{\tau\delta}\Lambda \right)  G_{\mu\nu\rho\sigma}G_{\alpha\beta\tau\delta}
    \\&
    \qquad = 16 \Lambda^2 - 4 R_{ab} R^{ab}  + 3 R_{abcd} R^{abcd}.
    \end{split}
\end{align}
The second term of \eqref{E squared} gives
\begin{align}
     \Tr \left(\frac{3}{2}g^{\alpha \beta} F_{\mu \nu} F^{\mu \nu} - \Lambda g^{\alpha\beta} \right)\left(\frac{3}{2}g^{\tau \delta} F_{\theta\phi} F^{\theta\phi} - \Lambda g^{\tau\delta} \right)
    =
    4\Lambda^2 -12 \Lambda F_{\alpha\beta}F^{\alpha\beta}+  9 \left(F_{\theta\phi} F^{\theta\phi} \right)^2,
\end{align}
where $g^{\alpha \beta}g^{\tau \delta} g_{\alpha\tau}g_{\beta\delta}=D=4$. Now, for the remaining terms in the trace in \eqref{E squared}, we need the following identities
\begin{align} \label{identities}
    \begin{split}
    (D_{\rho}F_{\mu\nu})(D^{\rho}F^{\mu\nu})&= {\color{black}}R_{\mu\nu\rho\sigma}F^{\mu\nu}F^{\rho\sigma}-R_{\mu\nu}R^{\mu\nu}+\frac{1}{2}R^2-\Lambda R -\frac{1}{2}R F_{\rho\sigma}F^{\rho\sigma}~, 
    \\
    (D_{\mu}\tensor{F}{_{\rho}^{\nu}})(D_{\nu}F^{\rho\mu}) &= \frac{1}{2}(D_{\rho}F_{\mu\nu})(D^{\rho}F^{\mu\nu})~.
    \end{split}
\end{align}
These identities can be found by using the Bianchi identity \eqref{eqn: gravitybianchi app} and \eqref{eqn: gaugebianchi app} followed by an integration of parts, dropping the boundary terms along the way, and imposing the commutator relations \eqref{eqn: bosoncommutator} of the covariant derivatives acting on the gauge field. Note that \eqref{identities} are on-shell since we have explicitly imposed the Maxwell equations and Einstein equations. Then,
\begin{align}
    \begin{split}
    X\equiv~&\frac{1}{2}\Tr \left(\left(D^{\mu} F^{\alpha \nu}+D^{\nu} F^{\alpha \mu}\right) \left(D^{\rho} F^{\beta \sigma}+D^{\sigma} F^{\beta \rho}\right)\right)
    \\&+
    \frac{1}{2}\Tr \left(\left(D^{\mu} F^{\alpha \nu}+D^{\nu} F^{\alpha \mu}\right)\left(D^{\rho} F^{\beta \sigma}+D^{\sigma} F^{\beta \rho}\right)\right)
    \\=&
    \left(D^{\mu} F^{\alpha \nu}+D^{\nu} F^{\alpha \mu}\right) \left(D^{\rho} F^{\beta \sigma}+D^{\sigma} F^{\beta \rho}\right)g_{\alpha\beta}G_{\mu\nu\rho\sigma}~.
    \end{split}
\end{align}
Imposing the Bianchi identities simplifies our expression to
\begin{align}
    X=2(D_{\rho}F_{\alpha\sigma})(D^{\rho} F^{\alpha \sigma})+ 2(D_{\rho}F_{\alpha\sigma})(D^{\sigma} F^{\alpha \rho})~.
\end{align}
Finally, using \eqref{identities}, we find
\begin{align}
    \begin{split}
    X=& 2\left(-R^{\mu\nu}R_{\mu\nu}+\frac{1}{2}R^2 - \Lambda R - \frac{1}{2}R F_{\rho \sigma} F^{\rho  \sigma}+R_{\mu\nu\rho\alpha}F^{\mu\nu}F^{\rho\alpha} \right) 
    \\&
    + \left(R_{\mu\nu\rho\sigma}F^{\mu\nu}F^{\rho\sigma}-R_{\mu\nu}R^{\mu\nu}+\frac{1}{2}R^2-\Lambda R -\frac{1}{2}R F_{\rho\sigma}F^{\rho\sigma}+R_{\mu\nu\rho\alpha}F^{\mu\nu}F^{\rho\alpha}\right)
    \\=&
    3\left(-R^{\mu\nu}R_{\mu\nu}+\frac{1}{2}R^2 - \Lambda R - \frac{1}{2}R F_{\rho \sigma} F^{\rho  \sigma}+R_{\mu\nu\rho\alpha}F^{\mu\nu}F^{\rho\alpha} \right)~.
    \end{split}
\end{align}
Putting all the contributions together, the trace of the square of $E$ is therefore
\begin{align}
    \begin{split}
    \Tr E^2 =&  \left(16 \Lambda^2 - 4 R_{ab} R^{ab} + 3 R_{abcd} R^{abcd} \right) + \left( 4\Lambda^2 -12 \Lambda F_{\alpha\beta}F^{\alpha\beta}+ 9 \left(F_{\theta\phi} F^{\theta\phi} \right)^2 \right)\\
    &+3\left(-R^{\mu\nu}R_{\mu\nu}+\frac{1}{2}R^2 - \Lambda R - \frac{1}{2}R F_{\rho \sigma} F^{\rho  \sigma}+R_{\mu\nu\rho\alpha}F^{\mu\nu}F^{\rho\alpha} \right)\\
    =&32\Lambda^2-7R_{\mu\nu}R^{\mu\nu} + 3R_{\mu\nu\rho\sigma}R^{\mu\nu\rho\sigma}-18\Lambda F_{\rho\sigma}F^{\rho\sigma}+3R_{\mu\nu\rho\alpha}F^{\mu\nu}F^{\rho\alpha}+ 9 \left(F_{\theta\phi} F^{\theta\phi} \right)^2~.
    \end{split}
\end{align}
The necessary traces are summarised below:
\begin{align}
\label{eqn: bosonbulkI}
    \Tr (I) &= 14~,
    \\
\label{eqn: bosonbulkE}
    \Tr (R E) &= - 32 \Lambda^2 + 6 R F_{\mu \nu} F^{\mu \nu}~,
    \\
\begin{split}
    \label{eqn: bosonbulkE2}
    \Tr (E^2) &=
    32\Lambda^2-7R_{\mu\nu}R^{\mu\nu} + 3R_{\mu\nu\rho\sigma}R^{\mu\nu\rho\sigma}-18\Lambda F_{\rho\sigma}F^{\rho\sigma}
    \\& \qquad +3R_{\mu\nu\rho\sigma}F^{\mu\nu}F^{\rho\sigma}+ 9 \left(F_{\theta\phi} F^{\theta\phi} \right)^2~,
\end{split}\\
\begin{split}
    \label{eqn: bosonbulkOmega2}
    \Tr (\Omega^2) &= -224 \Lambda^2 + 60 \Lambda F_{\mu\nu} F^{\mu\nu} - 54 F_{\mu\nu} F^{\mu\nu} F_{\rho\sigma} F^{\rho\sigma} + 56 R_{\mu\nu} R^{\mu\nu}
    \\& \qquad -18 F^{\mu\nu} F^{\rho\sigma} R_{\mu\nu\rho\sigma} - 7 R_{\mu\nu\rho\sigma} R^{\mu\nu\rho\sigma}~.
\end{split}
\end{align}
Substituting the traces \eqref{eqn: bosonbulkI}-\eqref{eqn: bosonbulkOmega2} into \eqref{intro:a4}, we obtain the fourth heat kernel coefficient of Einstein-Maxwell AdS theory without ghosts
\begin{align}
\label{Eq: a4EM}
    (4\pi)^2 a_{4}^{\text{EM}}(x) &= -\frac{880}{180}\Lambda^2 + \frac{196}{180}R_{\mu\nu}R^{\mu\nu} + \frac{179}{180}R_{\mu\nu\rho\sigma}R^{\mu\nu\rho\sigma}~.
\end{align}

\subsubsection{Ghost contribution}

\label{app: bosonghost}

The addition of the gauge-fixing Lagrangian \eqref{eqn: EMgf app} introduces an action for the ghosts, given by
\begin{align}
\label{eqn: EMgh app}
    \mathcal{S}_{\mathrm{ghost}, b}=\frac{1}{2} \int d^{4} x \sqrt{g}\Big\{2 b_{\mu}\big(g^{\mu \nu} \Box+R^{\mu \nu}\big) c_{\nu}+2 b \, \Box \, c-4 b F^{\rho \nu} D_{\rho} c_{\nu}\Big\}~,
\end{align}
where $b_{\mu}$ and $c_{\mu}$ are vector fields and $b$ and $c$ are scalar fields. From these expressions, we can extract the matrices $E$ and $\Omega$ as
\begin{align}
    \begin{aligned}
        \phi_{n} E_{m}^{n} \phi^{m}=& b_{\mu}\left(R_{~\nu}^{\mu}\right) b^{\nu}+c_{\mu}\left(R_{~\nu}^{\mu}\right) c^{\nu} ,
        \\
        \phi_{n}\left(\Omega_{\alpha \beta}\right)_{m}^{n} \phi^{m}=& b_{\mu}\left(R_{~\nu \alpha \beta}^{\mu}\right) b^{\nu}+c_{\mu}\left(R_{~\nu \alpha \beta}^{\mu}\right) c^{\nu}-\frac{1}{2}\left(b_{\mu}-i c_{\mu}\right)\left(D^{\mu} F_{\alpha \beta}\right)(b+i c) 
        \\
        &+\frac{1}{2}(b+i c)\left(D_{\nu} F_{\alpha \beta}\right)\left(b^{\nu}-i c^{\nu}\right)~,
    \end{aligned}
    \label{eqn: bosonghost app}
\end{align}
Note that in the case of the ghost fields, we are raising and lowering the indices with $g^{\alpha\beta}$ and $\mathbf{1}$. The result for the heat kernel is
\be
a_4^\text{ghosts,EM}(x) = {13\/36} E_4 - {1\/4} W^2-{3\/4}R^2~.
\ee
where we have already included here the negative sign due to the opposite statistics.

\subsection{Logarithmic correction}

Adding the above results, the heat kernel for Einstein-Maxwell-AdS theory takes the form,
\be\label{a4B app}
(4\pi)^2 a_4^{\text{B}}(x) = -{53\/45}E_4 + {137\/60} W^2 - {13\/36} R^2 ~.
\ee
We can read off the coefficients from \eqref{a4general} to be
\be
a_\text{E} = {53\/45},\qq c = {137\/60},\qq b_1 = -{13\/36}~,\qq b_2= 0~.
\ee

\subsection{Neutral limit}\label{sec:neutrallimit}

As previously mentioned, if we properly truncate the fluctuations and the resulting curvature invariants, we recover the result obtained in \cite{Christensen:1979iy} for the theory of pure gravity with a negative  cosmological constant. Let us show this explicitly as a sanity check of our results. In this limit, we must truncate the fluctuation of $a_\alpha$ in \eqref{appeqn: bosonbulkE} 
\begin{equation}
    \phi_{m} E^{m n} \phi_{n}= h_{\mu \nu}\left(R^{\mu \alpha \nu \beta}+R^{\mu \beta \nu \alpha}-g^{\mu \nu} R^{\alpha \beta}-g^{\alpha \beta} R^{\mu \nu} + \Lambda g^{\mu\nu}g^{\alpha\beta} \right) h_{\alpha \beta}~.
\end{equation}
This yields the following traces
\begin{equation}
    \begin{split}
        &\Tr(RE)=-16\Lambda^2~, \\
        &\Tr(E^2)=16\Lambda^2-4R_{\mu\nu}R^{\mu\nu}+3R_{\mu\nu\rho\sigma}R^{\mu\nu\rho\sigma}~, \\
        &\Tr(I)=10~.
    \end{split}
\end{equation}
Note that the trace of $I$ is 10 instead of 14 because we no longer have the fluctuation $a_{\mu}$, this is the kind of intermediate result that makes a naive truncation of the final answer yield the wrong result. Moreover, the field strength $\Omega$ is simply the commutator of $\nabla$. Its trace is well-known and takes the value $\Tr(\Omega^2) = -6R_{\mu\nu\rho\sigma}R^{\mu\nu\rho\sigma}$. Combining these results, we have 
\begin{equation}
    \begin{split}
        180(4\pi)^2a_4^{\text{bulk},\Lambda}=&\frac{1}{2}\Big[60(-16\Lambda^2)+180(16\Lambda^2-4R_{\mu\nu}R^{\mu\nu}+3R^{\mu\nu\alpha\beta}R_{\mu\nu\alpha\beta})+30(-6)R^{\mu\nu\alpha\beta}R_{\mu\nu\alpha\beta}\\
        &\indent+10(5R^2+2R^{\mu\nu\alpha\beta}R_{\mu\nu\alpha\beta}-2R_{\mu\nu}R^{\mu\nu} \Big]\\
        =&-120\Lambda^2+190R^{\mu\nu\alpha\beta}R_{\mu\nu\alpha\beta}~.
    \end{split}
\end{equation}
In the second line, we used the field equation $R_{\mu\nu}=g_{\mu\nu}\Lambda$. For the ghost contribution, we do not have the ghost of the graviphoton, {\it  i.e.}, the scalar ghosts $b$ and $c$. 
Therefore, the matrix $E$ in the neutral limit remains the same and $\Omega$ reduces to 
\begin{align}
    \begin{aligned}
        \phi_{n}\left(\Omega_{\alpha \beta}\right)_{m}^{n} \phi^{m}=& b_{\mu}\left(R_{\nu \alpha \beta}^{\mu}\right) b^{\nu}+c_{\mu}\left(R_{\nu \alpha \beta}^{\mu}\right) c^{\nu}~.
    \end{aligned}
    \label{eqn: neutralbosonghost}
\end{align}
The only change in the trace is the $I$ operator
\begin{align}
    \Tr(I)=\Tr(g^{\mu\nu})+\Tr(g^{\mu\nu})=8~,
\end{align}
which yields the ghost contribution in the neutral limit
\begin{align}
\begin{split}
    -(4 \pi)^{2} a_{4}^{\text{ghost,B}}(x)&=\frac{1}{3}R^2 + R_{\mu\nu}R^{\mu\nu} -\frac{1}{6}R_{\mu\nu\rho\sigma}R^{\mu\nu\rho\sigma}+(\frac{1}{9}R^2+\frac{2}{45}R_{\mu\nu\rho\sigma}R^{\mu\nu\rho\sigma}-\frac{2}{45}R_{\mu\nu}R^{\mu\nu})
    \\&=
    \frac{4}{9}R^2 + \frac{43}{45}R_{\mu\nu}R^{\mu\nu}-\frac{43}{45}R_{\mu\nu\rho\sigma}R^{\mu\nu\rho\sigma}\\&=
    \frac{11}{90}R_{\mu\nu\rho\sigma}R^{\mu\nu\rho\sigma}-\frac{164}{15}\Lambda^2~,
\end{split}
\end{align}
where we use the field equation $R_{\mu\nu}=g_{\mu\nu}\Lambda$ once more. In the neutral limit, the heat kernel coefficient is therefore
\begin{equation}
    180(4\pi)^2a_{4}^{\Lambda}(x)=212R_{\mu\nu\rho\sigma}R^{\mu\nu\rho\sigma}-2088\Lambda^2~,
\end{equation}
which agrees with the result in \cite{Christensen:1979iy}. 

\section{Fermionic computation}\label{app:fermions}

In this appendix, we present the details of the computation for the gravitini of minimal gauged supergravity described in section \ref{Sec:N2Sugra}.

\subsection{Majorana Lagrangian}

The fermionic Lagrangian is given as \cite{Freedman:1976aw}
\bea
\cL \=  {1\/2} \bpsi_\mu\g^{\mu\nu\rho} D_\nu\psi_\rho + {i\/4}  F^\mn \bpsi_\rho \g_{\mu} \g^\rs \g_{\nu} \psi_\s -{1\/2\l} \bpsi_\mu \g^\mn \psi_\nu~,
\eea
where $\psi_\mu$ is a complex spinor with spin ${3\/2}$. It can be written as
\be
\psi_\mu = \psi_\mu^1 +  i \psi_\mu^2~,
\ee
where $\psi_\mu^A$ are Majorana spinors. For definiteness, we use the really real representation of the Clifford algebra given by explicitly as \cite{Freedman:2012zz}
\bea \label{gamma matrices}
\g^0 = \bpm 0 & {\bf 1} \\ -{\bf 1} & 0\epm,\qq \g^1 = \bpm {\bf 1} & 0 \\ 0 & -{\bf 1}\epm,\qq \g^2 = \bpm 0 & \s_1 \\ \s_1 & 0\epm,\qq \g^3 = \bpm 0 & \s_3 \\ \s_3 & 0\epm~.
\eea
In this case, the Majorana condition reduces to
\be
\psi^\ast=\psi,
\ee
which is just the reality condition.  We choose the gauge $\g^\mu\psi_\mu=0$. This can be implemented by the gauge-fixing term
\be
\cL_\r{g.f.}=- {1\/4} (\bpsi_{\mu} \g^\mu) \g^\nu D_\nu (\g^\rho \psi_\rho)~.
\ee
More details about the Faddeev-Popov procedure are given in the following subsection when we consider the ghost contribution. We then perform the field redefinition
\be
\psi_\mu = \sqrt{2}\le( \chi_\mu - {1\/2}\g_\mu \g^\nu \chi_\nu\ri)~,
\ee
which leads to the Lagrangian
\be
\cL_\r{Fermi}+\cL_\r{g.f.}= \cL_\r{kin}+\cL_{\bpsi F\psi} + \cL_{\bpsi\psi}~,
\ee
where
\bea
\cL_\r{kin}\= {1\/2} g^\mn \bpsi_\mu \g^\rho \n_\rho \psi_\nu ~,\-
\cL_{\bpsi F\psi} \= {i\/4}  F^\mn \bpsi_\rho \g_{\mu} \g^\rs \g_{\nu} \psi_\s ~,\-
\cL_{\bpsi\psi}\=-{1\/2\l} \bpsi_\mu \g^\mn \psi_\nu~.
\eea
For the kinetic term, we can use the fact that (see \cite{Banerjee:2010qc, Charles:2015eha} for details)
\be
\bchi_\mu\g^\nu D_\nu \chi^\mu= {1\/2} \bpsi_\mu \le(\g^{\mn\rho}D_\nu - {1\/2} \g^\mu\g^\nu\g^\rho D_\nu\ri)\psi_{\rho}~.
\ee
Note that gauge invariance guarantees that this identity holds also for the gauge connection. For the mass term, we have
\bea
\cL_{\bpsi\psi}\=-{1\/2\l}  \bpsi_\mu \g^\mn \psi_\nu~\-
\=-{1\/\l}  \le(\bchi_\mu -{1\/2} \bchi_\rho \g^\rho\g_\mu \ri) \g^\mn \le(\chi_\nu- {1\/2} \g_\nu \g^\s \chi_\s\ri) \-
\=-{1\/\l}  \le(\bchi_\mu \g^\mn\chi_\nu-{1\/2} \bchi_\rho \g^\rho\g_\mu \g^\mn\chi_\nu- {1\/2} \bchi_\mu  \g^\mn \g_\nu \g^\s \chi_\s + {1\/4} \bchi_\rho \g^\rho\g_\mu  \g^\mn  \g_\nu \g^\s \chi_\s\ri)  \-
\=-{1\/\l}  \le(\bchi_\mu \g^\mn\chi_\nu-3 \bchi_\mu \g^\mu \g^\nu\chi_\nu+ 3\bchi_\rho \g^\rho \g^\s \chi_\s\ri)  \-
\= -{1\/\l} \bchi_\mu \g^\mn\chi_\nu~,
\eea
where we have used $\g^\mn = \g^\mu \g^\nu - g^\mn$ so that $\g_\mu \g^\mn = 3\g^\nu$ and $\g_\mu \g^\mn \g_\nu=12$. Finally, we have
\bea \label{intermediate step in LF}
\cL_{\bpsi F \psi} \= {i\/4}  F^\mn \bpsi_\rho \g_{\mu} \g^\rs \g_{\nu} \psi_\s \-
\={i\/2}  F^\mn \le( \bchi_\rho- {1\/2}\bchi_\a\g^\a \g_\rho \ri) \g_{\mu} \g^\rs \g_{\nu} \le(\chi_\s- {1\/2} \g_\s \g^\b \chi_\b\ri) \-
\={i\/2}  F^\mn \le( \bchi_\rho \g_{\mu} \g^\rs \g_{\nu} \chi_\s- {1\/2}\bchi_\a\g^\a \g_\rho \g_{\mu} \g^\rs \g_{\nu}\chi_\s -{1\/2} \bchi_\rho \g_{\mu} \g^\rs \g_{\nu}  \g_\s \g^\b \chi_\b\ri.\-
&& \le.+ {1\/4}\bchi_\a\g^\a \g_\rho\g_{\mu} \g^\rs \g_{\nu}  \g_\s \g^\b \chi_\b \ri)  ~.
\eea
To simplify, we use the following gamma matrix identities
\begin{align}
\begin{split}
\g^\mu \g_\nu \g_\mu &=-2\g_\nu,
\\
\g_\rho \g_\mu \g^\rs  &= \g_\rho\g_\mu(\g^\rho\g^\s - g^\rs) = - \g_\mu \g^\s -  2 \d^\s_\mu~,
\\
\g_\rho \g_\mu \g^\rs \g_\nu\g_\s &= ( - \g_\mu \g^\s -  2 \d^\s_\mu)\g_\nu\g_\s = 4 \g_\mn ~.
\end{split}
\end{align}
For the second term in \eqref{intermediate step in LF}, we have
\bea \label{intermediate step in LF 2}
- {1\/2}\bchi_\a\g^\a \g_\rho \g_{\mu} \g^\rs \g_{\nu}\chi_\s\= {1\/2}\bchi_\a\g^\a ( \g_\mu \g^\s + 2 \d^\s_\mu)\g_{\nu}\chi_\s\-
\= {1\/2}\bchi_\a\g^\a  \g_\mu \g^\s\g_{\nu}\chi_\s + \bchi_\a\g^\a  \g_{\nu}\chi_\mu\-
\= {1\/2}\bchi_\a\g^\a \g_\mu(2\d^\s_\nu - \g_\nu\g^\s)\chi_\s + \bchi_\a\g^\a  \g_{\nu}\chi_\mu\-
\= \bchi_\a\g^\a  \g_\mu\chi_\nu -{1\/2}\bchi_\a\g^\a  \g_\mu \g_\nu\g^\s\chi_\s + \bchi_\a\g^\a  \g_{\nu}\chi_\mu~.
\eea
After contracting with $ F^\mn$, the first and last terms in \eqref{intermediate step in LF} cancel due to antisymmetry. Finally, symmetry arguments show that the third term in \eqref{intermediate step in LF} gives the same simplification as \eqref{intermediate step in LF 2}. At the end, we obtain
\bea
\cL_{ \bpsi F \psi}\={i\/2}  F^\mn \le( \bchi_\rho \g_{\mu} \g^\rs \g_{\nu} \chi_\s-\bchi_\a\g^\a  \g_\mu \g_\nu\g^\s\chi_\s+\bchi_\a\g^\a \g_\mn\g^\b \chi_\b \ri)  ~\-
\= {i\/2}  F^\mn  \bchi_\rho \g_{\mu} \g^\rs \g_{\nu} \chi_\s~.
\eea
We thus obtain the Lagrangian
\bea
\cL_f \=  g^{\mn}\bchi_\mu\g^{\rho} D_\rho\chi_\nu + {i\/2}  F^\mn \bchi_\rho \g_{\mu} \g^\rs \g_{\nu} \chi_\s -{1\/\l} \bchi_\mu \g^\mn \chi_\nu.
\eea
We introduce the complex spinor $\chi_{\mu}$ as
\be
\chi_\mu=\chi_\mu^1+ i \chi_\mu^2~,
\ee
in terms of Majorana spinors. We use the label $A=1,2$ for the two spinors and make use of Majorana flip identities \cite{Freedman:2012zz}
\be\label{MajoranaFlip}
\bla \g_{\mu_1\dots\mu_r} \chi = t_r \bchi \g_{\mu_1\dots\mu_r}\la~,
\ee
where 
\be
t_r = \begin{cases} 
-1 & r =1,2 ~\r{mod}~4\\
1 & r =3,4 ~\r{mod}~4
\end{cases}~.
\ee
The sign $t_r$ reflects the symmetry of the gamma matrices under charge conjugation. Another useful identity is \cite{Freedman:2012zz}
\be
\bar\la \g^{\mu_1}\g^{\mu_2}\dots \g^{\mu_p}\chi= (-1)^p \bar\chi \g^{\mu_p}\dots\g^{\mu_2}\g^{\mu_1}\la .
\ee
Each term in the Lagrangian can be simplified using Majorana flips and is either proportional to the identity matrix $\d_{AB}$ or the antisymmetric matrix $\ve_{AB}$ (wtih $\ve^{12}=1$). In the kinetic term, the cross-terms cancel
 \bea
\bchi_\mu^1 \g^\rho D_\rho \chi_\nu^2 \=- D_\rho\bar\chi_\nu^2 \g^\rho \chi_\mu^1\-
\= \bchi_\nu^2 \g^\rho D_\rho \chi_\mu^1
\eea
where we used a Majorana flip and integration by parts. Hence we have
\be
\cL_\r{kin}=\d_{AB} g^{\mn}\bchi_\mu^A\g^{\rho} D_\rho\chi_\nu^B ~.
\ee
We then have
\bea
  F^\mn \bpsi_\rho^1 \g_{\mu} \g^\rs \g_{\nu} \psi_\s^1 \= F^\mn ( \g_\mu\psi^1_\rho)^\dg  \g^\rs (\g_{\nu} \psi_\s^1) \-
 \=  - F^\mn ( \g_\nu\psi^1_\s)^\dg  \g^\rs (\g_{\mu} \psi_\rho^1) \-
 \= - F^\mn \bpsi_\rho^1 \g_{\mu} \g^\rs \g_{\nu} \psi_\s^1\-
 \=0
\eea
where we used a Majorana flip in the second line. This shows that
\be
\cL_{\bpsi F\psi}=-{1\/2}\ve_{AB}  F^\mn \bchi_\rho^A \g_{\mu} \g^\rs \g_{\nu} \chi_\s^B~.
\ee
Finally, we have
\bea
\bchi^1_\mu \g^\mn \chi_\nu^2 \=- \bchi_\nu^2 \g^\mn \chi_\mu^1 \-
\=\bchi_\mu^2 \g^\mn \chi_\nu^1 
\eea
where we used a Majorana flip and antisymmetry of $\g^\mn$. This shows that mass term is
\be
\cL_{\bpsi\psi} =-{1\/\l}\d_{AB} \bchi_\mu^A \g^\mn \chi_\nu^B~.
\ee
The final Lagrangian is then
\bea \label{Lagragian in chi}
\cL_f +\cL_\text{g.f.}\=  \d_{AB} g^{\mn}\bchi_\mu^A\g^{\rho} D_\rho\chi_\nu^B -{1\/2}\ve_{AB}  F^\mn \bchi_\rho^A \g_{\mu} \g^\rs \g_{\nu} \chi_\s^B -{1\/\l}\d_{AB} \bchi_\mu^A \g^\mn \chi_\nu^B~.
\eea
This Lagrangian could now be interpreted as a Euclidean Lagrangian by performing the Wick rotation and using $\bar\chi_A = \chi_A^\dg$. This can then be used in the algorithm to compute the logarithmic corrections.

\subsection{Symplectic Lagrangian}

We are ultimately interested in the fermionic Lagrangian in $(0,4)$ signature. It is known that Majorana spinors do not exist in $(0,4)$ signature \cite{VanProeyen:1999ni,Cortes:2003zd}. Instead, we should use symplectic Majorana spinors. Thus, we first convert our Lagrangian from Majorana to symplectic Majorana spinors in $(1,3)$ signature, where both  Majorana and symplectic Majorana spinors exist. We then perform the Wick rotation to obtain the Lagrangian in $(0,4)$ signature.

\sss{Symplectic Majorana spinors}

 The symmetries of the gamma matrices are captured by matrices $A,B$ and $C$. The matrix $A$ expresses the Hermitian conjugate of a gamma matrix as
\begin{equation}
\left(\gamma^{\mu}\right)^{\dagger}=(-1)^{t} A \gamma^{\mu} A^{-1}~
\end{equation}
and we can take $A = -\g^0$. The charge conjugation matrix gives the transpose as
\begin{equation}
(\gamma^{\mu})^t= t_0 t_1 C \gamma^{\mu} C^{-1},
\end{equation}
and satisfies $C^t = - t_0 C$. Here $t_0,t_1$ can take the values $\pm 1$. The matrix $B$ captures the complex conjugate
\be
(\g^\mu)^\ast = - t_0 t_1 B \g^\mu B^{-1}~.
\ee
and can be obtained as
\be
B = (C A^{-1})^t~.
\ee
We also define
\be
\g^5 = - i \g^0\g^1\g^2\g^3 = \bpm 0 & \s_2 \\ \s_2 & 0 \epm
\ee
given in the representation \eqref{gamma matrices}. Note that we have
\be
(\g^5)^2 = 1 ,\qq (\g^5)^\dg = \g^5 ,\qq (\g^5)^t = (\g^5)^\ast = -\g^5~.
\ee
There are two possible choice of charge conjugation matrix which we will denote $C_+$ and $C_-$. The $C_+$ matrix has $t_0=1,t_1=-1$. It gives a matrix $B_+ = (C_+ A^{-1})^t = \mathbf{1}$. The Majorana condition is then written as 
\be
\psi^\ast = B_+ \psi = \psi~.
\ee
To define symplectic Majoranas, we need to use another charge conjugation matrix $C_- = C_+ \g^5$ which has $t_0=t_1=1$. It gives the matrix 
\be
B_- = (C_-A^{-1})^t = \g^5~.
\ee
The symplectic Marojana condition can be written as 
\be
(\la_A^\mu)^\ast = B_- \ve_{AB} \la_B^\mu~.
\ee
The mapping between Majoranas and symplectic Majoranas in $(1,3)$ signature is given in  \cite{Cortes:2003zd} and takes the form
\bea\label{latochi}
\la_1^\mu \= \chi_1^\mu - i\chi_\mu^2~,\\
\la_2^\mu \= \g^5 (\chi_1^\mu + i \chi_2^\mu)~,
\eea
where we have used that $B_-=\g^5$. This gives
\bea
\chi_1^\mu \= {1\/2} (\la_1^\mu + \g^5 \la_2^\mu)~,\\
\chi_2^\mu \= {i\/2} (\la_1^\mu - \g^5 \la_2^\mu)~.
\eea
It is also useful to note that the Dirac conjugated defined as $\bar\chi_A^\mu = (\chi_A^\mu) i\g^0$ gives
\bea
\bar\chi_1^\mu \= {1\/2}(\bar\la_1^\mu - \bar\la_2^\mu \g^5)~, \\
\bar\chi_2^\mu \= - {i\/2} (\bar\la_1^\mu +\bar\la_2^\mu \g^5)~.
\eea
We will write the Majorana Lagrangian \eqref{Lagragian in chi} as
\be
\cL_f +\cL_\text{g.f.} = \cL_\r{kin} + \cL_{\chi F\chi} + \cL_{\chi\chi}
\ee
where
\bea
\cL_\r{kin} \= \d_{AB} g_{\mn}\bchi^\mu_A\g^{\rho} D_\rho\chi^\nu_B~,\-
\cL_{\chi F\chi}\=-{1\/2}\ve_{AB}  F^\mn \bchi^\rho_A \g_{\mu} \g_\rs \g_{\nu} \chi^\s_B ~,\-
\cL_{\chi\chi}\= -{1\/\l}\d_{AB} \bchi^\mu_A \g_\mn \chi^\nu_B~.
\eea
We will now convert these terms one by one.

\sss{Kinetic term}

For the kinetic term, we compute
\bea
g_\mn \bchi^\mu_1 \g^\rho\n_\rho \chi^\nu_1\= {1\/4} g_\mn (\bla_1^\mu - \bla_2^\mu\g^5) \g^\rho\n_\rho (\la_1^\mu +\g^5 \la_2^\mu) \-
\= {1\/4} g_\mn (\bla_1^\mu\g^\rho\n_\rho \la_1^\mu - \bla_2^\mu\g^5\g^\rho\n_\rho\g^5 \la_2^\mu + \bla_1^\mu\g^\rho\n_\rho\g^5 \la_2^\mu - \bla_2^\mu\g^5\g^\rho\n_\rho \la_1^\mu ) \-
\= {1\/4} g_\mn (\bla_1^\mu\g^\rho\n_\rho \la_1^\mu + \bla_2^\mu\g^\rho\n_\rho \la_2^\mu + \bla_1^\mu\g^\rho\n_\rho\g^5 \la_2^\mu - \bla_2^\mu\g^5\g^\rho\n_\rho \la_1^\mu ).
\eea
The other contribution is
\bea
g_\mn \bchi^\mu_2 \g^\rho\n_\rho \chi^\nu_2\= {1\/4} g_\mn (\bla_1^\mu + \bla_2^\mu\g^5) \g^\rho\n_\rho (\la_1^\mu -\g^5 \la_2^\mu) \-
\= {1\/4} g_\mn (\bla_1^\mu\g^\rho\n_\rho \la_1^\mu - \bla_2^\mu\g^5\g^\rho\n_\rho\g^5 \la_2^\mu - \bla_1^\mu\g^\rho\n_\rho\g^5 \la_2^\mu + \bla_2^\mu\g^5\g^\rho\n_\rho \la_1^\mu ) \-
\= {1\/4} g_\mn (\bla_1^\mu\g^\rho\n_\rho \la_1^\mu + \bla_2^\mu\g^\rho\n_\rho \la_2^\mu - \bla_1^\mu\g^\rho\n_\rho\g^5 \la_2^\mu + \bla_2^\mu\g^5\g^\rho\n_\rho \la_1^\mu ).
\eea
Therefore, we get
\be
\cL_\text{kin} = {1\/2} g_\mn (\bla_1^\mu\g^\rho\n_\rho \la_1^\mu + \bla_2^\mu\g^\rho\n_\rho \la_2^\mu ).
\ee
We now compute the contribution from the gauge connection. The Majorana spinors $(\chi_1^\mu \hspace{0.1cm} \chi_2^\mu)$ form a doublet under the $\r{SO}(2)\cong \r{U}(1)$ gauge symmetry. We have
\be
D_\mu \chi^\nu_A = \le(\d_{AB} \n_\mu +{1\/\l}\ve_{AB} A_\mu\ri) \chi^\nu_B~,
\ee
or more explicitly
\bea
D_\mu \chi^\nu_1 \=  \n_\mu\chi^\nu_1 +{1\/\l} A_\mu\chi^\nu_2 ~,\\
D_\mu \chi^\nu_2 \=  \n_\mu\chi^\nu_2 -{1\/\l} A_\mu\chi^\nu_1 ~.
\eea
Using \eqref{latochi}, we see that
\bea
D_\mu \la_1^\nu \= D_\mu (\chi_1^\nu - i  \chi_2^\nu) \-
\= \n_\mu\chi^\nu_1 +{1\/\l} A_\mu\chi^\nu_2  - i\le( \n_\mu\chi^\nu_2 -{1\/\l} A_\mu\chi^\nu_1 \ri)\-
\= \n_\mu \la_1^\nu + {i\/\l} A_\mu \la_1^\nu~,
\eea
and
\bea
D_\mu \la_2^\nu \= \g^5 D_\mu (\chi_1^\nu +i  \chi_2^\nu) \-
\= \g^5\le(\n_\mu\chi^\nu_1 +{1\/\l} A_\mu  \chi^\nu_2\ri)  + i \g^5 \le( \n_\mu\chi^\nu_2 -{1\/\l} A_\mu\chi^\nu_1\ri) \-
\= \n_\mu \la_2^\nu - {i\/\l} A_\mu \la_2^\nu~.
\eea
As a result, we see that $\la_1^\mu$ and $\la_2^\mu$ are singlet under the $\r{U}(1)$ gauge symmetry and have opposite charges. By gauge invariance, the kinetic term including the gauge connection is then
\be
\cL_\text{kin} = {1\/2} g_\mn (\bla_1^\mu\g^\rho D_\rho \la_1^\mu + \bla_2^\mu\g^\rho D_\rho \la_2^\mu  )~.
\ee

\sss{Mass term}

We have
\bea
\cL_{\chi\chi} = -{1\/\l} \d_{AB} \bchi^\mu_A \g_\mn \chi^\nu_B.
\eea
We compute
\bea
\bchi_1^\mu \g_\mn \chi_1^\nu \= {1\/4} (\bla_1^\mu - \bla_2^\mu\g^5) \g_\mn  (\la_1^\nu +\g^5 \la_2^\nu) \-
\= {1\/4}\le( \bla_1^\mu\g_\mn \la_1^\nu - \bla_2^\mu\g^5 \g_\mn \g^5 \la_2^\nu + \bla_1^\mu\g_\mn \g^5 \la_2^\nu - \bla_2^\mu\g^5 \g_\mn \la_1^\nu\ri)\-
\= {1\/4}\le( \bla_1^\mu\g_\mn \la_1^\nu - \bla_2^\mu\g_\mn  \la_2^\nu + \bla_1^\mu\g_\mn \g^5 \la_2^\nu - \bla_2^\mu\g^5 \g_\mn \la_1^\nu\ri),
\eea
as well as
\bea
\bchi_2^\mu \g_\mn \chi_2^\nu \= {1\/4} (\bla_1^\mu + \bla_2^\mu\g^5) \g_\mn  (\la_1^\nu -\g^5 \la_2^\nu) \-
\= {1\/4}\le( \bla_1^\mu\g_\mn \la_1^\nu - \bla_2^\mu\g_\mn  \la_2^\nu - \bla_1^\mu\g_\mn \g^5 \la_2^\nu +\bla_2^\mu\g^5 \g_\mn \la_1^\nu\ri).
\eea
We see that the cross terms cancel upon addition of the two contributions and we end up with
\be
\cL_{\chi\chi} =  -{1\/2\l}\le( \bla_1^\mu\g_\mn \la_1^\nu - \bla_2^\mu\g_\mn  \la_2^\nu \ri).
\ee

\sss{Gauge interaction term}

We compute
\begin{align}
    \begin{split}
        &F^\mn \bchi^\rho_1 \g_{\mu} \g_\rs \g_{\nu} \chi^\s_2 \\&= {i\/4} F^\mn  (\bla_1^\rho - \bla_2^\rho\g^5) \g_{\mu} \g_\rs \g_{\nu}(\la_1^\s -\g^5 \la_2^\s)
        \\&= {i\/4} F^\mn \le( \bla_1^\rho  \g_{\mu} \g_\rs \g_{\nu} \la_1^\s + \bla_2^\rho\g^5  \g_{\mu} \g_\rs \g_{\nu} \g^5 \la_2^\s -\bla_1^\rho  \g_{\mu} \g_\rs \g_{\nu}\g^5 \la_2^\s -\bla_2^\rho\g^5  \g_{\mu} \g_\rs \g_{\nu}\la_1^\s \ri)
        \\&= {i\/4} F^\mn \le( \bla_1^\rho  \g_{\mu} \g_\rs \g_{\nu} \la_1^\s + \bla_2^\rho  \g_{\mu} \g_\rs \g_{\nu}  \la_2^\s -\bla_1^\rho  \g_{\mu} \g_\rs \g_{\nu}\g^5 \la_2^\s -\bla_2^\rho\g^5  \g_{\mu} \g_\rs \g_{\nu}\la_1^\s \ri),
    \end{split}
\end{align}
and
\begin{align}
    \begin{split}
        &F^\mn \bchi^\rho_2 \g_{\mu} \g_\rs \g_{\nu} \chi^\s_1 
        \\&=
        -{i\/4} F^\mn (\bla_1^\rho + \bla_2^\rho\g^5)\g_{\mu} \g_\rs \g_{\nu}(\la_1^\s +\g^5 \la_2^\s)
        \\&=
        -{i\/4} F^\mn \le( \bla_1^\rho  \g_{\mu} \g_\rs \g_{\nu} \la_1^\s + \bla_2^\rho\g^5  \g_{\mu} \g_\rs \g_{\nu} \g^5 \la_2^\s +\bla_1^\rho  \g_{\mu} \g_\rs \g_{\nu}\g^5 \la_2^\s +\bla_2^\rho\g^5  \g_{\mu} \g_\rs \g_{\nu}\la_1^\s \ri)
        \\&=
        -{i\/4} F^\mn \le( \bla_1^\rho  \g_{\mu} \g_\rs \g_{\nu} \la_1^\s + \bla_2^\rho\g_{\mu} \g_\rs \g_{\nu} \la_2^\s +\bla_1^\rho  \g_{\mu} \g_\rs \g_{\nu}\g^5 \la_2^\s +\bla_2^\rho\g^5  \g_{\mu} \g_\rs \g_{\nu}\la_1^\s \ri).
    \end{split}
\end{align}
Finally, we get
\bea
\cL_{\chi F\chi}\=-{1\/2}\ve_{AB}  F^\mn \bchi^\rho_A \g_{\mu} \g_\rs \g_{\nu} \chi^\s_B  = -{i\/4}   F^\mn \le( \bla_1^\rho  \g_{\mu} \g_\rs \g_{\nu} \la_1^\s + \bla_2^\rho  \g_{\mu} \g_\rs \g_{\nu}  \la_2^\s \ri).
\eea
\sss{Final Lagrangian}

The final Lagrangian, written in terms of symplectic Majorana spinors, takes the form
\be
\cL_f = {1\/2}\d_{AB} g_\mn \bla_A^\mu\g^\rho D_\rho \la_A^\nu   -{i\/4}   F^\mn \le( \bla_1^\rho  \g_{\mu} \g_\rs \g_{\nu} \la_1^\s + \bla_2^\rho  \g_{\mu} \g_\rs \g_{\nu}  \la_2^\s \ri)  -{1\/2\l}\le( \bla_1^\mu\g_\mn \la_1^\nu - \bla_2^\mu\g_\mn  \la_2^\nu \ri).
\ee
We rescale $\la_A\ra  \sqrt{2}\la_A$ and write explicitly the gauge  covariant derivative. At the end,  the two symplectic Majorana spinors decouple and we can write
\be\label{LagSymsum}
\cL_f = \cL_1+\cL_2,
\ee
where
\bea
\cL_1 \= g_\mn \bla_1^\mu\g^\rho(\n_\rho + i\l^{-1} A_\rho)\la_1^\nu - {i\/2}   F^\mn  \bla_1^\rho  \g_{\mu} \g_\rs \g_{\nu} \la_1^\s  -{1\/\l} \bla_1^\mu\g_\mn \la_1^\nu~,\\
\cL_2 \= g_\mn \bla_2^\mu\g^\rho(\n_\rho - i\l^{-1} A_\rho)\la_2^\nu - {i\/2}   F^\mn  \bla_2^\rho  \g_{\mu} \g_\rs \g_{\nu} \la_2^\s  +{1\/\l} \bla_2^\mu\g_\mn \la_2^\nu~.
\eea
We can now reinterpret this Lagrangian to be in $(0,4)$ signature. To perform the Wick rotation, we define Euclidean gamma matrices
\be
\hg^1,\hg^2,\hg^3,\hg^4,
\ee
where $\hg^i = \g^i$ for $i=1,2,3$ and $\hg^4 = -i\g^0$. They satisfy
\be
(\hg^\mu)^\dg = \hg^\mu~.
\ee
We also take the Hermitian conjugate to be
\be
\bla_A^\mu = (\la_A^\mu)^\dg~.
\ee
Note that it is clear that both flavors give the same contribution because $\mathcal{L}_{1}$ and $\mathcal{L}_{2}$ are equal up to an exchange of $\l\leftrightarrow-\l$ and the four-derivative terms only involve $\l^2$. As a result, it is enough to do the computation for $\cL_1$ and multiply the final heat kernel by two.

\subsection{Ghosts}

In this section, we discuss the contributions from ghosts.

\subsubsection{Faddeev-Popov procedure}\label{sec:FPfermions}

Given the crucial role of a proper treatment of the ghosts, we include here details about the Faddeev-Popov procedure. This was first explained in \cite{Nielsen:1978mp} (see also \cite{Banerjee:2010qc}). The fermionic path integral is schematically of the form
\be
Z = \int D\bar{\psi}_\mu D\psi_\nu\, e^{-S[\Bar{\psi}_\mu,\psi_\nu]}.
\ee
The Faddeev-Popov procedure corresponds to inserting  in the path integral
\be
1 = \int D\e D\bar\e\,\d(\xi-\g^\mu \psi_\mu^{(\e)})\d(\bar\xi-\g^\mu \bar\psi_\mu^{(\e)}) \D_\r{FP}^{-1},
\ee
where the Faddeev-Popov determinant is
\be
\D_\r{FP}= \r{det}\le({ \d(\g^\mu\psi_\mu^{(\e)})\/\d \e}  \ri)\r{det}\le({ \d(\g^\mu\bar\psi_\mu^{(\bar\e)})\/\d \bar\e}  \ri),
\ee
and $\psi_\mu^{(\e)} = \psi_\mu +\cD_\mu \e$ is the infinitesimal transform of $\psi^\mu$ under a supersymmetry transformation. Here, $\xi$ is an arbitrary spinor. We then insert
\be
1 =  {1\/\r{det}\,\sD}\int D\xi D\bar\xi \, \exp\le(- \bar\xi\sD \xi\ri)~.
\ee
As a result, we have
\bea\nt
Z \={\D_\r{FP}^{-1}\/\r{det}\,\sD } \int D\bar{\psi}_\mu D\psi_\nu D\xi D\bar\xi D\e D\bar\e\, \exp\le(- \bar\xi\sD \xi\ri)\d(\xi-\g^\mu \psi_\mu^{(\e)}) \d(\bar\xi-\g^\mu \bar\psi_\mu^{(\e)})  \,e^{-S[\Bar{\psi}_\mu,\psi_\nu]}\\
\={\D_\r{FP}^{-1}\/\r{det}\,\sD } \int D\bar{\psi}_\mu D\psi_\nu D\e D\bar\e\, \exp\le(- \bar\psi_\nu\g^\nu\sD \g^\mu\psi_\mu\ri)  \,e^{-S[\Bar{\psi}_\mu,\psi_\nu]},\-
\eea
where we have performed the integral over $\xi,\bar\xi$ and performed the field redefinition $\psi_\mu\ra\psi_\mu -\cD_\mu\e$. By supersymmetry, the action is invariant under this redefinition. We see that the correct gauge-fixing term appears. Now, we can rewrite the prefactor in terms of $b,c$ ghosts and an additional $d$ ghost
\bea
\D_\r{FP}^{-1} \= \int Db Dc \,\exp\le(- b \g^\mu \cD_\mu c \ri),\-
{1\/\r{det}\,\sD} \=\int Dd D\bar{d} \,\r{exp}\le( -\bar{d}\g^\mu D_\mu d\ri),
\eea
where $b,c,d,\bar{d}$ are spin ${1\/2}$ ghosts with bosonic statistics.

\sss{Ghost Lagrangian}

The ghost Lagrangian
\be
\cL_\r{ghost} = \bar{b}_A\le(\g^\mu D_\mu +{2\/\l}\ri)c_A  +\bar{c}_A\le(\g^\mu D_\mu +{2\/\l}\ri)b_A  + \bar{e}_A \g^\mu D_\mu e_A~,
\ee
where $b_A,c_A,e_A$ are Majorana spinors. We map them to symplectic Majoranas using
\bea
b_1 \= {1\/2}(\b_1+\g^5 \b_2),\qq b_2 = {i\/2}(\b_1 -\g^5 \b_2)~,\\
c_1 \= {1\/2}(\eta_1+\g^5 \eta_2),\qq c_2 = {i\/2}(\eta_1 -\g^5 \eta_2)~,\\
e_1 \= {1\/2}(\e_1+\g^5 \e_2),\qq e_2 = {i\/2}(\e_1 -\g^5 \e_2)~.
\eea
We have
\bea
\bar{b}_1 \g^\mu D_\mu c_1 \= {1\/4}  (\bar\b_1-\bar\b_2\g^5) \g^\mu D_\mu (\eta_1 + \g^5 \eta_2) \-
\= {1\/4}  \le(\bar\b_1\g^\mu D_\mu \eta_1 -\bar\b_2\g^5\g^\mu D_\mu\g^5 \eta_2  + \bar\b_1\g^\mu D_\mu \g^5 \eta_2 -\bar\b_2\g^5\g^\mu D_\mu\eta_1\ri) \-
\= {1\/4}  \le(\bar\b_1\g^\mu D_\mu \eta_1 +\bar\b_2\g^\mu D_\mu\eta_2  + \bar\b_1\g^\mu D_\mu \g^5 \eta_2 -\bar\b_2\g^5\g^\mu D_\mu\eta_1\ri), \-
\eea
and
\bea
\bar{b}_2 \g^\mu D_\mu c_2 \= {1\/4}  (\bar\b_1+\bar\b_2\g^5) \g^\mu D_\mu (\eta_1 - \g^5 \eta_2) \-
\= {1\/4}  \le(\bar\b_1\g^\mu D_\mu \eta_1 +\bar\b_2\g^\mu D_\mu\eta_2  - \bar\b_1\g^\mu D_\mu \g^5 \eta_2 +\bar\b_2\g^5\g^\mu D_\mu\eta_1\ri). \-
\eea
We see that the cross terms cancel upon summation so that
\be
\bar{b}_A \g^\mu D_\mu c_A = {1\/2}(\bar\b_1\g^\mu D_\mu \eta_1 +\bar\b_2\g^\mu D_\mu\eta_2  )~.
\ee
Similarly,
\be
\bar{c}_A \g^\mu D_\mu b_A = {1\/2}(\bar\eta_1\g^\mu D_\mu \b_1 +\bar\eta_2\g^\mu D_\mu\b_2  )~,
\ee
and
\be
\bar{e}_A \g^\mu D_\mu e_A = {1\/2}(\bar\e_1\g^\mu D_\mu \e_1 +\bar\e_2\g^\mu D_\mu\e_2  )~.
\ee
The mass term gives
\bea
\bar{b}_1 c_1 \= {1\/4}  (\bar\b_1-\bar\b_2\g^5) (\eta_1 + \g^5 \eta_2) \-
\={1\/4}  (\bar\b_1 \eta_1-\bar\b_2\eta_2 + \bar\b_1 \g^5 \eta_2-\bar\b_2\g^5\eta_1),
\eea
and
\bea
\bar{b}_2 c_2 \= {1\/4}  (\bar\b_1+\bar\b_2\g^5) (\eta_1 - \g^5 \eta_2) \-
\={1\/4}  (\bar\b_1 \eta_1-\bar\b_2\eta_2 -\bar\b_1 \g^5 \eta_2+\bar\b_2\g^5\eta_1),
\eea
so that
\be
\bar{b}_A c_A = {1\/2}(\bar\b_1 \eta_1-\bar\b_2\eta_2 ),
\ee
and
\be
\bar{c}_A b_A = {1\/2}(\bar\eta_1 \b_1-\bar\eta_2\b_2 ).
\ee
We now rescale all ghosts by a factor $\sqrt{2}$. Finally, we obtain
\bea
\cL_\r{ghosts} \=  \bar\b_1 \le( \g^\mu D_\mu +  {2\/\l}\ri)\eta_1+ \bar\eta_1 \le( \g^\mu D_\mu +  {2\/\l}\ri)\b_1 \-
&&+ \bar\b_2 \le( \g^\mu D_\mu -  {2\/\l}\ri)\eta_2  + \bar\eta_2 \le( \g^\mu D_\mu -  {2\/\l}\ri)\b_2  \-
&&+ \bar\e_1\g^\mu D_\mu \e_1 +\bar\e_2\g^\mu D_\mu\e_2  
\eea
We now Wick rotate and interpret the Dirac conjugates as Hermitian conjugates in Euclidean signature
\be
\bar\b_1 = \eta_1^\dg,\qq \bar\b_2=\eta_2^\dg,\qq \bar\e_1 = \e_1^\dg~.
\ee
The above choice of Hermitian conjugate makes the kinetic term diagonal and suitable for the heat kernel computation. We can also use the more natural choice $\bar\b_1 = \b_1^\dg,  \bar\b_2 = \b_2^\dg$ and make the kinetic diagonal by a simple field redefinition.

\ss{Result}

The heat kernel can now be computed using the algorithm described in section \ref{app:algo}. The gravitini contribution is computed using the Lagrangian \eqref{LagSymsum}. The result is
\be
(4\pi)^2 a_4(x) = {139\/90} E_4-{32\/15}W^2 -{2\/9}R^2 + {8\/9} R\FF~.
\ee
In total we have one massless pair of ghosts and two massive pairs. A massless pair of ghosts gives the contribution
\be
(4\pi)^2a_4(x) = {11\/360}E_4 - {1\/20}W^2 - {1\/18} R \FF,
\ee
and each of the massive pair gives
\be
(4\pi)^2a_4(x) = {11\/360}E_4 - {1\/20}W^2 +{1\/9}R^2 - {1\/18} R \FF.
\ee 
In all these formulas, we have already included the minus sign due to the opposite statistics of ghosts. Finally, the total fermionic contribution gives
\be\label{appResGravitini}
(4\pi)^2a_4(x) = {589\/360}E_4 -{137\/60}W^2 + {13\/18} R F_\mn F^\mn~,
\ee
as reported in \eqref{ResGravitini}.

\sss{A different gauge-fixing term}

Another possible gauge-fixing term is to take
\bea
\cL_\r{g.f.}\=- {1\/4} (\bpsi_{\mu} \g^\mu) (\g^\nu D_\nu-m) (\g^\rho \psi_\rho)\-
\=- {1\/4} \d_{AB}(\bpsi_{\mu}^A \g^\mu) (\g^\nu D_\nu-m) (\g^\rho \psi_\rho^B)~.
\eea
This is natural because with $m={2\/\l}$, the three pairs of ghosts become identical. Of course the final result should not depend on this choice. This adds the term
\bea
\cL_\r{new} \= {1\/4} m \bar\psi_\mu \g^\mu \g^\nu \psi_\nu\-
\=  {1\/2} m \le(\bchi_\mu-{1\/2}\bchi_\a \g^\a \g_\mu\ri) \g^\mu \g^\nu \le( \chi_\nu-{1\/2} \g_\nu \g^\b\chi_\b\ri)\-
\= {1\/2} m \,\d_{AB}\bchi_\mu^A \g^\mu \g^\nu\chi_\nu^B.
\eea
Using that $g_\mn = \g^\mu \g^\nu-\g^\mn$, we can simplify the Majorana Lagrangian with the choice $m  ={2\/\l}$ so that it takes the form
\bea
\cL_f \=  \d_{AB} g^{\mn}\bchi_\mu^A\g^{\rho} D_\rho\chi_\nu^B -{1\/2}\ve_{AB}  F^\mn \bchi_\rho^A \g_{\mu} \g^\rs \g_{\nu} \chi_\s^B +{1\/\l}\d_{AB} \bchi_\mu^A g^\mn \chi_\nu^B~.
\eea
After converting to symplectic Majoranas and performing the computation, this gives the gravitini contribution
\be
(4\pi)^2 a_4(x) = {139\/90} E_4-{32\/15}W^2 -{1\/3}R^2 + {8\/9} R\FF ~.
\ee
The ghost contribution is also modified. Indeed, the Lagrangian of the $e$-ghost is determined by the gauge-fixing Lagrangian and hence acquires the same mass of the $b,c$ ghosts. So we end up with three identical pairs of charged ghosts for a total ghost contribution of
\be
(4\pi)^2 a_4(x) = -{11\/120} E_4-{3\/20}W^2 +{1\/3}R^2 -{1\/6} R\FF ~.
\ee
The total contribution is then 
\be
(4\pi)^2a_4(x) = {589\/360}E_4 -{137\/60}W^2 + {13\/18} R F_\mn F^\mn~,
\ee
which, as expected, is the same as \eqref{appResGravitini}.

\section{Renormalization of the couplings}\label{app:renorm}

Our focus in this paper has been on the Seeley-DeWitt coefficient $a_4$ which is responsible for the local contribution to the logarithmic corrections. The other Seeley-DeWitt coefficients $a_0$ and $a_2$  capture the one-loop renormalization of the couplings. Indeed, the effective Euclidean action takes the form
\be
S= S_\r{classical} + S_\text{1-loop} +\dots,
\ee
where the one-loop correction is
\be
S_\text{1-loop} = -{1\/2} \int_\e^{+\infty} {ds\/s} \int d^4 x\sqrt{g} \,K(x,s),\qq K(x,s)= s^{-2}a_0(x) + s^{-1}a_2(x) +\dots,
\ee
which gives
\be
S_\text{1-loop} =\int d^4 x\sqrt{g} \le( -{1\/4\e^2}a_0(x) - {1\/2 \e}a_2(x) +\dots\ri).
\ee
The coefficient $a_0$ is a constant while $a_2$ is a general two-derivative term
\be
(4\pi)^2 a_2(x) = d_1 R + d_2 F_\mn F^\mn~.
\ee
Thus, we get
\be
S_\text{1-loop} ={1\/16\pi^2}\int d^4 x\sqrt{g} \le( -{1\/4\e^2}a_0 - {1\/2 \e} (d_1 R + d_2 F_\mn F^\mn)+\dots\ri).
\ee
From this expression, we can see that $a_0$, $d_1$ and $d_2$ are respectively renormalizations of the cosmological constant, Newton's constant and the electric charge. Here $\e$ represents a UV cutoff. We can compute $a_0$ and $a_2$ using the formulas \cite{Vassilevich:2003xt}
\bea
(4\pi)^2 a_0 \= \r{Tr}\,1 ~,\\
(4\pi)^2 a_2 \= \r{Tr}\le( E + {1\/6}R\ri)~.
\eea
This allows us to compute the coefficients $a_0$, $d_1$, $d_2$ for the theories considered in this paper. The results are summarized in Table \ref{tab:a0a2} below. 

The renormalization of the cosmological constant, $a_0$, depends only on the number of fields and is as in flat space. However, our computations for the renormalization of Newton's constant, $d_1$, generalize previous flat space discussions in, for example,  \cite{Susskind:1994sm,Kabat:1995eq}. To compare with those papers we note that our $\epsilon$ above has dimensions of $[L]^2$. More precisely, considering a massless scalar in AdS, $\Delta=3$, leads to the same contribution as the one presented in \cite{Kabat:1995eq}: $d_1=1/6$. Similarly, the massless Dirac fermion ($\Delta=3/2$) leads to $d_1=1/3$ and the free vector to $d_1=-2/3$ which coincide with \cite{Kabat:1995eq}.

\begin{table}[h]
\centering\arraycolsep=4pt\def\arraystretch{1.4}
\begin{tabular}{|c|c|c|c|} 
\hline
Multiplet  & $a_0$  & $d_1$  & $d_2$   \cr
\hline
\hline
{\rm Free scalar} & $1$ & ${1\/12}(2-\D(\D-3))$ &  $0$ \cr
\hline
{\rm Free Dirac fermion} & $-4$ & $ -{1\/12}(5+4 \D(\D-3))$ & $0$  \cr
\hline 
{\rm Free vector} & $2 $ & $ -{2\/3}$ & $0$  \cr
\hline
{\rm Free gravitino} & $-2 $ & $  - {1\/2} $  & $0$  \cr
\hline
{\rm Einstein-Maxwell}  & $4 $ & $ -{10\/3}$ & $6$  \cr
\hline 
{\rm $\cN=2$ gravitini}  & $ -4 $ & $7$ &$ -8 $  \cr
\hline
{\rm $\cN=2$ gravity multiplet} &  $0  $ & ${11\/3}$  & $-2 $ \cr
\hline
\end{tabular} 
\caption{Seeley-DeWitt coefficients $a_0$ and $a_2$ for the theories studied in this paper}\label{tab:a0a2}
\end{table}

\section{Holographic renormalization and the Gauss-Bonnet-Chern theorem }\label{app:Euler}

Since the local contribution is given by an integral over the Euclidean spacetime, the  result for the logarithmic correction is sensitive to the choice of regularization procedure. In this work, we have used  holographic renormalization to regulate these integrals. This is natural because the logarithmic correction can be viewed as a term in the effective bulk action. We have found that this prescription always gives a finite and unambiguous result.

For the Euler density, a natural counterterm is provided by the Gauss-Bonnet-Chern theorem
\be
{1\/32\pi^2}\int_\mathcal{M} d^d x\sqrt{g}\, E_4 + {1\/32\pi^2} B  = \chi~,
\ee
where $B$ is is the boundary term found by Chern in \cite{10.2307/1969203} and $\chi$ is the Euler characteristic of spacetime, which is an integer. Our regularization prescription gives precisely the same counterterm. Indeed, we find that
\be
\lim_{r_c\to+\infty} \le[ {1\/32\pi^2}\int_\mathcal{M} d^d x\sqrt{g}\, E_4 +\int_{\p \mathcal{M}} d^3 y \sqrt{h}\, (c_1+ c_2 \cR) \ri]= \chi~.
\ee
where $c_1$ and $c_2$ are chosen to cancel the $r_c^3$ and $r_c$ divergences. This works for all the geometries considered in this paper. Note that a naive regularization procedure where we simply remove the divergent term would not give this result. In fact, it would lead to a non-topological result, depending on the black hole parameters. The holographic counterterm gives a finite contribution, necessary to obtain a topological result which is the same as the one appearing in the Gauss-Bonnet-Chern theorem. This gives us confidence that our regularization procedure is physically sensible.  In this appendix, we show explicitly the matching of the two counterterms for the AdS-Schwarzschild black hole.

\subsection{Chern's boundary term}

To illustrate the above points, we consider the application of the Gauss-Bonnet-Chern theorem \cite{10.2307/1969203} (see section 8 of \cite{Eguchi:1980jx} for a review and \cite{Larsen:2015aia} for a simple AdS application) to the AdS-Schwarzschild solution. The theorem takes the form
\be
\chi = {1\/32\pi^2} \int d^4x \sqrt{g}\,E_4 + {1\/32\pi^2} B,
\ee
where $B$ is the boundary term \cite{10.2307/1969203}
\begin{align}
  B\equiv -2\int \epsilon_{abcd}\theta^a_b\mathcal{R}^c_d+\frac{4}{3}\int\epsilon_{abcd}\theta^a_b\theta^c_e\theta^e_d~.
\end{align} 
where $\theta_{ab}$ is the second fundamental form, and $\mathcal{R}$ is the Riemann curvature tensor at the boundary. For AdS-Schwarzschild, the bulk contribution is
\bea
\label{eqn: Eulbulk}
{1\/32\pi^2}\int d^4 x\sqrt{g} \,E_4 \=        \int^{r_c}_{r_+}r^2dr\int d\Omega^2 \int_0^{\beta}d\tau ~\left(\frac{24}{\LFull^4}+\frac{48m^2}{r^6}\right)\-
\={\b\/\pi}\left[2m^2\left(\frac{1}{r_+^3}-\frac{1}{r_c^3}\right)+\frac{1}{\LFull^4}\left(r_c^3-r_+^3\right) \right]~,
\eea
where $r_+$ is the horizon radius and we have introduced a cutoff $r=r_c$ that should be taken to infinity at the end. For the boundary term, we first compute the fundamental form
\begin{align}
    \theta_{01}&=\frac{m\LFull^2+r_c^3}{r_c^2\LFull^2}\,d\tau~,\\
    \theta_{12}&=-\sqrt{\frac{r_c^3+r_c\LFull^2-2m\LFull^2}{r_c\LFull^2}}\,d\theta~,\\
    \theta_{12}&=-\sqrt{\frac{r_c^3+r_c\LFull^2-2m\LFull^2}{r_c\LFull^2}}\,\sin\,\theta\, d\phi~.
\end{align}
Note that the vielbeins are in general non-trivial linear combination of $dx^\mu$, but the fundamental forms here are simple monomials of $dx^\mu$. We can read out the coordinate component $\mathcal{R}^a_{~b\mu\nu}$ on the slice $r=r_c$ using that 
\begin{align}
    \mathcal{R}^a_b=\mathcal{R}^a_{~bcd}e^c\wedge e^d=\mathcal{R}^a_{b\mu\nu}dx^\mu\wedge dx^\nu~.
\end{align}
This gives
\begin{align}
    \mathcal{R}^0_{~2\tau\theta}=&-\frac{m\LFull^2+r_c^3}{\LFull^2r_c^2}\sqrt{\frac{r_c^3+r_c\LFull^2-2m\LFull^2}{r_c\LFull^2}}~,\\
    \mathcal{R}^0_{~3\tau\phi}=&-\frac{m\LFull^2+r_c^3}{\LFull^2r_c^2}\sqrt{\frac{r_c^3+r_c\LFull^2-2m\LFull^2}{r_c\LFull^2}}\sin\theta~,\\
    \mathcal{R}^0_{~3\theta\phi}=&\frac{2m\LFull^2-r_c^3}{r_c\LFull^2}~.
\end{align}
The boundary contribution is then
\bea\label{eqn: Eulbdry}
        B\= 8(4\pi\beta)\frac{(m\LFull^2+r_c^3)(2m\LFull^2-r_c^3)}{\LFull^4r_c^3}\-
        \= 4\pi\beta\left[-16m^2\frac{1}{r_c^3}+8m\frac{1}{\LFull^2}-\frac{8}{\LFull^4}r_c^3\right]~.
\eea
The total contribution of the Euler characteristic is 
\begin{equation}
    4\pi\beta\left[16m^2\frac{1}{r_+^3}+8m\frac{1}{\LFull^2}-\frac{8}{\LFull^4}r_+^3 \right]~.
\end{equation}
Using the fact that the periodicity of $\tau$ is the inverse Hawking temperature \eqref{eqn: AdSSST} and that $f(r_+)=0$ gives $2m=r_+\left(1+\frac{r_+^2}{\LFull^2}\right)$, we obtain 
\begin{equation}
\chi =   {\b\/\pi}\left[2m^2\frac{1}{r_+^3}+m\frac{1}{\LFull^2}-\frac{1}{\LFull^3}r_+^3 \right]=2 ~,
\end{equation}
which is the correct result for the Euler characteristic of a black hole.

\subsection{Holographic renormalization}

We now show that our regularization procedure, using the prescription of holographic renormalization, gives the same boundary term. This is already clear from the fact that the regularized Euler integral gives the correct $\chi$ but here we directly compare the boundary terms. The boundary geometry at $r=r_c$ is
\begin{equation}
    ds^2=\left(1+\frac{r_c^2}{\LFull^2}-\frac{2m}{r_c}\right)dt^2+r_c^2d\theta^2+r_c^2\,\sin^2\theta \,d\phi^2,
\end{equation}
and the Ricci scalar on the boundary is $\mathcal{R}=\frac{2}{r_c^2}$. The holographic counterterms are given by
\bea\label{eqn: Eulcounter}
        a^{\text{CT}}\= \int d^3 y \sqrt{h} (c_1+c_2\cR) \-
    \=\int_0^\beta dt\int^\pi_0 \sin\theta d\theta \int^{2\pi}_0 d\phi \sqrt{1+\frac{r_c^2}{\LFull^2}-\frac{2m}{r_c}}r_c^2\left(c_1+c_2\mathcal{R}\right)\\
        \=4 \pi  \beta  \left(c_1 r_c^2+2 c_2\right) \sqrt{\frac{r_c^2}{\LFull^2}-\frac{2 m}{r_c}+1}\\
        \=4\pi\beta\left[\frac{ c_1 }{\LFull}r_c^3 + \frac{\left(c_1 \LFull^2+4 c_2\right)}{2\LFull}r_c - c_1m \LFull\right]+\mathcal{O}(r_c^{-1}).
\eea
In order to remove the divergence in \eqref{eqn: Eulbulk}, we demand
\begin{equation}
    c_1=-\frac{8}{\LFull^3}~, \quad c_2=\frac{2}{\LFull}.
\end{equation}
Plugging back in \eqref{eqn: Eulcounter}, we see that the counterterm is exactly equal to Chern's boundary term. Note that this is non-trivial because the renormalization introduces a finite correction. It would be interesting to have a more geometrical understanding of this identification.

\section{Vanishing of boundary terms}
\label{subsubsec: Vbdry}

The proper application of the heat kernel expansion in AdS following \cite{Vassilevich:2003xt} requires the addition of boundary terms. They come from the fact that the computation needs to be done on a regularized geometry defined by a cutoff $r<r_c$. In this appendix, we show that these boundary terms vanish and thus can be ignored. One source of boundary terms is the fact that $a_4(x)$ actually contains total derivatives, highlighted below as
\be
\label{eqn: a4x}
a_4(x) = \dots + {1\/30} (5\Box E + \Box R )~.
\ee
This gives a contribution
\be\label{ClocalBdy1}
C_\r{local}= \dots+ {1\/(4\pi)^2 }  {1\/30} \int d^3 y\sqrt{h}\,n^\mu \n_\mu(5  \Tr\,E + R \,\r{Tr}\,1 )~.
\ee
Using the fact that $\r{Tr}\,E$ is a linear combination of two-derivative terms, it can be generally written as
 \be\label{TrEgen}
\Tr\,E = \a_1 R + \a_2 \FF,
 \ee
 for some coefficient $\a_1$ and $\a_2$. We can then compute the contribution \eqref{ClocalBdy1} on our background. It is of order $O(r_c^{-1})$ and hence vanishes in the limit $r_c\to +\infty$.

Another contribution comes from the formula of $a_4(x)$ on a manifold with boundaries, which include additional boundary terms. This can be written as an additional boundary contribution to $C$
\be
C_\r{bdy} =\int d^3 y\sqrt{h}\, a_4^\p(y),
\ee
where
\be
a_4^\p = B_1 \,\r{Tr}\,1 +B_2\,\r{Tr}\,E,
\ee
and $B_1$ and $B_2$ are geometric invariants of the boundary depending on both intrinsic and extrinsic data
\begin{align}
    \begin{split}
        B_1 &=  {1\/360}\le[ \vphantom{\frac12}  24 K_{:bb} + 20 R K  + 4 R_{anan} K -12 R_{anbn}K_{ab} + 4 R_{abcb} K_{ac} + 480 S^2 K + 480 S^3 \ri. 
        \\ &
        \le. \qquad \qquad +{1\/21} \le[(280\Pi_+ + 40\Pi_-) K^3 + (168\Pi_+ - 264 \Pi_-) K_{ab}K_{ab}K_{cc} \ri]  + 120 S_{:aa}\vphantom{\frac12} \ri.
        \\ & \qquad \qquad \left. + {1\/21} (224\Pi_+ +320\Pi_-) K_{ab}K_{bc}K_{ac} + 120 SR+ 144 S K^2  +48 S K_{ab}K_{ab}\ri]~,
        \\
        B_2 &=  {1\/3}( K + 6 S)~,
    \end{split}
\end{align}
where $K_{ab}$ is the extrinsic curvature of the boundary. Here $\Pi_\pm$ and $S$ capture the choice of boundary conditions for the fields at infinity. For normalizable boundary conditions in AdS$_4$, it can be checked that they are constants.

We can evaluate this term on our background using the general expression \eqref{TrEgen} for $\r{Tr}\,E$, which diverges and hence, we use the same regularization prescription as in holographic renormalization. Once the dust settles, we find that this contribution vanishes. Hence, no boundary term of this type gives a contribution to the logarithmic correction.

\section{Black hole curvature invariants}

In the main text, we emphasize the role of universality considering by the Euler characteristic and the square of the Weyl tensor. It is also common to express the curvature invariants in terms of $R_{\mu\nu\rho\sigma}R^{\mu\nu\rho\sigma}$, $R_{\mu\nu}R^{\mu\nu}$ and $R^2$. In this appendix, we explicitly write out the curvature invariants for the Kerr-Newman-AdS black hole studied in section \ref{sec:KN} and show how they are related to their flat space counterparts. 

The curvature invariants are
\bea \nt
    R_{\mu\nu\alpha\beta}R^{\mu\nu\alpha\beta}\=
    \frac{24}{\LFull^4} + \frac{8}{4 (r^2+a^2 \cos ^2\t)^6}\left[-24 m^2 \left(a^2 \cos ^2\theta +r^2\right)^3+ 192 r^4 \left((q_e^2+q_m^2)-2 m r\right)^2
    \right.\-
    && -192 r^2 \left((q_e^2+q_m^2)-3 m r\right) \left((q_e^2+q_m^2)-2 m r\right) \left(a^2 \cos ^2\theta +r^2\right)
    \-
    &&  \left.+4 \left((q_e^2+q_m^2)-6 m r\right) \left(7(q_e^2+q_m^2)-18 m r\right) \left(a^2 \cos ^2\theta +r^2\right)^2 \right]
    \-
    \=\frac{24}{\LFull^4} + \tilde{R}_{\mu\nu\alpha\beta} \tilde{R}^{\mu\nu\alpha\beta}~,
    \\\label{KN nonextremal curvature invariants}
    R_{\mu\nu}R^{\mu\nu}\= \frac{36}{\LFull^4} + \frac{4(q_e^2+q_m^2)^2}{\left(r^2+a^2 \cos^2 \theta\right)^4}
    \-
    \= \frac{36}{\LFull^4} + \tilde{R}_{\mu\nu}\tilde{R}^{\mu\nu}~,
    \-
    R^2 \= \frac{144}{\LFull^4}~,
    \-
    F_\mn F^\mn \= -\frac{2\big(q_e^2-q_m^2\big)\left(r^4-6 a^2 r^2 \cos ^2\theta +a^4 \cos ^4\theta\right)+16q_eq_mra~\cos \theta\big(r^2-a^2\cos^2\theta\big)}{\left(r^2 +a^2 \cos ^2\theta\right)^4}~.
\eea
Each of the invariants are a sum of two terms. The first term is proportional to $\ell^{-4}$ and the second term which has no $\ell$ dependence agrees with the analogous invariants $\Tilde{R}_{\mu\nu\alpha\beta}\Tilde{R}^{\mu\nu\alpha\beta}$ and $\Tilde{R}_{\mu\nu}\Tilde{R}^{\mu\nu}$ of asymptotically flat Kerr-Newman black holes. Thus, taking $\ell \to \infty$, we smoothly recover the same expressions for the invariants in \cite{Bhattacharyya:2012wz,Bhattacharyya:2012ye}. The invariants for Reissner-Nordstr\"om and Schwarzschild can be obtained by specializing the parameters. We can find the expressions for $E_4$ using \eqref{E4 and W2 definition}:
\begin{equation}
    \begin{split}
        E_4=\frac{24}{\ell^4}+\frac{8}{\left(r^2+a^2 \cos^2\theta\right)^6}\Big(&6 m^2 \left(r^6-15 a^2 r^4 \cos^2\theta+15 a^4 r^2 \cos^4\theta-a^6 \cos^6\theta\right)\\
          &-12 m r (q^2_e+q_m^2)  \left(r^4-10 a^2 r^2 \cos^2\theta+5 a^4 \cos^4\theta\right)\\
          &+(q^2_e+q_m^2)^2 \left(5 r^4-38 a^2 r^2 \cos^2\theta+5 a^4 \cos^4\theta\right)\Big)
    \end{split}
\end{equation}
Moreover, upon integration, the $\ell^{-4}$ term diverges because of the divergent AdS$_4$ volume, and we tame this divergence by holographic renormalisation. On the other hand, the second term is finite and does not require holographic renormalization. The integrated curvature invariants after proper renormalization take the form of 
\begin{align}
    \begin{split}
    &\hspace{-1cm}{1\/(4\pi)^2} \int d^4 x\sqrt{g}\, R_{\mu\nu\rho\sigma} R^{\mu\nu\rho\sigma}\\
    \label{KN nonextremal to extremal curvature invariants}
    =&-\frac{3 \left(a^2-r_{+}^2\right) \left(a^2+r_{+}^2\right)^2 \left(a^2 \left(r_{+}^2-\LFull^2\right)+r_{+}^2 \left(\LFull^2+3 r_{+}^2\right)\right) \arctan\left(\frac{a}{r_{+}}\right)}{a^5 \LFull^2 \Xi  r_{+}^3}
    \\& +  \left(a^2 \left(\LFull^2-r_{+}^2\right)-r_{+}^2 \left(\LFull^2+3 r_{+}^2\right)\right)^2 
    \\&  \times \frac{\beta\left(3 a^5 r_{+}+2 a^3 r_{+}^3+3 \left(a^2-r_{+}^2\right) \left(a^2+r_{+}^2\right)^2 \arctan\left(\frac{a}{r_{+}}\right)+3 a r_{+}^5\right)}{8 \pi  a^5 \LFull^4 \Xi  r_{+}^4 \left(a^2+r_{+}^2\right)}
    \\& +\frac{2 \pi  \left(a^2+r_{+}^2\right) \left(3 a^5 r_{+}+2 a^3 r_{+}^3+3 \left(a^2-r_{+}^2\right) \left(a^2+r_{+}^2\right)^2 \arctan\left(\frac{a}{r_{+}}\right)+3 a r_{+}^5\right)}{a^5 \beta  \Xi  r_{+}^2}
    \\& +\frac{a r_{+} \left(a^6 \left(3 \LFull^2-7 r_{+}^2\right)+a^4 \left(3 \LFull^2 r_{+}^2-11 r_{+}^4\right)+a^2 r_{+}^4 \left(\LFull^2-9 r_{+}^2\right)-3 r_{+}^6 \left(\LFull^2+3 r_{+}^2\right)\right)}{a^5 \LFull^2 \Xi  r_{+}^3}~,\\
    &\hspace{-1cm}{1\/(4\pi)^2}\int d^4 x\sqrt{g} \,R_{\mu\nu} R^{\mu\nu}\\
    =& \frac{96 \beta  r_{+}^4 \left(\beta  r_{+} \left(a^2+\LFull^2+2 r_{+}^2\right)-2 \pi  \LFull^2 \left(a^2+r_{+}^2\right)\right)-96 \beta ^2 r_{+}^5 \left(a^2+r_{+}^2\right)}{32 \pi  \beta  \LFull^4 \Xi  r_{+}^4}
    \\& + \left(a^2 \left(\beta  r_{+}^2-\LFull^2 (\beta +4 \pi r_{+})\right)+\LFull^2 r_{+}^2 (\beta -4 \pi  r_{+})+3 \beta  r_{+}^4\right)^2
    \\&  \times \frac{\left(3 a^5 r_{+}+2 a^3 r_{+}^3+3 (a-r_{+}) (a+r_{+}) \left(a^2+r_{+}^2\right)^2 \arctan\left(\frac{a}{r_{+}}\right)+3 a r_{+}^5\right)}{32 \pi  \beta  \LFull^4 \Xi  r_{+}^4 \left(a^5 \left(a^2+r_{+}^2\right)\right)}~,
    \\
    &{1\/(4\pi)^2}\int d^4 x\sqrt{g}\, R^2 = \frac{12 \left(\beta  r_{+} \left(\LFull^2+r_{+}^2\right)-2 \pi  \LFull^2 \left(a^2+r_{+}^2\right)\right)}{\pi  \LFull^4 \Xi }~,
    \\
    &{1\/(4\pi)^2}\int d^4 x \sqrt{g} F_\mn F^\mn  = 2r_+^2- \frac{\beta  r_+ \left(3r_+^4+(a^2+\LFull^2)r_+^2-a^2\LFull^2-2\LFull^2q_m^2\right)}{2\pi\LFull^2 \left(a^2+r_+^2\right)}~.
    \end{split}
\end{align}

\bibliographystyle{utphys}

\bibliography{references}

\end{document}